\documentclass[12pt]{article}
\usepackage{amsmath}
\usepackage{graphicx}
\usepackage{enumerate}
\usepackage{natbib}
\usepackage{url} 

\usepackage{color}
\usepackage{amsmath}
\usepackage{amssymb}
\usepackage{amsthm}
\usepackage{mathrsfs}
\usepackage{bm}

\usepackage{multirow}
\usepackage{array}
\usepackage{booktabs}
\usepackage[font=small,labelfont=bf]{caption}

\usepackage{graphicx}
\usepackage{psfrag}
\usepackage{epsf}
\usepackage{float}

\usepackage{url} 

\usepackage{setspace}
\newtheorem{theorem}{Theorem}

\newtheoremstyle{exampstyle}
{\topsep} 
{\topsep} 
{} 
{} 
{\bfseries} 
{.} 
{.5em} 
{} 

\theoremstyle{exampstyle}

\usepackage{mathtools}
\newcommand{\expect}{\operatorname{E^{\pi}}\expectarg}
\DeclarePairedDelimiterX{\expectarg}[1]{[}{]}{%
  \ifnum\currentgrouptype=16 \else\begingroup\fi
  \activatebar#1
  \ifnum\currentgrouptype=16 \else\endgroup\fi
}
\newcommand{\innermid}{\nonscript\;\delimsize\vert\nonscript\;}
\newcommand{\activatebar}{%
  \begingroup\lccode`\~=`\|
  \lowercase{\endgroup\let~}\innermid 
  \mathcode`|=\string"8000
}

\newcommand{\cmark}{
\raisebox{0.6ex}{\scalebox{0.7}{$\sqrt{}$}}
}
\newcommand{\xmark}{ 
\scalebox{0.85}[1]{$\times$}
}

\newcommand{\blind}{1}

\usepackage{xcite}
\usepackage{xr}
\usepackage{xr-hyper}
\makeatletter
\newcommand*{\addFileDependency}[1]{
\typeout{(#1)}
\@addtofilelist{#1}
\IfFileExists{#1}{}{\typeout{No file #1.}}
}
\makeatother

\newcommand*{\myexternaldocument}[1]{%
\externaldocument{#1}%
\addFileDependency{#1.tex}%
\addFileDependency{#1.aux}%
}

\myexternaldocument{supp_unblinded}

\begin{document}

\def\spacingset#1{\renewcommand{\baselinestretch}%
{#1}\small\normalsize} \spacingset{1}


\if1\blind
{
  \title{\bf Off-Policy Evaluation with Irregularly-Spaced, Outcome-Dependent Observation Times }
  \author{Xin Chen
    \hspace{.2cm}\\
    School of Statistics and Mathematics,\\
Shanghai Lixin University of Accounting and Finance\\
    Wenbin Lu \\
    Department of Statistics, North Carolina State University\\
    Shu Yang \\
    Department of Statistics, North Carolina State University
    \\
     and \\
       Dipankar Bandyopadhyay\\
    Department of Biostatistics, Virginia Commonwealth University
    }
  \maketitle
} \fi

\if0\blind
{
  \bigskip
  \bigskip
  \bigskip
  \begin{center}
    {\LARGE\bf Off-Policy Evaluation with Irregularly-Spaced, Outcome-Dependent Observation Times }
\end{center}
  \medskip
} \fi

\bigskip

\begin{abstract} 
While the classic off-policy evaluation (OPE) literature commonly assumes decision time points to be evenly spaced for simplicity, in many real-world scenarios, such as those involving user-initiated visits,  decisions are made at irregularly-spaced and potentially outcome-dependent time points.  For a more principled evaluation of the dynamic policies,  this paper constructs a novel OPE framework,  which concerns not only the state-action process but also an observation process dictating the time points at which decisions are made. The framework is closely connected to the Markov decision process in computer science and with the renewal process in the statistical literature. Within the framework, two distinct value functions,  derived from cumulative reward and integrated reward respectively, are considered,  and statistical inference for each value function is developed under revised Markov and time-homogeneous assumptions. The validity of the proposed method is further supported by theoretical results, simulation studies, and a real-world application from electronic health records (EHR) evaluating periodontal disease treatments. 
\end{abstract}

\noindent%
\vfill

\newpage
\spacingset{1.9} 

\section{Introduction}
\label{sec:intro}
A dynamic policy 
is a sequence of decision rules guiding an agent to select time-varying actions based on updated information at each decision stage throughout its interaction with the environment. 
The frameworks for analyzing dynamic policies can be broadly categorized into two settings: (a) finite-horizon and (b) long/infinite-horizon, based on how they handle the number of decision stages. 
Traditional research in dynamic policies predominantly focused on the finite-horizon settings with a limited number of decision stages  \citep{ Murphy2001marginal, Murphy2003Optimal, Robins2004optimal, Zhang2013robust, Zhao2015new},  
where the potential outcome framework is commonly employed to describe the target of interest, 
and the classical causal inference assumptions, such as unconfoundedness and consistency, are sufficient to guarantee target estimability. 
Applications of these methods appear in multi-stage treatment settings for cancer, 
HIV infection,  
depression,   
and other chronic conditions.
By contrast, infinite-horizon dynamic policies involve a relatively large number of decision stages.  For example, consider our motivating periodontal disease (PD) treatment data extracted from the HealthPartners (HP) electronic health records (EHR), where patients are generally recommended for a regular check-up every 6 months \citep{Clarkson2020risk}. 
In this dataset, half of the patients visit their dentists for dental treatments more than 8 times, with the maximum value of treatment stages reaching 31 during the 8-year follow-up period. 
Similar decision policies also arise in the long-haul treatments of chronic diseases, 
intensively interventions provided during the stay in the ICU,
and other sequential decision-making applications, 
such as in mobile health, gaming, robotics, and ride-sharing. 
In such cases, 
the backward induction algorithms used in the finite horizon settings may suffer from the curse of the horizon 
\citep{ 
Uehara2022review}. 
To mitigate this, off-policy evaluation methods developed within the reinforcement learning (RL) framework 
\citep{ 
Sutton2018reinforcement} are preferred.
Specifically, instead of directly estimating the target of interest, such as the value of a dynamic policy,  researchers reformulate and identify the value function through an action-value function. Under time-homogeneous and Markov assumptions, the action-value function can be estimated by solving the Bellman Equation,  followed by the estimation of the quantity of interest. 
Recently, interest has shifted to 
dynamic policies in long- or infinite-horizon settings, 
and statistical inferences for off-policy evaluations have also been developed \citep{Ertefaie2018constructing, Luckett2019estimating, Kallus2020double,
Shi2021statistical, Shi2022statistically}.  

This paper considers the evaluation of dynamic policies in an infinite horizon setting. 
As a variation to the aforementioned literature, our framework includes not only the sequential decision-making process but also an observation process 
which dictates the time points $\{ T_k\}_{ k\geq 1}$ at which decisions are to be made.
In the existing literature, it is common to assume that decisions are made at regular, pre-assigned time points $T_k = k$ for simplicity.
However, in many real-world scenarios,  such as those involving user-initiated visits, 
the observation process is often irregularly spaced and may depend on both the decision-making process and the outcomes \citep{Yang2019semi, Yang2021semi}. 
For example, in the PD treatment study,  the gap times between the dental revisits, even for the same patient, can range from 3 to 24 months.
Additionally, 
the length of recall intervals is correlated with not only the actions made at the previous visit (such as the clinic visit recommendations made by dentists) 
but also with the outcomes (such as the patient's oral health status after treatment). 
The necessity of including such an irregularly spaced and outcome-dependent observation process in the analysis of dynamic policies arises for at least two reasons: 
(a) the gap times between decision stages might be useful for the decision-maker in choosing actions and, therefore, could be included as the inputs of dynamic policies along with the state; 
and more importantly, 
(b) replacing the regular decision stage number $k$ with its specific decision time point $T_k$  allows for more sensible and principled  evaluations of dynamic policies (as will be discussed throughout the paper), 
regardless of whether the dynamic policy involves gap times.

Within the realm of developing dynamic policies with irregularly spaced and outcome-dependent observation times, this work advances the field in multiple ways. 
Our first contribution is offering two options for evaluating dynamic policies: one based on cumulative reward and the other on integrated reward. 
Following the OPE approaches with regular decision points,  
a simple way to define the value of a dynamic policy is to use the expected discounted cumulative reward, 
assuming all actions are taken according to the policy.
However, 
because the value of a cumulative sum 
is typically highly correlated with the number of accumulators (especially when rewards are consistently non-negative or non-positive), 
the policy value tends to increase or decrease with the number of decision stages.
This observation motivates us to consider a more principled evaluation, where the value of dynamic policies could be independent of the number of observed decision stages.
Therefore, 
by treating the rewards observed at $\{ T_k\}_{k \geq 1}$ 
as discrete observations from 
a continuous reward process potentially available throughout the entire follow-up period, 
we define an alternative value function for the dynamic policies based on the integration of the reward function over the full follow-up time.
The value function, when derived from integrated reward, differs in magnitude, significance, and application contexts compared to one based on cumulative reward. 
This difference necessitates additional effort in the policy evaluation process. 
Our next key contribution is developing a novel RL framework that addresses both the state-action and observation processes. The complex and often unknown relationship between these two processes implies that the value of a dynamic policy, whether based on cumulative or integrated reward, may vary with the time $T_k$ at which the policy is evaluated. 
To address this complexity, we introduce a revised assumption of time-homogeneity and Markov properties for states, actions, and gap times. 
Under these assumptions, we demonstrate that the value function at $T_k=t$ can be independent of both $k$ and $t$, provided the current state and gap time are known. Furthermore, we validate the broad applicability of our assumptions using two data-generating schemes. 
These schemes show that our framework is closely related to Markov decision process models widely used in computer science and renewal process models developed in statistics. 
Finally, under the proposed RL framework and for each of the value functions, we develop two policy evaluation methods: one based on the classic Bellman equation
and the other on a Bellman equation modulated by an observation process model. 
We demonstrate that the proposed observation process model is consistent with  the revised time-homogeneity and Markov assumptions for states and gap times.
The model also allows us to formalize the connection between the value function based on cumulative reward and that based on integrated reward, thus playing a crucial role in the policy evaluation procedure based on integrated reward.

The rest of the paper is organized as follows. 
In Section \ref{sec:def}, 
we introduce the notations, value functions, assumptions about the data structure, and two data-generating schemes. 
Sections \ref{sect:OPE_cumulative} and \ref{sect:OPE_integrated} 
present the off-policy evaluation methods for the value function based on cumulative and integrated reward, respectively. 
Section \ref{sect:simulation} conducts simulation studies to evaluate the empirical performance of the proposed methods. 
An application to the PD treatment dataset is given in Section \ref{sect:application}. 
Section \ref{sect: discuss} concludes the paper with a discussion.

\section{
Markovian decision process with irregularly-spaced and outcome-dependent observation times
} 

\label{sec:def}

\subsection{ Notations and value functions}
\label{ssec:def_notations}

\begin{figure}\small
	\graphicspath{{figures/}}
	\begin{center} 
	\includegraphics[width=8.5 cm,height = 3cm]{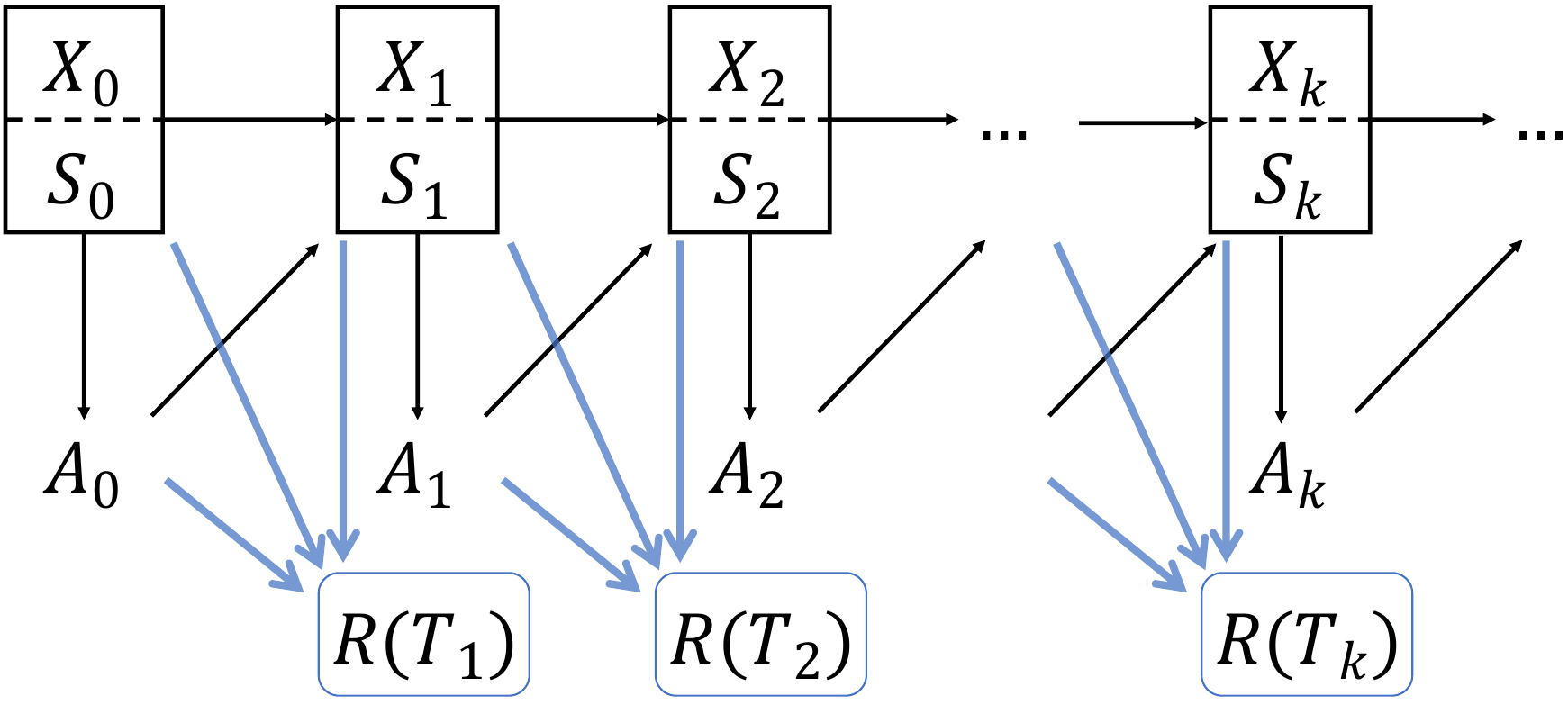} 
	\end{center}
	\caption{ Basic structure of data}
	\label{fig:data-structure}
\end{figure}

Consider a study where treatment decisions are made at irregularly-spaced time points  $\{T_0=0\,, T_1\,, T_2, \ldots, T_k,\ldots\}$.  For each $k\geq 1$, let $X_k  = T_k - T_{k-1}$ denote the gap times, 
and let $\bm{S}_k\in \mathbb{S}$ be a vector of state variables that summarizes the subject's information collected up to and including time $T_k$. At each $T_k$,
investigators can observe the status of subject $\bm{S}_k = S(T_k)$ and the gap time $X_k = T_k - T_{k-1}$, and then  take an action $A_k = A(T_k) \in \mathbb{A} $ based on $\bm{S}_k$ and $X_k$. The reward of action $A_k$ is observed at $T_{k+1}$, and thus denoted as  $R(T_{k+1})$.
For simplicity, we assume that the state space $ \mathbb{S} $ is a subspace of $\mathbb{R}^d$ where  $d$ is the number of state vectors, and 
 the action space  $\mathbb{A}$  is a discrete space $\{0, 1\,, \ldots\,, m-1\}$ with $m$
denoting the number of actions. We also assume that  $\bm{S}_0$, $X_0$, and $A_0$ exist and are observable at $T_0$. An illustration of the data structure is given in Figure \ref{fig:data-structure}.

Let $\pi( \cdot | \cdot,\cdot)$ denote a policy that maps the state and gap time $( \bm{s} , x) \in \mathbb{S} \times [0,\infty]$ to a probability mass function $\pi(\cdot|  \bm{s}, x)$ on  $\mathbb{A}$.
Under such a policy $\pi$, a decision maker will take action $A_k = a$ with probability $\pi(a | \bm{S}_k, X_k)$  for $a \in \mathbb{A}$ and $k \geq 0$.
Aiming at off-policy evaluation for any given dynamic policy $\pi$, we first define the corresponding value functions. 
Unlike the existing literature 
 \citep{   
Shi2021statistical}, we treat $\{T_k\}_{k\geq 0}$ as jump points of a counting process $N(t)$ and 
propose the following two value functions
\begin{align}
V_k^{\pi}( \bm{s}, x, t) 
=&  \
\expect*{
            \int_t^{\infty} \gamma^{u-t}  R(u) \ \mathrm{d} N(u)    
                |  \bm{S}_t = \bm{s} \,,X_t = x\,, T_k = t
                }
\label{V1}
\\
V_{\mathcal{I},k}^{\pi}( \bm{s}, x, t) 
=& \
	\expect*{
		\int_t^{\infty} \gamma^{u-t}  R(u) \ \mathrm{d}u   \
                |  \bm{S}_t = \bm{s}\,,X_t = x\,, T_k = t
		}\,.
\label{V2}
\end{align}
Here,  $0< \gamma <1$ is a discounted factor that balances the immediate and long-term effects of treatments; $R(t)$, an extension of the aforementioned notation $R(T_k)$,
is redefined as a reward process only observed at $\{T_k\}_{k\geq 1}$, but could be potentially available at time $t \geq 0$; and $E^{\pi}$ denotes the expectation when all the actions follow policy $\pi$. By the definition of $N(t)$, $V_k^{\pi}( \bm{s}, x, t) $ defined in \eqref{V1} can be rewritten as an expectation of discounted cumulative reward, given as
\begin{align}
V^{\pi}_{k}(  \bm{s}, x, t )
&\equiv
 \expect*{	\sum_{j \geq  k+1 } 
	\gamma^{T_j - T_k}  R(T_j)  
	 | 
	 \bm{S}_k = s, X_k = x, T_k = t }
\label{V_1'}
\end{align}
Therefore, we refer to $V^{\pi}_{k}(  \bm{s}, x, t )$  as the value function based on cumulative reward 
and $V^{\pi}_{\mathcal{I},k}(  \bm{s}, x, t )$  as the value function based on integrated reward. 
Similarly, 
we define $
Q_k^{\pi}( \bm{s}, x, a,t) 
= \expect*{
		\sum_{j \geq  k+1 } 
		\gamma^{T_j - T_k}  R(T_j)  
                |  \bm{S}_k = \bm{s}\,,X_k = x\,, A_k = a\,,T_k = t
        }
$
as the action-value function (or Q-function) based on cumulative reward,
while the one based on integrated reward is 
 defined as
$
     Q_{\mathcal{I},k}^{\pi}(  \bm{s}, x, a, t) 
 =
   \expect*{ \int_t^{\infty}
                       \gamma^{u-t}  R(u) \ \mathrm{d}u \    
                |  \bm{S}_k = \bm{s}\,,X_k = x\,, A_k = a\,, T_k = t
      }\,.
 $

{\it Remark 1.}
Both value functions, whether based on cumulative or integrated reward, may depend not only on the current value of state $\bm{S}_k = \bm{s}$,  but also on the time $T_k = t$ at which the policy is evaluated. This differs from the value functions developed under regularly-spaced observations 
and introduces challenges to the policy evaluation process.
To address this issue,
we incorporate $X_k$ into the value functions 
and will demonstrate in subsequent sections that, 
under modified Markov and time-homogeneous assumptions,  the value function at $T_k = t$ can be independent of $k$ and $t$, given the current state $\bm{S}_k$ and gap time $X_k$.

{\it Remark 2.}
The two value functions differ in terms of quantity, interpretation, application scenarios, and estimation procedures. 
First,
the value function based on cumulative reward, $V^{\pi}_{k}(  \bm{s}, x, t )$,
can be highly correlated with the rate of observation process $N(t)$. 
For example,  
assuming $R(t)$ is a non-negative function, a higher rate of 
$N(t)$ would yield more decision stages $T_j$, leading to a higher policy value. 
In contrast, 
$V^{\pi}_{\mathcal{I},k}(  \bm{s}, x, t )$, 
which integrates the reward function over the follow-up period, 
provides a more principled evaluation of the policy as it remains independent of the number of observed decision stages. 
As a result, 
the two value functions are suited to different application scenarios, 
and the choice of value function should depend on how the reward function is defined. 
Taking our motivating PD treatment study as an example, 
when the outcome of interest is whether a patient's probing pocket depths (PPD) are reduced after treatment, 
the reward $R(T_k)$ is defined as an indicator of PPD reduction, making the value function based on cumulative reward preferable. 
Conversely, 
when the reward function $R(t)$ is defined as the PPD measurement at time $t$ for $t \in [0,\infty]$, 
policy evaluation should be independent of the measurement times, 
and the value function based on integrated reward provides a comprehensive assessment of the patient's oral health status. 
Finally, since $R(t)$ is unobservable when $t \neq T_j$, 
additional assumptions about $R(t)$ and $N(t)$ are required to ensure that 
the expectation of $ \int_t^{\infty}  \gamma^{u-t}  R(u) \mathrm{d}u$ is estimable. This makes the evaluation process for $V^{\pi}_{\mathcal{I},k}(  \bm{s}, x, t )$ 
different from that for $V^{\pi}_{k}(  \bm{s}, x, t )$; further details are provided in Section \ref{sect:OPE_integrated}.

\subsection{Assumptions}
\label{ssect: assumption}
Now, we posit assumptions on the states, actions, rewards, and gap times.
\begin{itemize}
\item[(A1)] (Markov and time-homogeneous) 
For all $k \geq 0$, 
and for any measurable set $\mathcal{B} \subset \mathbb{S}\times \mathbb{R}^{+}$ with $\bm{S}_{k}$ and $X_{k}$ as the current state and gap times, respectively, the next state variable $\bm{S}_{k+1}$ and gap time $X_{k+1}$ satisfy
\begin{align*}
 \label{ass:A1}
 P\bigg[  \begin{pmatrix}
   \bm{S}_{k+1} \\
   X_{k+1} 
   \end{pmatrix}
   \in \mathcal{B}
      \bigg| \{\bm{S}_j\}_{ 0\leq j \leq  k}, \{X_j\}_{ 0\leq j \leq  k},
      \{A_j\}_{ 0\leq j \leq  k}, \{T_j \}_{ 0\leq j \leq  k}
        \bigg]
 = 
 \mathcal{P}( \mathcal{B} ; \bm{S}_k , X_k, A_k )\,.
\end{align*}
Here, $\mathcal{P}$ denotes the  joint distribution of the next state and next gap time 
conditional on the current state, action, and gap time.

\item[(A2)](Conditional mean independence) 
For all $k \geq 0$ and for some bounded function $r$, the reward $R(T_{k+1})$ satisfies
\begin{align*}
  E\left[ R(T_{k+1})| \{R(T_j), \bm{S}_j,  X_j , A_j\}_{1\leq j \leq k},
                              \bm{S}_{k+1} \,,  X_{k+1}
        \right]
  = r( \bm{S}_k, X_k, A_k ,\bm{S}_{k+1}, X_{k+1})\,.
\end{align*}
\end{itemize}

Assumption A1 implies that, given the current state, action, and gap time, the joint distribution of $(\bm{S}_{k+1}, X_{k+1})$ does not depend on the historical observations $\{ \bm{S}_j, X_j, A_j, T_j\}_{ j < k} $, or the current observation time $T_k$.
Instead of assuming the marginal distributions of the states and the gap times
separately, we present assumption A1 based on the joint distribution of the next state and the next gap time. 
This indicates that, 
not only the marginal distributions of $\bm{S}_{k+1}$ and $X_{k+1}$, 
but also the dependence between $\bm{S}_{k+1}$ and $X_{k+1}$, 
should be independent of the historical observations and the current observation time $T_k$, conditionally on $\{ \bm{S}_k, X_k, A_k\}$. For instance, in the PD treatment study, the recall interval $X_{k+1}$ often correlates strongly with the PD status at the subsequent visit $\bm{S}_{k+1}$. This correlation, along with the values of $\bm{S}_{k+1}$ and $X_{k+1}$, may be influenced by the current PD status of patients and the treatments provided by the dentists. However, it's reasonable to assume conditional independence of these factors from historical states, treatments, and recall intervals, given the latest observations $\{ \bm{S}_k, X_k, A_k\}$.

By denoting $\widetilde{\bm{S}}_{k+1} = (\bm{S}_{k+1}^{\top}, X_{k+1})^{\top}$  as an extended vector of state variables, it is not hard to see that A1 takes a similar form as the Markov and time-homogeneous assumption commonly adopted in the RL literature. 
However, in the following sections, we will demonstrate  that the gap times $\{X_k\}_{k\geq 1}$ 
and the state variables $\{\bm{S}_k\}_{k \geq 1}$ play different roles in the estimation procedure and theoretical properties. This distinction explains why we do not simply
include $X_{k+1}$ in the state vector. 

\subsection{Example schemes}
\label{ssect:example}
For illustration, we present two data-generating schemes that satisfy assumption A1, 
while A1 itself does not specify any structure on the relationship between $\bm{S}_{k+1}$ and $X_{k+1}$.  
The two schemes bridge the proposed framework with modulated renewal process models and Markov decision process models, as well as providing guidelines for constructing specific models on the states, actions, and gap times.

{\bf Scheme 1.}
Consider a stochastic process where, given $\bm{S}_k= \bm{s}$, $X_k = x$ and $A_k = a$,
the next state $\bm{S}_{k+1}$ is generated as 
\begin{align*}
\text{(S1-s)} \quad 
   P\left( \bm{S}_{k+1} \in \mathcal{B}_s
      | \bm{S}_k= \bm{s} ,X_k = x,A_k = a, T_k,
        \{\bm{S}_j,X_j,A_j, T_j\}_{ j < k}
        \right)
= 
 \mathcal{P}_{S}( \mathcal{B}_s;  \bm{s}, x, a )  
\end{align*}
for any measurable set $\mathcal{B}_s\subset \mathbb{S}$, while given the value of next state $\bm{S}_{k+1} = \bm{s'}$, the next gap time $X_{k+1}$ is generated as 
\begin{align*}
\text{(S1-x)} \quad  
       P\left(
      X_{k+1} \leq x' | 
\bm{S}_{k+1} = \bm{s'}, \bm{S}_k = s, X_k = x, A_k = a,  T_k, \{\bm{S}_j,X_j,A_j, T_j\}_{ j < k} 
\right) 
= \mathcal{P}_X ( x'; \bm{s'}, \bm{s}, x, a)  
\end{align*}
for any $x' \in \mathbb{R}^{+}$. Here, $\mathcal{P}_{S}$ denotes the transition function of the next state conditional on the current state, action, and gap time, and $\mathcal{P}_{X}$ denotes the cumulative distribution function of next gap time conditional on the current state, gap time, action, as well as the next state.

{\it Remark 3.} 
As a special case, when $\mathbb{S}$ consists of finite or countable number of states  and  $\mathcal{P}_{S}(\mathcal{B}_s;  \bm{s}, x, a )$ is independent of the value of current gap time $x$, (S1-s) and (S1-x) together form a time-homogenous semi-Markov decision process 
\citep{Bradtke1994reinforcement,Parr1998hierarchical,Du2020model},
where $\{\bm{S}_{k} \}_{k \geq 0}$ is the Markov chain and $X_{k+1}$ is the sojourn time 
in the $k$th state $\bm{S}_{k}$.

{\it Remark 4.} 
Considering that $\{X_{k} \}_{k \geq 1}$ are the gap times of a counting process, 
we can restate the generation procedure in the model (S1-x) using the notations of the counting process. Let $\{N(t)\}_{t\geq0}$ denote a right-continuous counting process with jumps at $\{T_k\}_{k\geq 1}$. Then, we have $T_{N(t)} = T_k$, $X_{N(t)} = X_k$, $A_{N(t)} = A_k $, $ \bm{S}_{N(t)}  = \bm{S}_{k} $, and $\bm{S}_{N(t)+1} = \bm{S}_{k+1}  $ for all $T_{k} \leq t < T_{k+1}$.
According to (S1-s), the distribution/generation of next state $\bm{S}_{N(t)+1}$ is fully determined by $\{\bm{S}_{N(t)}, X_{N(t)}, A_{N(t)}\}$, 
and therefore 
could be assumed available at time $ t $.
Let $\mathcal{F}(t) = \left\{ \{N(u)\}_{u \leq t}, \{\bm{S}_{N(u)} \}_{u \leq t}, \{X_{N(u)} \}_{u \leq t}, \{A_{N(u)} \}_{u \leq t},  \bm{S}_{N(t)+1} \right\}$
denote all  information available at $t$.
It is not hard to observe that if $N(t)$ satisfies 
\begin{align*}
\text{(S1-x$'$)}  \qquad 
  E\left[ \mathrm{d} N(t) | \mathcal{F}(t-) \right]   
=& \lambda\{ t - T_{N(t-)}; \bm{S}_{N(t-)+1}, \bm{S}_{N(t-)}, X_{N(t-)},A_{N(t-)} \} \mathrm{d} t \\
=& \lambda\{ t - T_k; \bm{S}_{k+1}, \bm{S}_k, X_k,A_k \}  \mathrm{d} t\,,  \text{for} \ k = \max\{j; T_j < t \} 
\quad
\end{align*}
or some intensity function $\lambda$, then for each $k \geq 0$, the gap time $X_{k+1}$ with the hazard function 
$h(x) = \lambda\{x; \bm{S}_{k+1}, \bm{S}_k, X_k, A_k \}$
satisfies model (S1-x) too. On the other hand, if the distribution function $\mathcal{P}_X$ in the model (S1-x) is absolutely continuous wrt. the Lebesgue measure, then the counting process defined by $N(t) = \sum_{j \geq 1} I(T_j \leq t)$  also satisfies (S1-x$'$).
As a matter of fact, by viewing $\{ \bm{S}_k , A_k ,  \bm{S}_{k+1} \}_{k \geq 0 }$ as the covariates available at $\{ T_k \}_{k \geq 0}$, 
(S1-x$'$) takes a similar form to the modulated renewal process model. The assumption is natural for user-initiated visits and has been popularly adopted in the statistical literature
\citep{Cox1972statistical, Lin2013robust, Chen2018semiparametric}.

{\bf Scheme 2.}
As an alternative, 
scheme 2 generates the gap time first and then determines the value of the next state. Specifically, assume that the next gap time $X_{k+1}$ satisfies  
$$
\text{(S2-x)} \quad  
      P\left(
            X_{k+1}\leq x' | 
\bm{S}_k = s, X_k = x, A_k = a,  T_k, \{\bm{S}_j,X_j,A_j, T_j\}_{ j < k} 
\right)
= \widetilde{\mathcal{P}}_X ( x' ; \bm{s}, x, a) 
\qquad  $$
for $x' \in \mathbb{R}^+$.
Then, given the  value  $X_{k+1} = x'$,  
the next state $\bm{S}_{k+1}$ satisfies 
$$
\text{(S2-s)} \quad 
   P\left( \bm{S}_{k+1} \in \mathcal{B}_s
      | X_{k+1}= x',  \bm{S}_k= \bm{s} ,X_k = x,A_k = a, T_k,
        \{\bm{S}_j,X_j,A_j, T_j\}_{ j < k}
        \right) 
=
\widetilde{ \mathcal{P}}_{S}( \mathcal{B}_s; x', \bm{s}, x, a )  \,,
$$
for any measurable set  $\mathcal{B}_s\subset \mathbb{S}$.
Here, $\widetilde{\mathcal{P}}_{X}$ denotes the cumulative distribution function of the next gap time conditional on the current state, gap time, and action,  
while $\widetilde{\mathcal{P}}_{S}$ 
denotes the transition function of the next state conditional on the current state, action, gap time, as well as the next gap time.
Let $\mathcal{\widetilde{F}}(t) = \left\{ \{N(u)\}_{u \leq t}, \{\bm{S}_{N(u)} \}_{u \leq t}, \{X_{N(u)} \}_{u \leq t}, \{A_{N(u)} \}_{u \leq t}\right\}$.
With arguments similar to Remark 4, it can be shown that (S2-x) is  equivalent to
$$
\text{(S2-x$'$)} \quad  
      E\left[ \mathrm{d} N(t) \big| \mathcal{\widetilde{F}}(t-) \right] 
= \lambda( t - T_{N(t-)};  \bm{S}_{N(t-)}, X_{N(t-)},A_{N(t-)} ) 
= \lambda( t - T_k;  \bm{S}_k, X_k,A_k )\,, 
$$
for $k = \max\{j; T_j < t \}$,
and takes the form of a modulated renewal process.

\section{ Policy evaluation based on cumulative reward}
\label{sect:OPE_cumulative}
In the following subsections, we initially propose a policy evaluation procedure, requiring only assumptions A1 and A2. Then, under a counting process model for the observation process, we
develop an alternative policy evaluation method.
The asymptotic properties of all the proposed estimates are given in the last subsection.

\subsection{ Utilizing the classic Bellman equation } 
\label{ssect:classic}
To estimate $V^{\pi}_{k}(  \bm{s}, x, t )$ and $Q^{\pi}_{k}(  \bm{s}, x, a, t )$,  we first present the following two lemmas, whose proof is available in Section \ref{ssect:proof_lemma} of the Supplementary Material.

{\it Lemma 1.
Under assumptions A1 and A2, 
the action-value function based on cumulative reward $Q^{\pi}_{k}(  \bm{s}, x, a, t )$ exists and 
satisfies $Q^{\pi}_{k}(  \bm{s}, x, a, t ) =  Q^{\pi}_{0}(  \bm{s}, x, a, 0)$ for all $k \geq 0$, $t\geq 0$, $s \in \mathbb{S}$, $x \in \mathbb{R}^+$ and $a \in \mathbb{A}$. }

By Lemma 1, the value of $Q^{\pi}_{k}(  \bm{s}, x, a, t )$ is time-homogeneous, meaning the expected value of a policy $\pi$ does not depend on the number of past observations $k$ or the current observation time $t$. Therefore, we can omit both $k$ and $t$, 
and denote the action-value function simply as $Q^{\pi}(  \bm{s}, x, a )$ for brevity.  
Similarly, the state-value function of policy $\pi$ can also be presented as 
$ 
V^{\pi}( \bm{s}, x)  =  \sum_{a \in \mathbb{A}} Q^{\pi}(  \bm{s}, x, a) \pi(a| \bm{s}, x) \,,
$  
and the value function of policy $\pi$ under some reference distribution $\mathcal{G}$ can be presented as 
$ 
V^{\pi}(\mathcal{G})  = 
\int   \sum_{a \in \mathbb{A}} Q^{\pi}(  \bm{s}, x, a) \pi(a| \bm{s}, x)
d \mathcal{G}( \bm{s}, x) \,,
$  
both of which are independent of $k$ and $t$. 
Furthermore, we derive a Bellman equation as formulated in Lemma 2.

{\it Lemma 2.
Under assumptions A1 and A2, 
the action-value function  based on cumulative reward  $Q^{\pi}(  \bm{s}, x, a )$ satisfies a Bellman equation 
\begin{align*}
         Q^{\pi}( \bm{S}_k, X_k ,  A_k )
= &
E\left[   \gamma^{X_{k+1}} R(T_{k+1})
 +     \gamma^{X_{k+1}}
 	 \sum_{a \in \mathbb{A}} 
                   Q^{\pi}( \bm{S}_{k+1}, X_{k+1}, a) 
                   \pi(a | \bm{S}_{k+1}, X_{k+1})  
         \bigg|
                           \bm{S}_k, X_k , A_k        
          \right] 
\end{align*}
for all $k \geq 0$. }

To solve the Bellman equation, let $Q( \bm{s}, x, a; \bm\theta)$ 
denote a model for $Q^{\pi}(  \bm{s}, x, a)$ indexed by $\bm\theta \in \Theta$,  and satisfies 
$ Q(  \bm{s}, x, a; \bm\theta_{\pi}) = Q^{\pi}(  \bm{s}, x, a)$ 
for some $\bm\theta_{\pi}\in \Theta$. Define 
\begin{align*}
       \delta_{k+1}(\bm\theta) 
=&  
       \gamma^{X_{k+1}} R(T_{k+1}) 
    +  
       \gamma^{X_{k+1}}
       \sum_{a \in \mathbb{A}} 
                     Q( \bm{S}_{k+1}, X_{k+1}, a;\bm\theta  )
                      \pi(a | \bm{S}_{k+1}, X_{k+1})
        - Q( \bm{S}_k,X_k, A_k ;\bm\theta)\,.
\end{align*}
According to Lemma 2, we have 
$E[ \delta_{k+1}(\bm{\theta}_{\pi}) | \bm{S}_k, X_k, A_k ] = 0$ for all  $k \geq 0$.
Assume the map 
$\bm{\theta} \rightarrow Q( \bm{s}, x, a; \bm{\theta})$ is differentiable everywhere for each fixed $s$, $x$ and $a$, 
and let $\bm{\dot{Q}}_{\bm{\theta}}$ denote the gradient of $Q$ with respect to $\bm{\theta}$. Then, we construct an estimating function for $\bm{\theta}_{\pi}$ as 
\begin{equation}
    	 \bm{U}(\bm{\theta}) 
= 
	\frac{1}{\sum_i K_i  } \sum_{i = 1}^{n}   \sum_{k =0}^{K_i-1} 
	\bm{\dot{Q}}_{\bm{\theta}}( \bm{S}_{i,k},X_{i,k}, A_{i,k};\bm{\theta} ) \widehat \delta_{i,k+1}(\bm{\theta}) \,,
\label{U_theta1}                       
\end{equation}
where 
 $\{     X_{i,k}\,,
       \bm{S}_{i,k}\,,
       A_{i,k}\,,
       R_i(T_{i,k+1});  1\leq k  \leq K_i, 1\leq i \leq n \} 
$ are $n$ independent and identically distributed trajectories of 
$\{\bm{S}_k, X_k, A_k,R(T_{k+1}) \}_{k \geq 0}$.
$ K_i \in \mathbb{Z}_{+}$ denotes the total number of decisions observed on the $i$th sequence, and 
$
      \delta_{i, k+1}(\bm{\theta}) 
=    
	 \gamma^{X_{i, k+1}}  R_{i}(T_{i, k+1}) 
    +  
        \gamma^{X_{i, k+1}} 
        \sum_{a \in \mathbb{A}}         
        Q( \bm{S}_{i, k+1}, X_{i, k+1}, a; \bm{\theta})
        \pi(a|\bm{S}_{i, k+1}, X_{i, k+1}) 
      - Q( \bm{S}_{i, k},X_{i, k}, A_{i, k}; \bm{\theta}) \,.
$

Generally, \eqref{U_theta1} can be solved by numerical methods,
but in certain cases, 
it may have a closed form. 
For example, consider a B-spline model 
$ Q( \bm{s}, x, a;\bm{\theta}) 
= 	 \bm{\xi}( \bm{s}, x, a)^{\top} \bm{\theta} \,,
$
where 
$
      \bm{\xi}( \bm{s}, x, a)
=
     \{
           \Phi_L^{\top}( \bm{s}, x)\mathbb{I}(a=0),
            \Phi_L^{\top}( \bm{s}, x)\mathbb{I}(a=1),
            \ldots,
            \Phi_L^{\top}( \bm{s}, x)\mathbb{I}(a=m-1)
      \}^{\top}\,,
$ 
$
\bm{\theta}
= 
  (\bm{\theta}_{0}^{\top}, \bm{\theta}_{1}^{\top}, \cdots, \bm{\theta}_{m-1}^{\top})^{\top} \,,
 $ 
$\Phi_L(\cdot) = \{ \Phi_{L,1}(\cdot) ,\cdots,\Phi_{L,L}(\cdot) \}^{\top}$ is a vector consisting of $L$ B-spline basis functions, and 
$\bm{\theta}_{a}$ is a $L$-dimensional parameter vector for each $a\in \mathbb{A}$. 
Denote
 $\bm{\xi}_{i,k}
= \bm{\xi}(\bm{S}_{i,k}, X_{i,k},A_{i,k})
 $
 and 
 $\bm{\zeta}_{\pi,i,k+1}
 = 
 \sum_{a \in \mathbb{A}}         
 \bm{\xi}(\bm{S}_{i,k+1}, X_{i,k+1}, a)
 \pi(a| \bm{S}_{i,k+1},X_{i,k+1})\,.
 $
Then, 
the estimating equation takes the form
$	 \bm{U}(\bm{\theta})
  = 
	 \frac{1}{\sum_i K_i } 
	 \sum_{i = 1}^n  \sum_{k = 0}^{K_i-1} 
	 	\bm{\xi}_{i,k} 
		 \Big\{ 
			\gamma^{ X_{i,k+1}}R_i(T_{i,k+1}) 
		       + 
		 	(\gamma^{ X_{i,k+1}} \bm{\zeta}_{\pi,i,k+1}
			  -  
			  \bm{\xi}_{i,k}
		        )^{\top}
		         \bm{\theta}
	\Big\} = 0  \,,
$ 
and can be solved as
\begin{align*}
	\widehat{\bm{\theta}}_{\pi}
	=
	\biggl\{
	           		\frac{1}{\sum_i K_i } 
				\sum_{i=1}^n \sum_{k= 0}^{K_i-1}
	 				\bm{\xi}_{i,k}
					 (\bm{\xi}_{i,k}- 
					 \gamma^{ X_{i,k+1}} \bm{\zeta}_{\pi,i,k+1}
					 )^{\top}  
	\biggr\}^{-1} 
	\bigg\{ 
		\frac{1}{\sum_i K_i }  
		\sum_{i=1}^n \sum_{k= 0}^{K_i-1} 
		 	\bm{\xi}_{i,k} 
			\gamma^{ X_{i,k+1}}R_i(T_{i,k+1}) 
	 \bigg\}\,.
\end{align*}
Therefore,  we propose to 
estimate $Q^{\pi}( \bm{s}, x, a )$  by 
 $
 	 \widehat Q^{\pi}( \bm{s}, x, a ) 
=	  Q ( \bm{s}, x, a; \widehat{\bm{\theta}}_{\pi} ) 
=
	 \bm{\xi}( \bm{s}, x, a)^{\top} \widehat{\bm{\theta}}_{\pi} \,,
$
estimate state-value function $V^{\pi}( \bm{s}, x)$  by 
$
\widehat{V}^{\pi}( \bm{s}, x) 
 =    
  \sum_{a \in \mathbb{A}}  \pi(a |  \bm{s}, x )     
\bm{\xi}(  \bm{s}, x, a)^{\top}   \widehat{\bm{\theta}}_{\pi} 
$,
and estimate  $V^{\pi}(\mathcal{G})$ by
$
	\widehat{V}^{\pi}(\mathcal{G})
=
	\bm{\zeta}_{\pi, \mathcal{G}}^{\top} 
	\widehat{\bm{\theta}}_{\pi}\,,
$
where 
$
	\bm{\zeta}_{\pi, \mathcal{G}} 
	=  
	\sum_{a \in \mathbb{A}} 
	\int  
	 \pi(a |  \bm{s}, x )    
	\bm{\xi}(  \bm{s}, x, a)
	\mathrm{d} \mathcal{G}( \bm{s}, x) 
$.

\subsection{Utilizing the modulated Bellman equation via an observation process model }\label{ss:modulated}

In this subsection, we further propose an alternative policy evaluation procedure that relies on a specified model for the observation process. 
Here, the motivation for modeling the observation process arises from two aspects. First, as demonstrated in Subsection \ref{ssect:example}, under the revised time homogeneity and Markov assumptions for states and gap times, the irregularly spaced and outcome-dependent observation times can be represented as the occurrence times of a renewal process modulated by the state and action sequences \citep{Cox1972statistical}. Therefore, positing a modulated renewal process model on the observation process is consistent with the structure of the data and helps us make more accurate inferences. Second, by employing an intensity model for the observation process,
the integration of rewards can be transformed into a cumulative form. This enables us to leverage policy evaluation techniques based on cumulative reward for estimating the value function based on integrated reward. More details are given in Section \ref{sect:OPE_integrated}.

For brevity, we focus on Scheme 1 given in Section \ref{ssect:example}; similar derivations can also be applied to Scheme 2. 
The modification begins with the Bellman equation outlined in Lemma 2. Under assumption (S1-x), we can express the equation as  
{\small
\begin{align*}
    &     Q^{\pi}( \bm{S}_k, X_k ,  A_k )
\\
= &
E\left[  
 \gamma^{X_{k+1}} R(T_{k+1})  
 +    
 E\big[   \gamma^{X_{k+1}}
 	\sum_{a \in \mathbb{A}}
	\pi(a | \bm{S}_{k+1}, X_{k+1})  
        Q^{\pi}(\bm{S}_{k+1}, X_{k+1}, a)
  \big |   \bm{S}_{k+1}, \bm{S}_k, X_k , A_k        
  \big] 
         \bigg|
                           \bm{S}_k, X_k , A_k        
 \right]
 \\
= &
E\left[  
 \gamma^{X_{k+1}} R(T_{k+1})  
 +    
  \int  \gamma^{x'}
  	\sum_{a \in \mathbb{A}}
	\pi(a | \bm{S}_{k+1}, x') 
        Q^{\pi}( \bm{S}_{k+1}, x',  a)  
   d\mathcal{P}_X(x'; \bm{S}_{k+1}, \bm{S}_k, X_k , A_k  )     
        \big|
                           \bm{S}_k, X_k , A_k        
          \right]\,.
\end{align*}
}
Let $\widehat{\mathcal{P}}_X$ denote an estimate of the distribution function 
$\mathcal{P}_X$. Similar to \eqref{U_theta1}, we construct 
$\widetilde{ \bm{U}}(\bm{\theta}) 
= 
	\frac{1}{\sum_i K_i  } \sum_{i = 1}^{n} \sum_{k =0}^{K_i-1} \bm{\dot{Q}}_{\bm{\theta}}( \bm{S}_{i,k},X_{i,k}, A_{i,k};\bm{\theta} )
	\widetilde \delta_{i,k+1}(\bm{\theta}) \,,
$
with 
$
      \widetilde \delta_{i, k+1}(\bm{\theta}) 
=   
	 \gamma^{X_{i, k+1}}  R_{i}(T_{i, k+1}) 
    +  
     \int  \gamma^{x'}
     	\sum_{a \in \mathbb{A}}
	\pi(a | \bm{S}_{i,k+1}, x') 
        Q(\bm{S}_{i,k+1}, x', a ;  \bm{\theta}) 
    \mathrm{d} \widehat{\mathcal{P}}_X(x'; \bm{S}_{i,k+1}, \bm{S}_{i,k}, X_{i,k} , A_{i,k}  )     
    - 
    Q( \bm{S}_{i, k},X_{i, k}, A_{i, k}; \bm{\theta}) \,.
$
Further approximating action-value functions with B-splines
$Q( \bm{s}, x, a;\bm{\theta}) 
= \bm{\xi}( \bm{s}, x, a)^{\top} \bm{\theta} 
$, 
then the solution to $ \widetilde{ \bm{U}}(\bm{\theta}) = 0 $ takes the form of 
\begin{align}
	\widetilde{\bm{\theta}}_{\pi}
	=
	\biggl\{
	           		\frac{1}{\sum_i K_i } 
				\sum_{i=1}^n \sum_{k= 0}^{K_i-1}
	 				\bm{\xi}_{i,k}
					 (\bm{\xi}_{i,k}- 
					 \widetilde{\mathcal{U}}_{\pi,i,k+1}
					 )^{\top} 
	\biggr\}^{-1} 
	\bigg\{  
		\frac{1}{\sum_i K_i }  
		\sum_{i=1}^n \sum_{k= 0}^{K_i-1} 
		 	\bm{\xi}_{i,k} 
			\gamma^{ X_{i,k+1}}R_i(T_{i,k+1}) 
	 \bigg\}
\end{align}
with 
$\widetilde{\mathcal{U}}_{\pi,i,k+1} = 
 \int
        \gamma^{x'} 
        \sum_{a \in \mathbb{A}}
	\pi(a | \bm{S}_{i,k+1}, x') 
        \bm{\xi}(\bm{S}_{i,k+1},x',a)
         d\widehat{\mathcal{P}}_X(x'; \bm{S}_{i,k+1},\bm{S}_{i,k},X_{i,k}, A_{i,k})\,.
$

Then, the estimator for $Q^{\pi}( \bm{s}, x, a )$ is $ \widetilde Q^{\pi}( \bm{s}, x, a ) = \bm{\xi}( \bm{s}, x, a)^{\top} \widetilde{\bm{\theta}}_{\pi} $, 
the estimator for state-value function $V^{\pi}( \bm{s}, x) $ is
 $
	\widetilde{V}^{\pi}( \bm{s}, x) 
	= 
	\sum_{a \in \mathbb{A}}
	\pi(a |  \bm{s}, x) \bm{\xi}(  \bm{s}, x, a)^{\top} 
	\widetilde{\bm{\theta}}_{\pi} 
$
and
the estimator for the value function  $V^{\pi}(\mathcal{G})$ under given reference distribution $\mathcal{G}$ is
$
	\widetilde{V}^{\pi}(\mathcal{G})
	=
	\bm{\zeta}_{\pi, \mathcal{G}}^{\top} 
	\widetilde{\bm{\theta}}_{\pi}$.
Here, extra effort is required in estimating $\mathcal{P}_X$, the distribution of the next gap time $X_{k+1}$ conditional on $\bm{S}_{k+1}, \bm{S}_k, X_k$ and $A_k$. As an illustration, we propose a Cox-type counting process model that satisfies (S1-$x'$), and takes the form
\begin{align}
  E[ \mathrm{d} N(t) | \mathcal{F}(t-) ] 
  = \lambda(t - T_k; \bm{S}_{k+1},\bm{S}_k, X_k,A_k )  \ \mathrm{d} t 
=
 \lambda_0( t - T_k) 
 \exp( \bm{\beta}_0^{\top} \bm{Z}_k ) \ \mathrm{d} t\,, 
\label{model-cox}
\end{align}
Here,
$k = \max\{j; T_j < t \}$,
$\bm{Z}_k = \bm{\phi}(\bm{S}_{k+1}, \bm{S}_k, X_k, A_k ) \in \mathbb{R}^q$ for some pre-specified $q$-dimensional function $\bm{\phi}$, $\bm{\beta}_0$ is a vector of unknown parameters and $\lambda_0$ is an unspecified baseline intensity function. Let $\{N_i(t) \}_{1\leq i \leq n}$  denote $n$ independent and identically distributed trajectories of $N(t)$. 
For $i = 1, \ldots, n$ and $k = 1,\ldots, K_i$, 
define $N_{i,k}(x) = N_i(T_{i,k-1}+x) - N_i(T_{i,k-1})$, 
$\bm{Z}_{i,k} = \bm{\phi}(\bm{S}_{i,k+1},\bm{S}_{i,k}, X_{i,k}, A_{i,k} ) $.
Then, we propose to estimate $\bm{\beta}_0$ by solving the equation
\begin{align}
	\bm{U}_z(\beta) 
=
	 \frac{1}{\sum_i K_i} \sum_{i = 1}^n \sum_{k = 0}^{K_i-1} \int_0^{\tau} \{ \bm{Z}_{i,k} - \overline{\bm{Z}}(x;\bm{\beta}) \}  \ \mathrm{d}  N_{i,k+1}(x) = 0 \,,
\label{U_x}
\end{align}
where 
$
	 \overline{\bm{Z}}(x;\bm{\beta}) 
= 
	{  \{   \sum_{i,k}  
	 I(X_{i,k+1} \geq x) \exp(\bm{\beta}^{\top} \bm{Z}_{i,k})\bm{Z}_{i,k}  \}^{-1}
	     }{
	   \{ \sum_{i,k}   
	            I(X_{i,k+1} \geq x) \exp(\bm{\beta}^{\top} \bm{Z}_{i,k}) \}
	     } \,,
$
and $\tau$ is a positive constant satisfying $P(X_k \geq \tau) > 0$ for each $k$. 
Let  $\widehat{\bm{\beta}}$ denote the obtained estimator. Then, we can further 
estimate $\Lambda_0(x)= \int_0^x \lambda_0(u) du$ 
by
$
	\widehat \Lambda_0(x) 
=
      \int_0^x  
	\{
	 	 \sum_{i} \sum_{k}   I(X_{i,k+1} \geq u) \exp(\widehat{\bm{\beta}}^{\top} \bm{Z}_{i,k})   
	   \}^{-1}
	   \{
	 	 \sum_{i} \sum_{k} \mathrm{d}N_{i,k+1}(u)
	  \}
$,  
and can estimate  
$ \mathcal{P}_X$ by  
$	\widehat{\mathcal{P}}_X(x'; \bm{s'}, \bm{s}, x, a) 
=
	1 - \exp\{ - \widehat{\Lambda}_0(x') \exp( \widehat{\bm{\beta}}^{\top}  \bm{z}) \} 
$
for $ \bm{z} = \bm{\phi}(\bm{s'}, \bm{s}, x,a)$.

\subsection{Asymptotic properties}
\label{ssect:properties}

We first introduce the notations and impose necessary assumptions.
Denote $r(\bm{s}, x,a) = E[ \gamma^{X_{k+1}} R(T_{k+1})|\bm{S}_k = s, X_k = x, A_k = a]$.
Let $p(\bm{s'}, x';  \bm{s}, x, a)$ be the conditional density function
satisfying 
$ \mathcal{P}(\mathcal{B} |  \bm{s}, x, a) = 
\int_{\mathcal{B}} p(\bm{s'},x';  \bm{s}, x, a) \mathrm{d} \bm{s'} \mathrm{d} x'$ 
for the joint distribution $\mathcal{P}(\cdot |  \bm{s}, x, a)$ defined in (A1).
Then, (A3) is a regularity condition ensuring the action function $Q^{\pi}( \bm{s}, x, a)$ to be continuous \cite[Lemma 1]{Shi2021statistical}.

\begin{itemize}
\item[(A3)]  There exists some $p_0>0$, such that  $r(\cdot, \cdot, a)$ and $p(\cdot,\cdot;  \bm{s}, x, a) $ are   H\"{o}lder continuous functions with exponent $p_0$ and 
with respect to $(\bm{s}, x) \in \mathbb{S}\times \mathbb{R}^+$ . 
\end{itemize}

As discussed in Section \ref{ssect: assumption},  the transition trajectories of states and gap times  $\{ \widetilde{S}_{k} = (\bm{S}_k^{\top}, X_k)^{\top} \}_{k \geq 0}$ 
form a time-homogeneous Markov chain under assumption (A1).  Suppose $\{ (\bm{S}_k^{\top}, X_k)^{\top} \}_{k \geq 0}$ has a unique invariant joint distribution with density function 
$\mu(\cdot, \cdot)$ under some behavior policy $b$. 
Similarly,  when the reference distribution $\mathcal{G}$ is absolutely continuous with respect to the Lebesgue measure, there exists a density function $v_0( \bm{s}, x)$ satisfying 
$  \mathrm{d} \mathcal{G}( \bm{s}, x) =  v_0( \bm{s}, x) \mathrm{d}s \mathrm{d}x$,  
which is also a joint distribution for the states and gap time.
We posit the following assumptions on $\mu$ and $v_0$.
\begin{itemize}
\item[(A4)] 
$\mu(\cdot, \cdot)$ and $v_0(\cdot,\cdot)$ are uniformly 
bounded away from 0 and $\infty$. 
\end{itemize}

Recall that $\widehat{\bm\theta}_{\pi}$ and $\widetilde{\bm\theta}_{\pi}$  are obtained by solving 
$\bm{U}(\bm\theta)=0$ and $\widetilde{\bm{U}}(\bm\theta)=0$.
Define 
$\mathcal{U}_{\pi,i,k+1} = 
 \int
        \gamma^{x'} 
        \sum_{a \in \mathbb{A}}
         \pi(a |\bm{S}_{i,k+1},x') \bm{\xi}(\bm{S}_{i,k+1},x', a )
         d\mathcal{P}_X(x'; \bm{S}_{i,k+1},\bm{S}_{i,k},X_{i,k}, A_{i,k})
$,
$
\bar{\bm{\zeta}}_{\pi}( \bm{s}, x, a )
=
E[ \bm{\zeta}_{\pi, i, k+1} | \bm{S}_{i,k} = s, X_{i,k} = x, A_{i,k} = a]
$,
and
$
\bar{\bm{u}}_{\pi}( \bm{s}, x, a )
=
E[ \mathcal{U}_{\pi, i, k+1} | \bm{S}_{i,k} = s, X_{i,k} = x, A_{i,k} = a ] \,.
$

The following assumption ensures the existence of 
$\widehat{\bm\theta}_{\pi}$ and $\widetilde{\bm\theta}_{\pi}$.

\begin{itemize}
\item[(A5)] 
There exist constants $\bar{c}$, such that 
$ \lambda_{min}( 
	\sum_{k = 0}^{K-1} 
	E[
	\bm{\xi}_{i,k}^{\otimes 2}- 
	\gamma^{2 X_{i,k+1}} 
	\bar{\bm{\zeta}}_{\pi}(\bm{S}_{i,k}, X_{i,k}, A_{i,k} )^{\otimes 2}
	]
	)
	\geq K \bar{c}\,,
$
and 
$ \lambda_{min}(
	\sum_{k = 0}^{K-1} 
	E[
	\bm{\xi}_{i,k}^{\otimes 2}- 
	\bar{\bm{u}}_{\pi}(\bm{S}_{i,k}, X_{i,k}, A_{i,k} )^{\otimes 2} 
	]
	)
	\geq K \bar{c},
$
where $\lambda_{min}(\cdot)$ denotes 
the minimum eigenvalue of a matrix.
\end{itemize}

The following assumption is required in constructing asymptotic approximation when $K_i \rightarrow \infty$. Specifically,  the geometric ergodicity in A6(i)  is part of the sufficient conditions for the Markov chain central limit theorem to hold, and  A6(ii) guarantees the validity of estimating equation  \eqref{U_x} under renewal process model \eqref{model-cox}.
\begin{itemize}
\item[(A6)] When $K_i \rightarrow \infty$, 
(i) the Markov chain $\{ (\bm{S}_{ik}^{\top}, X_{ik})^{\top} \}_{k \geq 0}$  is geometrically ergodic, 
and (ii) there exists a probability measure $v(\cdot)$  such that for any measurable 
set $B_z $ on $\mathbb{R}^q$, $\{\bm{Z}_{i,k} \}_{k \geq 0}$ satisfying 
$\frac{1}{K_i} \sum_{k = 0}^{K_i-1} I(\bm{Z}_{i,k} \in B_z) 
\stackrel{p}{\rightarrow } v( B_z) \,.$
\end{itemize}

Now, we provide bidirectional asymptotic properties  for the estimated values
$\widehat{V}^{\pi}( \mathcal{G})$ and  $\widetilde{V}^{\pi}( \mathcal{G})$.
The property of the estimated state-value functions follows directly by setting $\mathcal{G}$
as a Dirac delta function at fixed $( \bm{s}, x)\in \mathbb{S}\times \mathbb{R}^+$. For any $s \in \mathbb{S}$,  $x\in \mathbb{R}^+$, $a \in \mathbb{A}$, define
$\omega_{\pi}(\bm{s}, x,a) 
=
 E [
       \{
       \gamma^{X_{1}} R(T_{1}) 
    +  
       \gamma^{X_{1}}
       \sum_{a \in \mathbb{A}}
       \pi(a | \bm{S}_{1}, X_{1})
       Q^{\pi}( \bm{S}_{1}, X_{1}, a)
        - Q^{\pi}( \bm{S}_0,X_0, A_0 ) 
        \}^2 
   |
   \bm{S}_{0} = s, X_0 = x, A_0 = a	
 ]
 $.
 
\begin{theorem}
Assume assumptions $A1$-$A6$ hold. Suppose  $	L = o\{ n_K^{1/2}/log(n_K) \}$, $	L^{\frac{2p}{d+1}} 
	\gg
	n_K ( 1+ \| \bm{\zeta}_{\pi, \mathcal{G}} \|_2^{-2} )
$
for $n_K = \sum_{i = 1}^n K_i$,
and there exists some constant $c_R\geq 1$, such that 
$
P(\max_{ k} | R(T_{k}) | \leq c_R) = 1
$,
and
$
\omega_{\pi}(\bm{s}, x,a) \geq c_R^{-1}
$
for any 
$
s \in \mathbb{S}\,,
x \in \mathbb{R}^+\,,
a \in \mathbb{A}
$
.
Then, as $n_K = \sum_{i = 1}^n K_i \rightarrow \infty$, 
we have:
$n_K^{1/2} \sigma_{\pi, \mathcal{G},1}^{-1}
\{ \widehat{V}^{\pi}( \mathcal{G}) - 
	V^{\pi}(\mathcal{G})\}
\stackrel{d}{\rightarrow} N(0,1)\,,
$
where
$\sigma_{\pi, \mathcal{G},1}$ is given in Subsection \ref{ssect:proof_thm} of the Supplementary Material.
\end{theorem}

\begin{theorem}
Assume assumptions $A1$-$A6$, conditions $C1$-$C6$, and model \eqref{model-cox} hold.
Suppose 
$	L = o\{ n_K^{1/2}/log(n_K) \}$, 
$	L^{\frac{2p}{d+1}} 
	\gg
	n_K ( 1+ \| \bm{\zeta}_{\pi, \mathcal{G}} \|_2^{-2} )
$
and there exists some constant $c_R\geq 0$ 
such that 
$
P(\max_{k} | R(T_{k}) | \leq c_R) = 1
$,
for any 
$
s \in \mathbb{S}\,,
x \in \mathbb{R}^+\,,
a \in \mathbb{A}
$.
Then as $n_K \rightarrow \infty$, 
we have
$
	n_K^{1/2} 
	\sigma_{\pi, \mathcal{G},2}^{-1}
	\{
		 \widetilde{V}^{\pi}( \mathcal{G}) 
		 - 
		V^{\pi}(\mathcal{G})
	\}
	\stackrel{d}{\rightarrow} 
	N(0,1)\,,
$
where $\sigma_{\pi, \mathcal{G},2}$  is given in Subsection \ref{ssect:proof_thm} of the Supplementary Material.
\end{theorem}

Here, conditions $C1$-$C6$ are regularity conditions for constructing asymptotic properties of $\mathcal{\widehat{P}}_X$ under model \eqref{model-cox}. They are commonly adopted in the modulated renewal process studies \citep{Pons1988cox, Lin2013robust},
and we relegate them, together with the proofs,  
in Subsection \ref{ssect:proof_thm} of the Supplementary Material for brevity.

\section{Policy evaluation based on integrated reward} \label{sect:OPE_integrated}

As remarked in Section \ref{ssec:def_notations}, additional efforts are required for policy evaluation based on integrated reward. 
Here, the key insight is to transform the integration in $Q_{\mathcal{I}}^{\pi}(  \bm{s}, x, a) $  into a cumulative reward, thereby enabling the application of the method developed in Section 4  to $Q_{\mathcal{I},k}^{\pi}(  \bm{s}, x, a, t)$. Assuming $N(t)$ satisfies (S1-x') and $R(t) \in \mathcal{F}(t-)$, we have 
\begin{align*}
& \mathbb{E}^{\pi} \left[   \int_{T_k}^{\infty}  \gamma^{u - T_k}  R(u) \mathrm{d}u    
                             \bigg |  \bm{S}_k \,,X_k \,, A_k \,, T_k \right] 
\\
=&  
  \mathbb{E}^{\pi} \left[ 
           \int_{T_k}^{\infty} 
           \frac{
                  \gamma^{u - T_k}  R(u)   \  
                  E[ \mathrm{d} N(u)  | \mathcal{F}(u-) ]
           }{
                  \lambda( u - T_{N(u-)}; 
                  \bm{S}_{N(u-)+1}, \bm{S}_{N(u-)}, X_{N(u-)},A_{N(u-)})
           }        
          \bigg|  \bm{S}_k \,,X_k \,, A_k \,, T_k \right] 
\\
=&  
      \mathbb{E}^{\pi}
       \left[ 
           \int_{T_k}^{\infty} 
           E\big[ 
           \frac{
                  \gamma^{u - T_k} R(u)  \  \mathrm{d} N(u)  
           }{
                  \lambda(u - T_{N(u-)}; 
                  	\bm{S}_{N(u-)+1}, \bm{S}_{N(u-)}, X_{N(u-)},A_{N(u-)} )
           }  
          \big | \mathcal{F}(u-)  
          \big]      
          \bigg|  \bm{S}_k \,,X_k \,, A_k \,, T_k 
        \right ]       
\\
=&  
      \mathbb{E}^{\pi} \left[  
            \int_{T_k}^{\infty} 
           \frac{
                  \gamma^{u - T_k}   R(u)  \  \mathrm{d} N(u)  
           }{
                  \lambda( u - T_{N(u-)}; \bm{S}_{N(u-)+1}, \bm{S}_{N(u-)}, X_{N(u-)},A_{N(u-)} ) 
           }  
          \bigg| \bm{S}_k \,,X_k \,, A_k \,, T_k \right ]       
\\
=&  
      \mathbb{E}^{\pi} \left[  
           \sum_{j \geq k+1}
           \frac{
                  \gamma^{T_j - T_k}  R(T_j)  
           }{
                  \lambda( X_j; \bm{S}_{j}, \bm{S}_{j-1}, X_{j-1},A_{j-1} )
           }  
         \bigg | \bm{S}_k \,,X_k \,, A_k \,, T_k \right ]       
\\
=&  
      \mathbb{E}^{\pi} \left[  
           \sum_{j \geq k+1}
            \gamma^{T_j - T_k}  R_{\mathcal{I}}(T_j)  
           \bigg |  \bm{S}_k \,,X_k \,, A_k \,, T_k \right ]  \,, 
\end{align*}                                
where 
$
 R_{\mathcal{I}}(T_j)   =  
    R(T_j)  / \lambda( X_j; \bm{S}_{j}, \bm{S}_{j-1}, X_{j-1},A_{j-1} )
$
is an inverse-intensity-weighted reward.

It is not hard to verify that $\{ R_{\mathcal{I}}(T_k)\}_{k \geq 1}$ satisfies assumption $(A2)$ and 
thus, the policy evaluation methods proposed in Section 3  are applicable.  
For example, by Lemma 1, 
$Q_{\mathcal{I},k}^{\pi}(  \bm{s}, x, a, t)$ is also free of $k$ and $t$,  and hence, we can rewrite the Q-function as 
$Q_{\mathcal{I}}^{\pi}(  \bm{s}, x, a)$, the value function as 
$ 
	V^{\pi}_{\mathcal{I}}( \bm{s}, x)  
=  
	\sum_{a \in \mathbb{A}} 
	Q^{\pi}_{\mathcal{I}}(  \bm{s}, x, a) \pi(a| \bm{s}, x) \,,
$ 
and taking the expression 
$
	V^{\pi}_{\mathcal{I}}(\mathcal{G})  
= 
	\int   \sum_{a \in \mathbb{A}} 
	Q^{\pi}_{\mathcal{I}}(  \bm{s}, x, a) \pi(a| \bm{s}, x)
	d \mathcal{G}( \bm{s}, x) \,
$ under some reference distribution $\mathcal{G}$, free of $k$ and $t$. 
Furthermore, by assuming  $Q_{\mathcal{I}}^{\pi}( \bm{s}, x, a ) = \xi( \bm{s}, x, a)^{\top} \bm{\theta}_{\pi,\mathcal{I}}$ for some B-spline basis 
$\{ \xi( \bm{s}, x, a)\,, s\in \mathbb{S}, x\in \mathbb{R}^{+}, a\in \mathbb{A}\}$, 
we can also obtain two estimates for  $ \bm{\theta}_{\pi,\mathcal{I}}$ as  
\begin{align*}
	\widehat{\bm{\theta}}_{\pi,\mathcal{I}}
	=
	\biggl\{
	           		\frac{1}{\sum_i K_i } 
				\sum_{i=1}^n \sum_{k= 0}^{K_i-1}
	 				\bm{\xi}_{i,k}
					 (\bm{\xi}_{i,k}- 
					 \gamma^{ X_{i,k+1}} \bm{\zeta}_{\pi,i,k+1}
					 )^{\top}   
	\biggr\}^{-1} 
	\bigg\{ 
		\frac{1}{\sum_i K_i }  
		\sum_{i=1}^n \sum_{k= 0}^{K_i-1} 
		 	\bm{\xi}_{i,k} 
			\gamma^{ X_{i,k+1}} \widehat{R}_{\mathcal{I},i}(T_{i,k+1}) 
	 \bigg\}\,,
\end{align*}
and 
\begin{align*}
	\widetilde{\bm{\theta}}_{\pi, \mathcal{I}}
	=
	\biggl\{
	           		\frac{1}{\sum_i K_i } 
				\sum_{i=1}^n \sum_{k= 0}^{K_i-1}
	 				\bm{\xi}_{i,k}
					 (\bm{\xi}_{i,k}- 
					  \widetilde{\mathcal{U}}_{\pi,i,k+1}
					 )^{\top}  	\biggr\}^{-1} 
	\bigg\{  
		\frac{1}{\sum_i K_i }  
		\sum_{i=1}^n \sum_{k= 0}^{K_i-1} 
		 	\bm{\xi}_{i,k} 
			\gamma^{ X_{i,k+1}} \widehat{R}_{\mathcal{I},i}(T_{i,k+1}) 
	 \bigg\} \,,
\end{align*}
respectively. Here, 
$ \widehat{R}_{\mathcal{I},i}(T_{i,k+1}) =  
R_i(T_{i,k+1})  / \widehat{\lambda}( X_{i,k+1}; \bm{S}_{i,k+1}, \bm{S}_{i,k}, X_{i,k},A_{i,k})$ 
and $\widehat{\lambda}(x';\bm{s'}, \bm{s}, x, a)$ denotes some estimator of 
$\lambda(x';\bm{s'}, \bm{s}, x, a)$
for $  \bm{s'}, \bm{s} \in \mathbb{S}, x', x\in \mathbb{R}^{+}$ 
and $a\in \mathbb{A}$.
For example, under the Cox-type model \eqref{model-cox}, 
the estimated intensity function is 
$
\widehat{\lambda}(x';\bm{s'}, \bm{s}, x, a)
 = 
 \widehat{\lambda}_0(x') \exp( \widehat{\bm{\beta}}^{\top} \bm{z} )
$ with 
$\bm{z} = \bm{\phi}(\bm{s'}, \bm{s}, x,a)$
and 
$
\widehat \lambda_0(x) 
= \frac{1}{b_n} \int_0^{\tau} K( \frac{u - x}{b_n})  \mathrm{d} \widehat \Lambda_0(u), 
$
where 
$\widehat{\bm{\beta}}$ and $\widehat \Lambda_0(u)$ is given in SubSection \ref{ss:modulated}, 
 $K(\cdot)$ is a kernel function, and $b_n$ is the kernel bandwidth.

Furthermore, we derive two estimates for action-value function $Q_{\mathcal{I}}^{\pi}( \bm{s}, x, a )$ 
as 
$ \widehat Q_{\mathcal{I}}^{\pi}( \bm{s}, x, a ) 
 =
  \bm{\xi}( \bm{s}, x, a)^{\top} \widehat{\bm{\theta}}_{\pi,{\mathcal{I}}} 
$ 
and
$ \widetilde Q_{\mathcal{I}}^{\pi}( \bm{s}, x, a ) 
 =
  \bm{\xi}( \bm{s}, x, a)^{\top} \widetilde{\bm{\theta}}_{\pi,{\mathcal{I}}} 
 $.
The estimates for value function $V_{\mathcal{I}}^{\pi}( \bm{s}, x)$
are
$
\widehat{V}_{\mathcal{I}}^{\pi}( \bm{s}, x) 
 = 
 \sum_{a \in \mathbb{A}}
 \pi(a |  \bm{s}, x) \bm{\xi}(  \bm{s}, x, a )^{\top} \widehat{\bm{\theta}}_{\mathcal{I},\pi} 
$
and
$
\widetilde{V}_{\mathcal{I}}^{\pi}( \bm{s}, x) 
 = 
 \sum_{a \in \mathbb{A}}
 \pi(a |  \bm{s}, x)
 \bm{\xi}(  \bm{s}, x, a)^{\top} \widetilde{\bm{\theta}}_{\mathcal{I},\pi} 
$,
and
the estimates for $V_{\mathcal{I}}^{\pi}(\mathcal{G})$
are
$
\widehat{V}^{\pi}_{\mathcal{I}}(\mathcal{G})
=
\bm{\zeta}_{\pi, \mathcal{G}}^{\top} \widehat{\bm{\theta}}_{\pi,\mathcal{I}}
$ and 
$\widetilde{V}^{\pi}_{\mathcal{I}}(\mathcal{G})
=
\bm{\zeta}_{\pi, \mathcal{G}}^{\top} \widetilde{\bm{\theta}}_{\pi,\mathcal{I}}\,.
$
The following theorem presents the asymptotic properties of 
$\widehat{V}^{\pi}_{\mathcal{I}}(\mathcal{G})
$
and
$\widetilde{V}^{\pi}_{\mathcal{I}}(\mathcal{G})
$, whose proof is given in Subsection \ref{ssect:proof_thm} of the Supplementary Material.

\begin{theorem}
Consider assumptions $A1$-$A6$, conditions $C1$-$C6$, and model \eqref{model-cox} hold.
Suppose 
$	L = o\{ n_K^{1/2}/log(n_K) \}$, 
$	L^{\frac{2p}{d+1}} 
	\gg
	n_K ( 1+ \| \bm{\zeta}_{\pi, \mathcal{G}} \|_2^{-2} )
$
$n_K b_n^4 \rightarrow 0$,
 $n_K b_n^2 \rightarrow \infty$,
and there exists some constant $c_R\geq 0$, 
such that 
$
P(\max_{ k} | R(T_{k}) | \leq c_R) = 1
$,
for any 
$
s \in \mathbb{S}\,,
x \in \mathbb{R}^+\,,
a \in \mathbb{A}
$.
 Then as $n_K = \sum_{i = 1}^n K_i \rightarrow \infty$, 
we have
(i)
$
	n_K^{1/2} 
	\sigma_{\pi, \mathcal{G},3}^{-1}
	\{
		 \widehat{V}_{\mathcal{I}}^{\pi}( \mathcal{G}) 
		 - 
		V_{\mathcal{I}}^{\pi}(\mathcal{G})
	\}
	\stackrel{d}{\rightarrow} 
	N(0,1)\,,
$
and
(ii) $
	n_K^{1/2} 
	\sigma_{\pi, \mathcal{G},4}^{-1}
	\{
		 \widetilde{V}_{\mathcal{I}}^{\pi}( \mathcal{G}) 
		 - 
		V_{\mathcal{I}}^{\pi}(\mathcal{G})
	\}
	\stackrel{d}{\rightarrow} 
	N(0,1)\,.
$
Here
$\sigma_{\pi, \mathcal{G},3}$
and
$\sigma_{\pi, \mathcal{G},4}$
are given in section \ref{ssect:proof_thm} of the Supplementary Material.
 \end{theorem}

\section{Simulation}\label{sect:simulation}

In this section, we conduct simulation studies to examine the finite sample performance of the proposed methods. We begin by describing our setup. 
First, for the initial values,  
we generate $S_0$ from a Uniform distribution on $[-3/2,3/2]$, $X_0$ from an Exponential distribution with intensity 1/2, and $A_0$ from a Bernoulli distribution with expectation 1/2. Then, given the current state $S_k$, gap time $X_k$, and action $A_k$, the next state is generated from one of the three models: 
(S1) $S_{k+1} = \frac{3}{4} ( 2 A_k - 1 ) S_k  +\epsilon_k$;
(S2) $S_{k+1} = 
\{ \frac{3}{4}  - \frac{1}{4} I(X_k < \frac{1}{2}) \} 
( 2 A_k - 1 )  S_k  + \epsilon_k$; 
(S3) $S_{k+1} = 
\{ \frac{3}{4}  - \frac{1}{4} I(X_k < \frac{1}{2}) 
 + \frac{1}{4} I(X_{k+1} > 1) \} 
( 2 A_k - 1 ) S_k  + \epsilon_k $,
where, $\{ \epsilon_k\}_{k \geq 1} \stackrel{i.i.d.}{\sim} \mathcal{N}(0, \frac{1}{4})$.
The next gap time $X_{k+1}$ is generated from of the following intensity models:
(X1) $ \lambda(x; S_k,X_k, A_k ) =  \lambda_0(x)$;
(X2) $  \lambda(x; S_k,X_k, A_k ) = \lambda_0(x)\exp( 
               - S_k + \frac{1}{2} X_k + A_k - \frac{1}{2} S_k A_k  )$;
(X3) $ \lambda(x; S_k,X_k, A_k,S_{k+1})
	   = \lambda_0(x)\exp(
               - S_k + \frac{1}{2} X_k + A_k - \frac{1}{2} S_k A_k + \frac{1}{2} S_{k+1} )
$.
All the actions $\{ A_{k}, k \geq 1\}$  are independently generated from a Bernoulli distribution with mean 0.5,  
and the reward of each action $A_k$, which is observed at $T_{k+1}$, is defined as 
$R(T_{k+1})  =  
\{ S_{k+1} - S_k -  \frac{1}{2}( 2A_k-1)\}X_{k+1} + r_k 
$
with 
$\{ r_k\}_{k \geq 1} \stackrel{i.i.d.}{\sim} \mathcal{N}(0, \frac{1}{4})$.

\begin{figure}\small
	\graphicspath{{figures/}}
	\begin{center}
       \begin{tabular}{cc}
                 Scenario 1 & Scenario 2\\
	 \includegraphics[width= 6 cm, height = 2.5cm ]{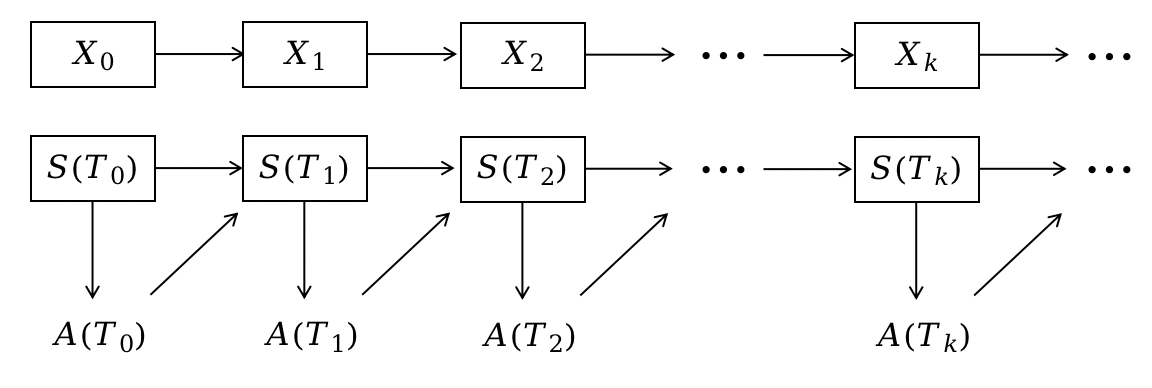} &
	 \includegraphics[width= 6cm, height = 2.5cm ]{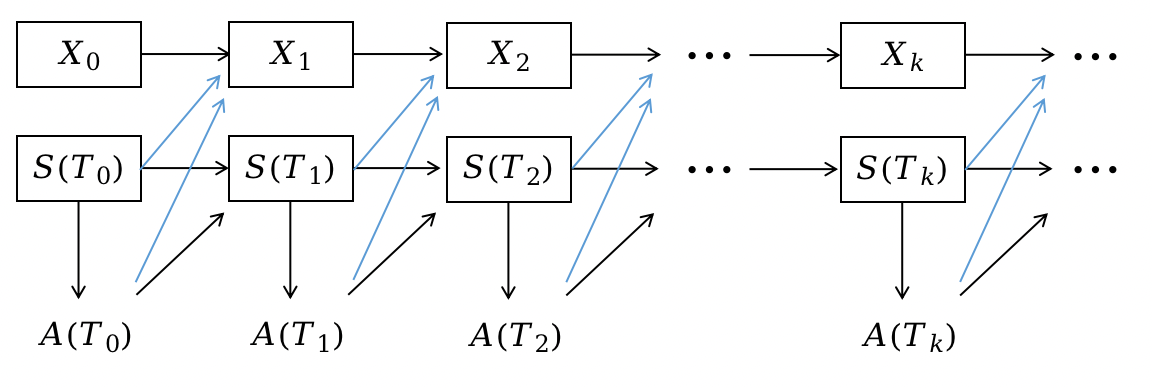}\\
	Scenario 3  & Scenario 4  \\
	 \includegraphics[width= 6 cm, height = 2.5cm ]{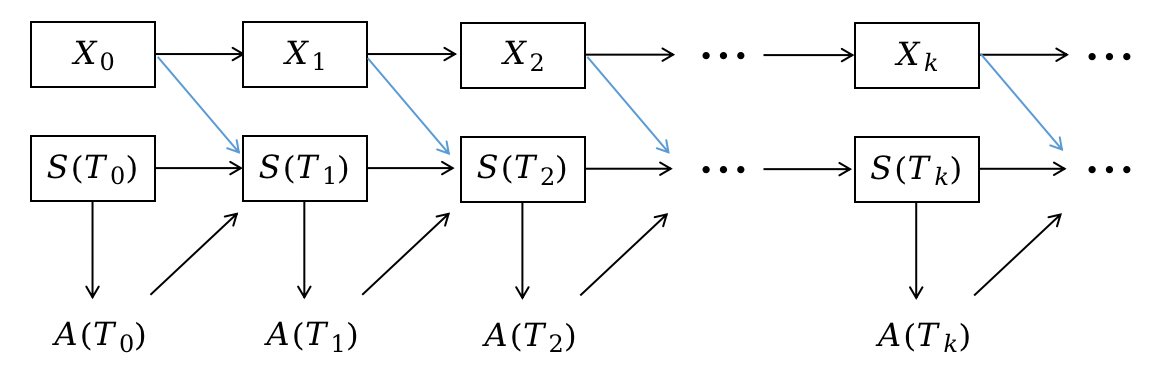}&
	 \includegraphics[width= 6 cm, height = 2.5cm ]{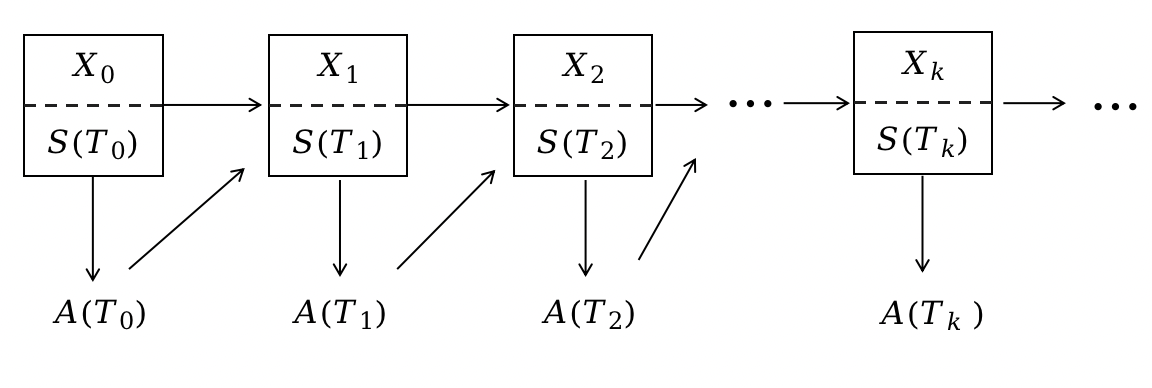}\\
	\end{tabular}
        \end{center}
	\caption{ Data structure under scenarios 1-4.}
	\label{fig:simulation1}
\end{figure}

Although there could be numerous combinations of the generative models, 
here we only choose four representative scenarios for a brief illustration. 
The settings of the four scenarios are listed in Table \ref{tab-setting}
and 
corresponding data structures are plotted in Figure \ref{fig:simulation1}.
In general, all four datasets satisfy the 
the time-homogeneous and Markov assumptions given in (A1), 
but they vary in the dependence structure among the states, actions, and gap times. 
In Scenario 1, the generation of $X_{k+1}$ does not depend on the values of $S_k$ or $A_k$, 
and the value of $X_k$ does not affect the generation of $( S_{k+1},A_{k+1})$ either.
This implies that the observation process $N(t)$ 
is independent of the state-action transition process.
In Scenario 2, we allow the generation of gap times to be correlated with the actions and states, 
but $X_{k+1}$ and $S_{k+1}$ are conditionally independent with each other, given $(S_k, X_k, A_k)$.
More general scenarios are given in scenarios 3 and 4. As noted in section \ref{ssect: assumption}, 
the Markov assumption in (A1) allows $X_{k+1}$ and $S_{k+1}$ to be correlated with each other, 
only with the restriction that their dependence structure satisfies the Markov property. Two example schemes are also developed in Section \ref{ssect:example}  illustrating two different dependence structures satisfying (A1). Here, Scenarios 3 and 4 represent Schemes 1 and 2,
respectively. Specifically, in Scenario 3,  we first generate $X_{k+1}$, 
and then generate $S_{k+1}$ based on $(S_k,X_k,A_k,X_{k+1})$;
while in Scenario 4, we generate $S_{k+1}$ first, and then $X_{k+1}$ based on $(S_k,X_k,A_k,S_{k+1})$.
A comparison of the data structures in Scenarios 1-4 is provided 
in Table $\ref{tab-setting}$.


{\small
\begin{table} [!ht]
\caption{\label{tab-setting} Settings in Scenarios 1-4.}
\begin{center}
\vspace*{-12pt}
\def\temptablewidth{1 \textwidth}
\begin{tabular*}
{\temptablewidth}{@{\extracolsep{\fill}}c|cc|cccc}
\hline
      & \multicolumn{2}{c}{Settings} &  \multicolumn{4}{c}{Properties} \\
 \cline{2-3} \cline{4-7} 
      &  State &  Gaptime &   A1\& A2 & $X_{k+1} \perp S_{k+1}$  &
      $X_{k+1} \perp S_{k+1} | \mathcal{F}(T_k)$ & model \eqref{model-cox} \\
\hline 
Scenario 1 & $S1$  & $X1$   & \cmark  & \cmark & \cmark & \cmark  \\
Scenario 2 & $S2$  & $X2$   & \cmark  & \xmark & \cmark & \cmark  \\
Scenario 3 & $S2$  & $X3$   & \cmark  & \xmark & \xmark & \cmark  \\
Scenario 4 & $S3$  & $X2$   & \cmark  & \xmark & \xmark & \xmark  \\
\hline
\end{tabular*}
\end{center}
\end{table}
}

In the policy evaluation procedure,  the discount factor is set as $\gamma = 0.7$ for all settings. 
For the reference distribution, we consider $\mathcal{G}(s, x) = (s +1) x/4$, with $-1 \leq s \leq 1$ and $0 \leq x \leq 2$ for policy evaluation under cumulative rewards, and  $\mathcal{G}(s, x) = (5s +1) x/2$, with $-1/5 \leq s \leq 1/5$ and $0 \leq x \leq 1$ for policy evaluation under integrated rewards. For the target policy, we considered a class of linear deterministic policies 
$
\pi(a; s, x ) =  I( \alpha_0 +\alpha_1 s +\alpha_2 x \leq 0 )\,,
$
where
$\alpha = ( \alpha_0, \alpha_1, \alpha_2)$  is set as $(0, -1, 0)$ in Scenario 1, and 
set as $(-1,-1,1)$ in Scenarios 2-4. 
To calculate the true values,
we simulate $N = 5 \times 10^5$ independent trajectories with $(S_0,X_0)$  distributed according to $\mathcal{G}$, and  $\{A_k\}_{k \geq 0}$ chosen according to target policy 
$\pi$. Then, $V^{\pi}(\mathcal{G})$ is approximated by $\frac{1}{N} \sum_{j = 1}^N \sum_{l = 1}^{K_j}  \gamma^{T_{jl}}  R_j(T_{j,l}) $ and  $V^{\pi}_{\mathcal{I}}(\mathcal{G})$ is approximated by 
$\frac{1}{N} \sum_{j = 1}^N \sum_{l = 1}^{K_i}  \gamma^{T_{jl}}  
R_{\mathcal{I},j}(T_{j,l}) $.

Since both state $S_k$ and gap time $X_k$ might not have bounded supports in our settings, 
we transform the data by 
$S_k^* = \Phi(S_k) $ and $X_k^* = 1- \exp(X_k)$, 
where $\Phi(\cdot)$ is the cumulative distribution function of a standard normal random variable. 
Then, the basis functions are constructed from the tensor product of  
two cubic B-spline sets, whose knots are placed at equally spaced sample quantiles of the transformed states and gap times. 
When estimating the B-spline coefficients $\bm\theta$, three methods were considered, 
namely, the naive method, the standard method, and the modulated method.
When the value of the policy is defined by cumulative reward, the standard method leads to the estimate 
$\widehat{\bm\theta}_{\pi}$. The modulated method, which depends on model \eqref{model-cox} for the observation process, leads to the estimate $\widetilde{\bm\theta}_{\pi}$. Finally, the naive estimate is
$
\widehat{\bm{\theta}}_{\pi,N}
=
	\bigl[
				\sum_{i,k} 
	 				\bm{\xi}_{i,k}
					\{ \bm{\xi}_{i,k}- 
				 \gamma 
					 \bm{\xi}(S_{i,k+1},X_{i,k+1},\pi)	 \}^{\top}   
	\bigr]^{-1} 
	\bigl\{ 
			\sum_{i,k} 
		 	\bm{\xi}_{i,k} 
		\gamma R_i(T_{i,k+1})  
	 \bigr\}\,, 
$
which is directly obtained from \cite{Shi2021statistical} by treating $(S_k, X_k)$ as the state variable.
Similarly, when the value of the policy is defined by integrated reward, 
we adopt the standard estimate as $\widehat{\bm\theta}_{\pi, \mathcal{I}}$,
the modulated estimate as 
$\widetilde{\bm\theta}_{\pi, \mathcal{I}}$, and 
the naive estimate as 
$$
\widehat{\bm{\theta}}_{\pi, \mathcal{I},N}
=
	\bigl[
				\sum_{i,k} 
	 				\bm{\xi}_{i,k}
					\{ \bm{\xi}_{i,k}- 
					 \gamma 
					 \bm{\xi}(S_{i,k+1},X_{i,k+1},\pi)
					 					 \}^{\top}   
	\bigr]^{-1} 
	\bigl\{ 
			\sum_{i,k} 
		 	\bm{\xi}_{i,k} 
		\gamma 
		\widehat{R}_{\mathcal{I},i}(T_{i,k+1})
	 \bigr\}\,.
$$

The performances of the three estimation methods are summarized in Tables  \ref{tab-Sim1} and \ref{tab-Sim2}.
Within each scenario, we further consider 6 cases by setting 
sample sizes $n = 100, 200, 400$ with the length of  trajectory $K = 10$, as well as
setting $K = 100, 200, 400$ when the number of  trajectory $n = 10$.
In each case, we generate 1000 simulation replicates to approximate the bias and standard deviation of the naive estimators (Bias$_N$, SD$_N$), and the bias, standard deviation, estimated standard error, and empirical coverage probability of $95\%$  confidence intervals for standardized estimators (Bias$_S$, SD$_S$, SE$_S$, CP$_S$) and modulated estimators (Bias$_M$, SD$_M$, SE$_M$, CP$_M$). From the results, we observe that both the standard method and the modulated method perform well in Scenarios 1-3:
the estimators are asymptotically unbiased,  the estimated standard errors are close to the standard deviation of the estimators, and the coverage probabilities of $95\%$ confidence interval match the nominal level when either $n$ or $K$ is large enough. As a comparison, the naive estimate is biased across all scenarios, including Scenario 1, where the observations follow a Poisson process of rate one.
This implies the observation process is informative and should be included in the policy evaluation, 
even if it is independent of the state-action process. 
Moreover, 
by comparing $SD_S$ and $SD_M$ in Scenarios 2-3, 
we observe that the modulated estimates are more efficient than the standard ones when $n_K = nK$ is relatively small. 
This suggests that the value function estimated using the modulated Bellman equation 
may lead to reduced variance in policy evaluation 
under small sample sizes, 
especially when the distribution of gap times depends on the state-action process.

{\tiny
\begin{table} [!ht]
\caption{ \label{tab-Sim1} 
Policy evaluation results based on cumulative reward, including the bias and standard deviation of the naive estimators (Bias$_N$, SD$_N$), and  the bias, standard deviation, estimated standard error, and empirical coverage probability of $95\%$ confidence intervals for standardized estimators (Bias$_S$, SD$_S$, SE$_S$, CP$_S$) and 
modulated estimators 
(Bias$_M$,SD$_M$,SE$_M$,CP$_M$).}
\begin{center}
\def\temptablewidth{1 \textwidth}
\begin{tabular*}
{\temptablewidth}{@{\extracolsep{\fill}}ccrcrcccrccc}
 \hline
  && \multicolumn{2}{c}{Naive}&  
          \multicolumn{4}{c}{Standard}&
         \multicolumn{4}{c}{Modulated } \\
 \cline{3-4} \cline{5-8} \cline{9-12}
   $n$  & K   &$Bias_N$& $SD_N$& $Bias_S$ &   $SD_S$ & $SE_S$ & $CP_S$ & $Bias_M$ &  $SD_M$ & $SE_M$  & $CP_M$\\            
\hline 
 &&\multicolumn{10}{c}{ Scenario 1 \quad \quad  True value =  -0.641}
\\
 \cline{3-4} \cline{5-8} \cline{9-12}  
 100 &  10 & -0.099 &  0.079 & -0.002  & 0.063 & 0.059 & 0.958 &  0.002 & 0.062 & 0.063 & 0.954 \\
 200 &  10 & -0.096 &  0.053 &  0.000  & 0.039 & 0.041 & 0.960 &  0.002 & 0.041 & 0.045 & 0.968 \\
 400 &  10 & -0.095 &  0.036 & -0.002  & 0.028 & 0.029 & 0.952 & -0.001 & 0.028 & 0.032 & 0.958 \\[2mm]
  10 & 100 & -0.075 &  0.100 & -0.030  & 0.073 & 0.065 & 0.950 &  0.001 & 0.080 & 0.071 & 0.938 \\
  10 & 200 & -0.091 &  0.065 &  0.001  & 0.049 & 0.044 & 0.934 &  0.003 & 0.047 & 0.049 & 0.952 \\
  10 & 400 & -0.094 &  0.043 & -0.002  & 0.030 & 0.031 & 0.944 & -0.001 & 0.030 & 0.034 & 0.968 \\
\hline
&&\multicolumn{10}{c}{Scenario 2 \quad \quad  True value = 0.637 }\\
 \cline{3-4} \cline{5-8} \cline{9-12}
  100 &  10 & -0.388 &  0.105 &  0.002 & 0.094 & 0.082 & 0.930 & -0.010 & 0.092 & 0.078 & 0.934 \\
  200 &  10 & -0.397 &  0.064 & -0.003 & 0.058 & 0.055 & 0.936 & -0.009 & 0.053 & 0.054 & 0.940 \\
  400 &  10 & -0.397 &  0.042 & -0.002 & 0.038 & 0.039 & 0.952 & -0.005 & 0.034 & 0.038 & 0.956 \\[2mm]
  10  & 100 & -0.387 &  0.147 & -0.002 & 0.151 & 0.117 & 0.896 & -0.016 & 0.135 & 0.106 & 0.906 \\
  10  & 200 & -0.385 &  0.091 & -0.002 & 0.077 & 0.072 & 0.924 & -0.008 & 0.074 & 0.066 & 0.932 \\
  10  & 400 & -0.389 &  0.068 & -0.004 & 0.056 & 0.049 & 0.928 & -0.006 & 0.051 & 0.046 & 0.936 \\
\hline
&& \multicolumn{10}{c}{Scenario 3 \quad \quad  True value = 0.510 }\\
  \cline{3-4} \cline{5-8} \cline{9-12}
 100 &  10 & -0.441 &  0.151  &  0.004 & 0.110 & 0.085 & 0.894 & -0.006 & 0.093 & 0.079 & 0.898 \\
 200 &  10 & -0.442 &  0.093  & -0.002 & 0.064 & 0.055 & 0.932 & -0.006 & 0.055 & 0.053 & 0.930 \\
 400 &  10 & -0.437 &  0.063  &  0.000 & 0.042 & 0.039 & 0.936 & -0.002 & 0.039 & 0.038 & 0.956 \\[2mm]
  10 & 100 & -0.410 &  0.234  & -0.011 & 0.200 & 0.141 & 0.884 &  0.002 & 0.159 & 0.124 & 0.888 \\
  10 & 200 & -0.423 &  0.126  & -0.001 & 0.092 & 0.079 & 0.902 & -0.005 & 0.084 & 0.071 & 0.920 \\
  10 & 400 & -0.431 &  0.095  & -0.004 & 0.062 & 0.053 & 0.924 & -0.005 & 0.058 & 0.048 & 0.938 \\
\hline
\end{tabular*}
\end{center}
\end{table}
}

{\tiny
\begin{table} [!ht]
\caption{ \label{tab-Sim2} 
Policy evaluation results based on integrated reward, including the bias and standard deviation of the naive estimators (Bias$_N$, SD$_N$), and 
the bias, standard deviation, estimated standard error, and empirical coverage probability of $95\%$ confidence intervals for standardized estimators (Bias$_S$, SD$_S$, SE$_S$, CP$_S$) and 
modulated estimators 
(Bias$_M$,SD$_M$,SE$_M$,CP$_M$).}
\begin{center}
\def\temptablewidth{1 \textwidth}
\begin{tabular*}
{\temptablewidth}{@{\extracolsep{\fill}}ccrcrcccrccc}
 \hline
     && \multicolumn{2}{c}{Naive}&  
          \multicolumn{4}{c}{Standard}&
         \multicolumn{4}{c}{Modulated } \\
 \cline{3-4} \cline{5-8} \cline{9-12}
      $n$  & K   &$Bias_N$& $SD_N$& $Bias_S$ &   $SD_S$ & $SE_S$ & $CP_S$ & $Bias_M$ &  $SD_M$ & $SE_M$  & $CP_M$\\            
\hline 
 &&\multicolumn{10}{c}{ Scenario 1 \quad \quad  True value =  -0.413}
\\
   \cline{3-4} \cline{5-8} \cline{9-12}                                                         
 100  &  10 & -0.019 & 0.105 &  0.001 & 0.080 & 0.071 & 0.934 &  0.005 & 0.083 & 0.068 & 0.910 \\
 200  &  10 & -0.023 & 0.072 &  0.000 & 0.056 & 0.049 & 0.929 &  0.003 & 0.053 & 0.048 & 0.908 \\
 400  &  10 & -0.028 & 0.053 & -0.003 & 0.040 & 0.034 & 0.929 & -0.001 & 0.041 & 0.034 & 0.912 \\[2mm]
  10  & 100 & -0.020 & 0.099 & -0.001 & 0.079 & 0.070 & 0.935 & -0.012 & 0.078 & 0.066 & 0.904 \\
  10  & 200 & -0.023 & 0.072 &  0.002 & 0.054 & 0.048 & 0.929 &  0.004 & 0.054 & 0.046 & 0.898 \\
  10  & 400 & -0.027 & 0.049 &  0.000 & 0.038 & 0.034 & 0.925 &  0.001 & 0.038 & 0.033 & 0.924 \\
\hline
 &&\multicolumn{10}{c}{ Scenario 2 \quad \quad  True value =  0.383}
 \\
 \cline{3-4} \cline{5-8} \cline{9-12}
 100 &  10 & -0.045 & 0.098 & -0.002 & 0.122 & 0.104 & 0.943 & -0.003 & 0.111 & 0.096 & 0.938 \\
 200 &  10 & -0.044 & 0.066 &  0.001 & 0.091 & 0.078 & 0.948 &  0.000 & 0.085 & 0.074 & 0.960 \\
 400 &  10 & -0.048 & 0.051 & -0.004 & 0.069 & 0.061 & 0.947 & -0.004 & 0.066 & 0.059 & 0.944 \\[2mm]
  10 & 100 & -0.047 & 0.094 &  0.000 & 0.120 & 0.107 & 0.949 & -0.006 & 0.110 & 0.098 & 0.934 \\
  10 & 200 & -0.044 & 0.066 & -0.001 & 0.086 & 0.076 & 0.934 & -0.003 & 0.084 & 0.071 & 0.934 \\
  10 & 400 & -0.046 & 0.048 & -0.002 & 0.064 & 0.058 & 0.949 & -0.004 & 0.063 & 0.056 & 0.957 \\
\hline
 &&\multicolumn{10}{c}{ Scenario 3 \quad \quad  True value =  0.460}
\\
\cline{3-4} \cline{5-8} \cline{9-12}
 100 &  10 & -0.138 & 0.069 &  0.004 & 0.076 & 0.073 & 0.950 & -0.001 & 0.076 & 0.069 & 0.940 \\
 200 &  10 & -0.150 & 0.051 &  0.002 & 0.050 & 0.051 & 0.945 &  0.000 & 0.049 & 0.049 & 0.949 \\
 400 &  10 & -0.160 & 0.042 & -0.001 & 0.037 & 0.036 & 0.944 & -0.002 & 0.035 & 0.035 & 0.945 \\[2mm]
  10 & 100 & -0.134 & 0.066 &  0.007 & 0.077 & 0.071 & 0.956 &  0.001 & 0.073 & 0.067 & 0.945 \\
  10 & 200 & -0.140 & 0.047 &  0.003 & 0.053 & 0.049 & 0.946 &  0.001 & 0.054 & 0.048 & 0.939 \\
  10 & 400 & -0.142 & 0.034 &  0.003 & 0.037 & 0.035 & 0.936 &  0.002 & 0.037 & 0.034 & 0.931 \\
\hline
\end{tabular*}
\end{center}
\end{table}
}
    
Finally, since the modulated estimates require modeling the observation process $N(t)$, we also investigate the robustness of the proposed estimates when model \eqref{model-cox} is misspecified under Scenario 4. Specifically,  we generated data following Scheme 2, calculated the estimates using methods developed under Scheme 1, and report the results in Table \ref{tab-Sim3} of the Supplementary Material. 
The results show that the standard estimator under cumulative reward $\widehat{V}^{\pi}(\mathcal{\mathcal{G}})$, which only requires A1 and A2, still performs well in all settings, and, what is beyond the expectation is that, the estimates $\widetilde{V}_{\pi}(\mathcal{\mathcal{G}})$, $\widehat{V}_{\pi, \mathcal{I}}(\mathcal{\mathcal{G}})$ and $\widetilde{V}_{\pi, \mathcal{I}}(\mathcal{\mathcal{G}})$ that are based on model \eqref{model-cox}, also demonstrate negligible bias and provide satisfactory variance estimation. Hence, the proposed methods are fairly robust to mild misspecification of the  observation process model.

\section{Application: HealthPartners Data}
\label{sect:application}

In this section, we illustrate our proposed methodology 
by applying it to the periodontal recall interval selection problem from a HealthPartners database (henceforth, HP data) of longitudinal electronic health records tracking PD status and progression, along with other covariates, among subjects around the Minneapolis, Minnesota area \citep{Guan2020bayesian}. 
The length of recall intervals continues to be a topic of debate and research 
while the current, insurance-mandated, standard of care is a recall interval of 6 months for all patients. 
However, 
recall frequencies are intimately related to patient outcomes, provider workload, and dental healthcare costs.
Recent NICE guidelines 
recommend \citep{clarkson2018interval} that recall intervals vary with time and be determined based on disease levels and oral disease risk. 
Overall,  the selection of an appropriate personalized recall interval is a multifaceted clinical decision 
that is challenging to evaluate mechanistically. 
Our HP dataset consists of 7654  
dental visit records from 1056 patients enrolled within the HP EHR from January 1, 2007, to December 31, 2014. 
For each visit record, we have the recall visit date, probed pocket depth (PPD, measured in mm) from all available tooth-sites as the PD assessment endpoint, and the clinician-recommended recall interval for the follow-up visit. The total follow-up time for these subjects has a median of 5.95 years, with a maximum of 7.97 years (resembling about 8 years of data). The frequency of recall visits has a median of 8, with a range from 3 to 29. The gap time (actual recall intervals) has a median of 181 days (approximately 6 months), with a range of 1 to 2598 days. Due to the unavailability of full-mouth, site-level measures of clinical attachment level (another important biomarker for PD) in the HP database, we refrain from using the ACES framework \citep{Holtfreter2024aces}, the modified 2018 periodontal status classification scheme to epidemiological survey data, and instead considered the mean PPD (mean across all tooth-sites at a specific visit) as a plausible subject-level PD endpoint, with a median of 2.51mm (range: 0.0--6.2mm).

To apply our methods, let $T_k$ denote the record time (in days), $X_k$ denote the gap time (in days) between $T_k$ and $T_{k-1}$, $S_k$ denote (mean) PPD measurements at $T_k$, and $A_k$ be an indicator of the recommended recall interval at $T_k$, where $A_k = 1$ implies the patient is advised to revisit within 6 months, and $A_k = 0$ otherwise. Within the HP database, the mean PPD measurements ($S_k$) are significantly correlated ($p$-value $<$0.001) with the (true) recommended recall intervals, suggesting that the mean PPD measurements play a crucial role in the dentists' decisions regarding recall intervals.
On the other hand, 
although other covariates 
were available in the dataset, only PPD and clinical attachment loss (CAL) had a significant correlation with the treatment decisions made by dentists. 
Given the high correlation between PPD and CAL, we selected PPD as the only state variable.
To measure the effects of the action $A_k$, we considered two definitions for the reward $R(T_{k+1})$. First, considering that the PPD around teeth would deepen in the presence of gum disease, and a reduction of PPD is one of the desired outcomes of periodontal therapy, 
we define the reward as an indicator of PPD reduction, i.e. $R(T_{k+1}) = I( S_{k+1} \leq S_k)$.
Based on this definition,  
the performance of any given policy $\pi$ could be evaluated in terms of the value function $V^{\pi}( s, x)$ defined in \eqref{V1}, which is based on the cumulated sum of rewards.
Moreover, since PPD $\leq$ 3 mm is usually considered a threshold to distinguish periodontally healthy sites from diseased ones, 
we also considered another definition of reward which takes the form: $R(T_{k+1}) = 3-S_{k+1}$.
In this case,  
$R(\cdot)$ could be viewed as a continuous function where $R(t)$ denotes the potential PPD measurement at time $t$, 
and therefore, a more reasonable evaluation of policies should be 
the value function $V^{\pi}_{\mathcal{I}}( s, x)$ defined in \eqref{V2},
which is based on integrated reward.

Under each value function, we considered three policies for comparison,  
namely, the observed stochastic policy $\pi_{os}$, 
the observed fixed policy $\pi_{of}$, and the optimal policy. 
Since the estimation of optimal policy depends on the choice of the value function, 
we considered two optimal policies, $\pi_{op}$ and $\pi_{op,\mathcal{I}}$, which correspond to $V^{\pi}$ and $V^{\pi}_{\mathcal{I}}$ respectively. The plots of the obtained policies are shown in Figure \ref{policies} of the Supplementary Material.
Typically, under the observed policies $\pi_{of}$ and $\pi_{os}$, patients with deeper pocket depth and more frequent recalls are more likely to be recommended a shorter recall interval,
while under the estimated optimal policies, there is no such linear relationship among actions, states, and gap times.
\begin{figure}		
\centering
 \begin{tabular}{c}
	\includegraphics[height= 4cm]{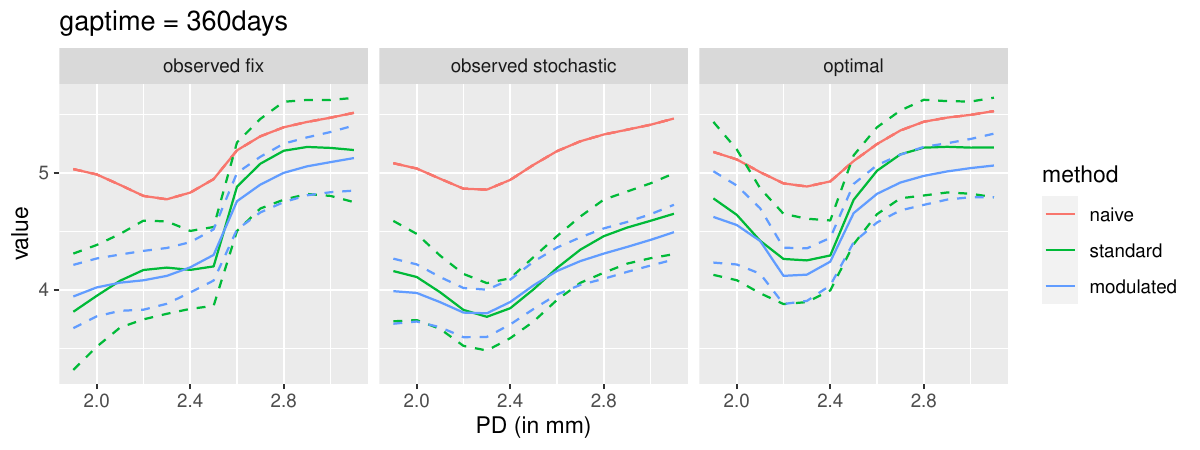}\\
         \includegraphics[height= 4cm]{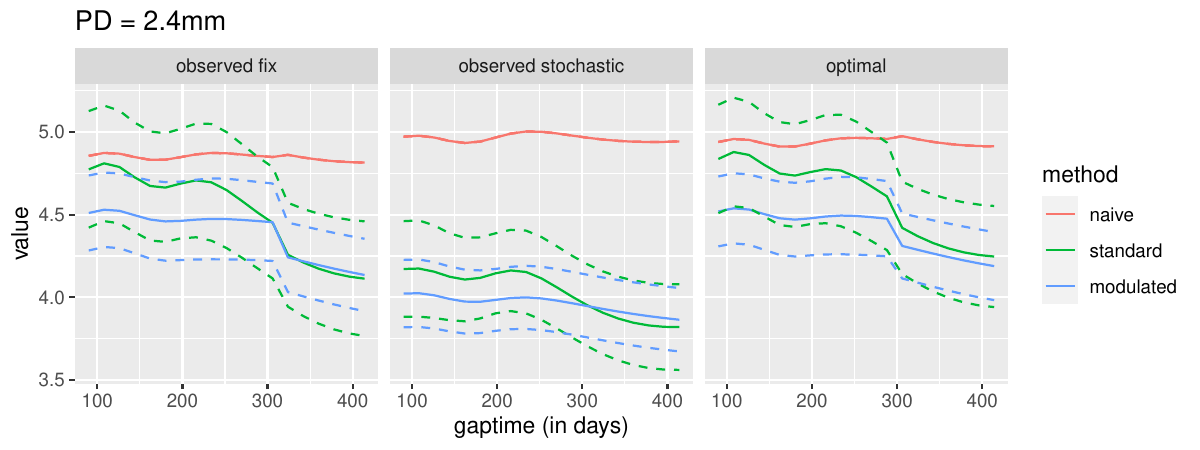}
 \end{tabular} 
 \caption{ 	
Plots of the estimated value (solid line) and confidence intervals (dashed lines) of 
the observed fix policy $\pi_{of}$, the observed stochastic policy $\pi_{os}$, and the estimated optimal policy $\pi_{op}$ with gap time fixed as 360 days (upper panel) and PPD fixed as 2.4 mm (lower panel), with the reward as an indicator of PPD reduction, i.e., $R(T_{k+1}) = I( S_{k+1} \leq S_k)$, obtained under the naive method (red line), standard method (green line), and the modulated method (blue line).  \label{values_method}}
\end{figure}

 \begin{figure}		
\centering
 \begin{tabular}{c}
	\includegraphics[height= 4cm]{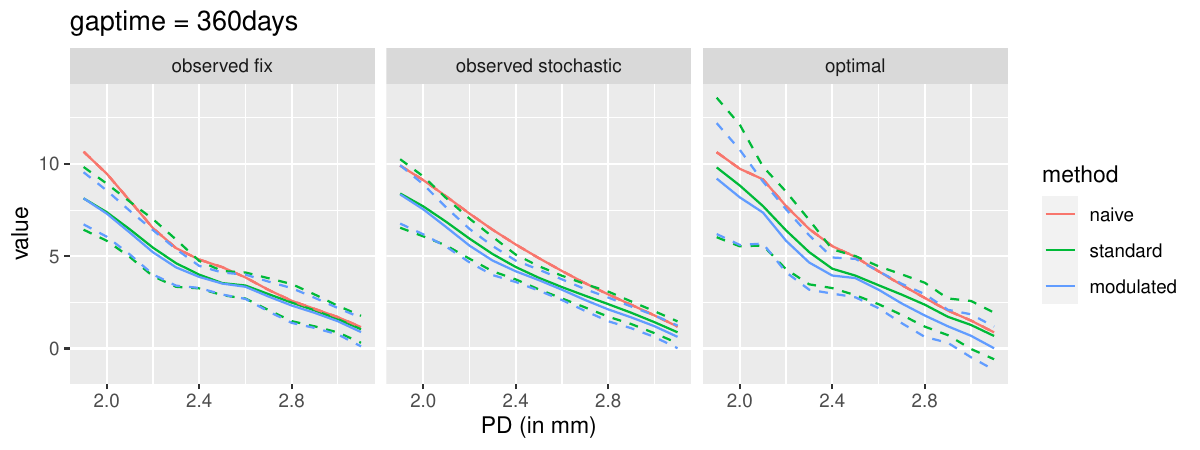}\\
         \includegraphics[height= 4cm]{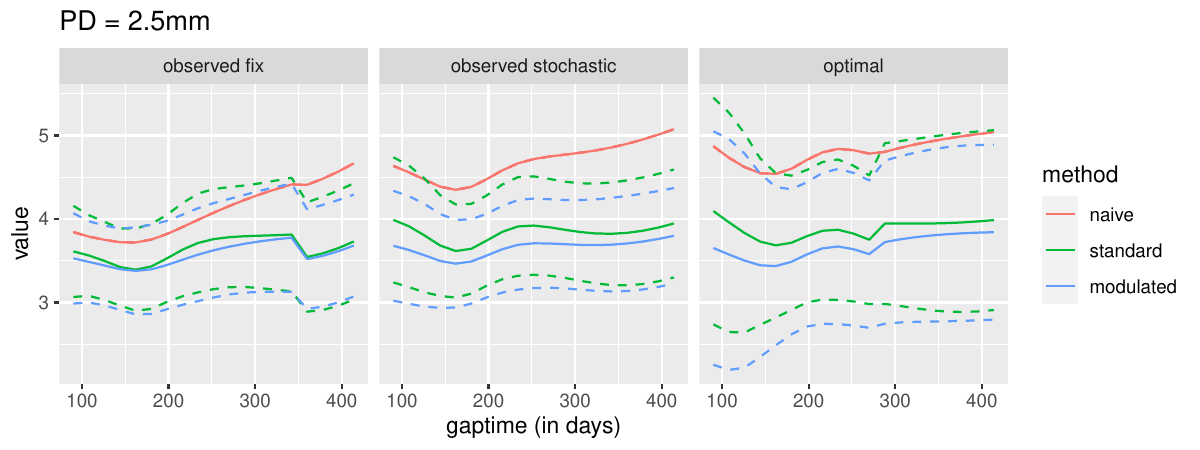}
 \end{tabular} 
 \caption{  \label{values_method_I}	
Plots of the estimated value (solid line) and confidence intervals (dashed lines) of 
the observed fix policy $\pi_{of}$, the observed stochastic policy $\pi_{os}$, and the estimated optimal policy $\pi_{op}$ with gap time fixed at 360 days (upper panel) and PPD fixed as 2.4 mm (lower panel), with the reward as PPD measurement, defined as  i.e., $R(T_{k+1}) = 3 - S_{k+1}$, obtained under the naive method (red line), standard method (green line), and the modulated method (blue line).
}
 \end{figure}

The data analysis results are summarized in Figures \ref{values_method}-\ref{values_method_I}.
In general, the estimated values change with not only the initial state but also 
the initial gap time,  which indicates that the frequency of recall visits does have effects on the  performance of the policies,  and should be taken into account in the evaluation process.
Moreover, by comparing the estimated results obtained from employing the standard method, modulated method, and naive method, we observe that the estimates from the standard method and modulated method are similar, while the results from the naive method exhibit significant differences from the former two in both the magnitude of values and the patterns of the function. For example, in Figure \ref{values_method} and when PPD is fixed at 2.4 mm, the value functions estimated by the naive method remain around 5 as the gap time varies, even though both the standard and modulated methods reveal a significant decrease in the value as the gap time values increase. Furthermore, it can also be seen that the confidence intervals provided by the modulated method are generally narrower than those given by the standard method (for example, when gap time = 360 days). This result is consistent with the findings in simulation studies.

We also compare the estimated value functions of the observed stochastic policy, observed fix policy, and the optimal policy, under the modulated method and based on cumulative reward and integrated reward respectively. The plots of the confidence intervals for the value of three policies are given in Figure \ref{values_policies} of the Supplementary Material.
From the results, it is evident that the alterations in the action assignment of the optimal policy, in comparison to the other two observed policies, effectively enhance the policy's value.
A more detailed discussion of the results is also given in section \ref{sect:addapplication} of the Supplementary Material.

\section{Discussion} \label{sect: discuss}

In this paper, we construct a framework for dynamic policies with irregularly spaced and 
outcome-dependent observation process and develop an off-policy evaluation procedure 
under revised time-homogeneous and Markov assumptions.
Similar Markov assumptions are commonly posited in the off-policy evaluation literature 
and, in real-world data applications, can be tested from the observed data using a forward-backwards learning algorithm \citep{Shi2020does}.
We also provide example schemes to demonstrate the connection between the proposed assumptions and existing ones.
In particular, we show that the proposed time-homogeneous and Markov assumption requires the observation process to satisfy some modulated renewal process assumption (see S1-x'),
and introduce an intensity-based modulated renewal process model \eqref{model-cox} in the estimation procedure. It would be an interesting but challenging task to generalize the intensity model to a rate model, under which the existing Markov assumption no longer holds, or generalize the modulated renewal process to a counting process, under which the time-homogenous assumption may not hold.
Finally, although our paper focuses on the off-policy evaluation of dynamic policies, it is not hard to adopt the proposed method to search for an optimal dynamic policy, as well as to make inferences on the value of the optimal dynamic policy.

\newpage
\appendix
{\bf Supplementary Material}

This supplementary material is organized as follows. In Section \ref{sect:proof}, we present proofs for Lemmas 1-2, Lemmas A.1-A.5 and Theorems 1-4. In Section \ref{sect:addsimulation}, we present additional simulation results. In Section \ref{sect:addapplication}, we present additional figures from the application, together with some discussion on the results.

\section{Proofs for lemmas and theorems}
\label{sect:proof}

For notational brevity, we assume $\pi$ to be a deterministic policy, with $\pi(a|s,x) \in \{  0,1\}$. Denote $\pi(s,x)\in \mathbb{A}$ as the selected action under policy $\pi$ given state $s$ and gap time $a$.
for any $a \in \mathbb{A}$, $s\in \mathbb{S}$ and $ x\in \mathbb{R}^+$. 
All derivations and results can be generalized to the evaluation of stochastic policies, straightforwardly.

\subsection{Proof of Lemmas 1-2}
\label{ssect:proof_lemma}

{\bf Proof of Lemma 1 :} 
Under assumptions A1 and A2, the existence of $Q_k^{\pi}(s, x, a, t)$ follows directly 
from the boundedness of $r$ and $\lim_{j \rightarrow \infty} T_j = \infty$.
Then, we can rewrite $Q_k^{\pi}(s, x, a, t)$ as 
{\small
\begin{align*}
&      \sum_{j \geq  k+1}
       E^{\pi} \left[
                         \gamma^{T_j - T_k} R(T_j) 
                        |
                        S_k = s, X_k = x, A_k = a ,T_k = t
                  \right]
\\
= &  
\lim_{M \rightarrow \infty }
 \sum_{j = k+1}^M
       E^{\pi} \left[
                         \gamma^{X_{k+1} + \cdots + X_j } R(T_j) 
                        |
                        S_k = s, X_k = x, A_k = a ,T_k = t
                  \right]
\\
= &  
\lim_{M \rightarrow \infty }
\bigg[  
       E\{
              \gamma^{X_{k+1}} R(T_{k+1}) |  S_k = s, X_k = x, A_k = a ,T_k = t
          \} 
 \\&
         + \sum_{j = k+2}^M  
               E^{\pi} \{
                                  \gamma^{X_{k+1} + \cdots + X_{j-1}}  \  \gamma^{X_j}R(T_j) 
                                  | S_k = s, X_k = x, A_k = a ,T_k = t
                           \}           
 \bigg]
 \\
= &  
\lim_{M \rightarrow \infty }
\bigg(  
       E\big[
           \gamma^{X_{k+1}} 
            E\{ R(T_{k+1}) | S_k, X_k, A_k , S_{k+1},X_{k+1} \}
           \big|   S_k = s, X_k = x, A_k = a ,T_k = t
          \big]
 \\&
         + \sum_{j = k+2}^M  
               E^{\pi} \big[
                             \gamma^{X_{k+1} + \cdots + X_{j-1}}  \  
                              \gamma^{X_i} 
                              E\{  R(T_i) 
                                         | S_{i-1}, X_{i-1}, A_{i-1}, X_i, S_i 
                                        \}
                            |
                   S_k = s, X_k = x, A_k = a ,T_k = t
                   \big]
  \bigg)
 \\
= &  
\lim_{M \rightarrow \infty }
\bigg( 
       E\{
                \gamma^{X_{k+1}} r( S_k, X_k, A_k , S_{k+1},X_{k+1})
           |   S_k = s, X_k = x, A_k = a ,T_k = t
          \} 
         + \sum_{j = k+2}^M  
               E^{\pi} \big[
                             \gamma^{X_{k+1} + \cdots + X_{j-1}}  \ 
  \\&                             E\{  
                                          \gamma^{X_j} 
                                          r(S_{j-1}, X_{j-1}, A_{j-1},S_j,X_j )
                                         | S_{j-1}, X_{j-1}, A_{j-1}
                                        \}
                            |
                   S_k = s, X_k = x, A_k = a ,T_k = t
                   \big]
  \bigg)
 \end{align*}
}%

\noindent Note, $\mathcal{P}$ denotes the joint distribution of the next gap time and next state. Under assumption $(A1)$, for all $k \geq 1$,  

\begin{align*}
  &E \left[
                 \gamma^{X_{k+1}} r( S_k, X_k, A_k, S_{k+1}, X_{k+1}  )
             |
                  S_k = s, X_k = x, A_k = a ,T_k 
        \right]
\\
=
&E \left[
                 \gamma^{X_{k+1}} r( S_k, X_k, A_k, S_{k+1}, X_{k+1}  )
             |
                  S_k = s, X_k = x, A_k = a 
       \right]
\\
=&
   \iint_{\mathbb{S}\times [0,\infty)}
       \gamma^{x'} r(s, x, a, s', x') d \mathcal{P}(s',x'; s,x,a)
=: \bm{g}_1(s, x, a) \,.
\end{align*}

\noindent Thus, 
\begin{align*}                       
  &Q_k^{\pi}(s, x, a, t)
 \\
 = &  
\lim_{M \rightarrow \infty }
\bigg( 
        \bm{g}_1(s, x, a)
     + \sum_{j = k+2}^M  
               E^{\pi} \big[
                             \gamma^{X_{k+1} + \cdots + X_{j-1}}  \ 
                            \bm{g}_1\{ S_{j-1}, X_{j-1}, \pi(S_{j-1},X_{j-1}) \} 
                   \big|
                    S_k = s, X_k = x,
\\
&                  A_k = a ,T_k = t
                   \big]
  \bigg)
 \\    
=&
 \lim_{M \rightarrow \infty }
 \bigg(
     \bm{g}_1(s, x, a)
   +
      E\big[ \gamma^{X_{k+1}} 
                \bm{g}_1\{  S_{k+1}, X_{k+1},  \pi( S_{k+1}, X_{k+1}) \}
                  \big|
                         S_k = s,
                        X_k = x, A_k = a ,T_k = t
                   \big] 
\\&   +
      \sum_{j =  k+3}^M   
               E^{\pi}\big[
                         \gamma^{X_{k+1} + \cdots + X_{j-2}} 
                          \gamma^{ X_{j-1}} 
                           \bm{g}_1\{  S_{j-1}, X_{j-1},  \pi(S_{j-1}, X_{j-1} ) \} 
                  \big|
                         S_k = s, X_k = x, A_k = a ,T_k = t
                   \big] 
                   \bigg)
\\
=&
 \lim_{M \rightarrow \infty }
 \bigg(
     \bm{g}_1(s, x, a)
   +
      E\big[ \gamma^{X_{k+1}} 
                \bm{g}_1\{  S_{k+1}, X_{k+1},  \pi( S_{k+1}, X_{k+1}) \}
                  \big|
                         S_k = s,
                        X_k = x, A_k = a ,T_k = t
                   \big] 
\\&   +
      \sum_{j =  k+3}^M   
               E^{\pi}\big[
                         \gamma^{X_{k+1} + \cdots + X_{j-2}}                                    
                          E\big[   
                          \gamma^{ X_{j-1}} 
                           \bm{g}_1\{  S_{j-1}, X_{j-1},  \pi(S_{j-1}, X_{j-1} ) \} 
                          \big|  
                           S_{j-2}, X_{j-2}, A_{j-2}, T_{j-2}
                          \big]
 \\&                 \big|
                         S_k = s, X_k = x, A_k = a ,T_k = t
                   \big] 
                   \bigg)                 
\end{align*}             

By similar arguments,  we have 
\begin{align*}
  &     E \left[
                 \gamma^{X_{k+1}} 
                 \bm{g}_1\{ S_{k+1}, X_{k+1}, \pi( S_{k+1}, X_{k+1})             |
                  S_k = s, X_k = x, A_k = a ,T_k
        \right]
\\
=&
   \iint_{\mathbb{S}\times [0,\infty)}
         \gamma^{x'}  \bm{g}_1\{ s', x', \pi(s',x')\} 
          d \mathcal{P}(s',x'; s, x, a) 
          =: \bm{g}_2^{\pi}(s, x, a) 
\end{align*}
which does not depend on $T_k$ or $k$ for all $k \geq 1$. For all $j\geq 2$, 
define
\begin{align*}
           \bm{g}_{j+1}^{\pi}(s, x, a) 
 \equiv
          \iint_{  \mathbb{S} \times [0,\infty) }
                \gamma^{x'} 
                \bm{g}_j^{\pi} \{ s',x',\pi(s',x') \} 
                d \mathcal{P}(s',x'; s,x,a)  \,.
\end{align*}

Then, we have 

{\small 
\begin{align*}
      &Q_k^{\pi}(s, x, a, t)
\\
=&
 \lim_{M \rightarrow \infty }
 \bigg(
     \bm{g}_1(s, x, a)  +  \bm{g}_2^{\pi}(s,x,a)
      +
      \sum_{j =  k+3}^M   
               E^{\pi}\big[
                         \gamma^{X_{k+1} + \cdots + X_{j-2}}                                    
                          \bm{g}_2^{\pi}\{S_{j-2}, X_{j-2},\pi( S_{j-2},X_{j-3})\}
 \\&                 \big|
                         S_k = s, X_k = x, A_k = a ,T_k = t
                   \big] 
                   \bigg)  
\\                   
 =&
  \lim_{M \rightarrow \infty }
 \bigg(
     \bm{g}_1(s, x, a)  +  \bm{g}_2^{\pi}(s, x, a) +  \bm{g}_3^{\pi}(s, x, a)
      +
      \sum_{j =  k+4}^M   
               E^{\pi}\big[
                         \gamma^{X_{k+1} + \cdots + X_{j-3}}                                    
                          \bm{g}_3^{\pi}\{S_{j-3}, X_{j-3},\pi( S_{j-3},X_{j-3})\}
 \\&                 \big|
                         S_k = s, X_k = x, A_k = a ,T_k = t
                   \big] 
                   \bigg)  
\\                   
=&\cdots
\\
 =&
  \lim_{M \rightarrow \infty } 
  \bigg\{
   \bm{g}_1(s, x, a) +
   \sum_{j =2}^{M-k}    \bm{g}_j^{\pi}(s, x, a)
  \bigg\}
=
   \bm{g}_1(s, x, a) +   \sum_{j \geq 2}    \bm{g}_j^{\pi}(s, x, a)\,.
   \end{align*}
   }
Therefore, $Q_k^{\pi}(s, x, a, t) =  Q_0^{\pi}(s, x, a, 0)$ 
for all $k \geq 0$, $t\geq 0$, $s \in \mathbb{S}$, $x \in \mathbb{R}^+$ and $a \in \mathbb{A}$.
 $\blacksquare$

{\bf Proof of Lemma 2 (Bellman Equation)}

Under assumptions A1-A2, $ Q^{\pi}( S_k, X_k, A_k) $ can be rewritten as 
{\small
\begin{align*}
&  E^{\pi}\big\{    
                          \sum_{ j \geq k+1} 
                                    \gamma^{X_{k+1} + \cdots + X_j} 
                                    R(T_j)
                   \big|
                           S_k, X_k , A_k
         \big\}
\\
=& 
       E\big\{   
                   \gamma^{X_{k+1}} R(T_{k+1})
         \big| 
                    S_k, X_k, A_k 
         \big\}
 +
        E^{\pi}\big[    
                    \gamma^{X_{k+1}}
                    E^{\pi}\big\{
                                       \sum_{j \geq k+2} 
                                            \gamma^{X_{k+2} + \cdots + X_j} 
                                      R(T_j)
                              \big|
                                      S_{k+1}, X_{k+1},  A_{k+1},
   \\&                                      S_k  , X_k  , A_k
                               \big\}
                   \big|
                           S_k  , X_k  , A_k       
                   \big]    
\\
=& 
       E\big\{   
                   \gamma^{X_{k+1}} R(T_{k+1})
         \big| 
                    S_k, X_k, A_k 
         \big\}
 +
        E^{\pi}\big[    
                    \gamma^{X_{k+1}}
                    E^{\pi}\big\{
                                       \sum_{j \geq k+2} 
                                            \gamma^{X_{k+2} + \cdots + X_j} 
                                      R(T_j)
                              \big|
                                      S_{k+1}, X_{k+1},  A_{k+1},
                              \big\}
             \\&       \big|
                           S_k  , X_k  , A_k       
                   \big]    
\\
=& 
       E\big\{  
                   \gamma^{X_{k+1}} R(T_{k+1})
         \big| 
                    S_k, X_k, A_k 
         \big\}
 +
        E^{\pi}\big[    
                    \gamma^{X_{k+1}}
                    Q^{\pi}\{ S_{k+1}, X_{k+1}, A_{k+1} \}
                  \big|
                           S_k , X_k , A_k     
                   \big]    
\\=& 
       E\big\{  
                   \gamma^{X_{k+1}} R(T_{k+1})
         \big| 
                    S_k, X_k, A_k 
         \big\}
 +
        E\big[    
                    \gamma^{X_{k+1}}
                    Q^{\pi}\{ S_{k+1}, X_{k+1}, \pi( S_{k+1}, X_{k+1} )\}
                  \big|
                           S_k , X_k , A_k     
                   \big]    
    \,.
\end{align*}}
Therefore, the Bellman Equation in Lemma 2 holds for all $k \geq 1$.
 $\blacksquare$

\newpage
\subsection{ Proofs of Theorems 1-4. }\label{ssect:proof_thm}
For notational brevity, let 
$Q^{\pi}(s,x,\pi)$  denote $Q^{\pi}\{ s,x,\pi(s,x)\}$, 
let $Q(s,x,\pi; \bm\theta)$ denote $Q\{ s,x,\pi(s,x); \bm\theta\}$, 
let $\xi(s,x,\pi; \bm\theta)$ denote $\bm{\xi}\{ s,x,\pi(s,x); \bm\theta\}$, 
and let 
$\sum_{i,k}$ denote $\sum_{i=1}^n \sum_{k= 0}^{K_i-1} $.
For a set of $l$-dimensional vectors, 
$ \mathcal{V} = \{ v_{i,k} \in \mathbb{R}^l\}_{1\leq i \leq n, 0\leq k \leq K_i-1}$,
let $array( \mathcal{V})$ denote a $n_K \times l$ dimension array, 
whose  $(\sum_{j <i }K_j +k+1)$th row is $n_K^{1/2}v_{i,k}$,
and taking the form
$$
array( \mathcal{V} )= n_K^{1/2}(v_{1,0},v_{1,1} \cdots, v_{i,k}, \cdots, v_{n, K_n-1})^{\top}.
$$

Moreover, the following notations and conditions are required in the  proofs of Theorems 2-4. 
For $j = 0,1,2$, define 
 $
	S_{z,j}(x;\bm{\beta}) 
=
	 \frac{1}{\sum_i K_i} \sum_{i = 1}^n\sum_{k = 0}^{K_i-1} I(X_{i,k+1}\geq x)\exp(\bm{\beta}^{\top} \bm{Z}_{i,k})\bm{Z}_{i,k}^{\otimes j} \,,
$
with 
$\bm{a}^{\otimes j} = 1, \bm{a}, \bm{a}\bm{a}^{\top}$.
The convergency of $ S_{z,j}(x;\bm{\beta})$ holds under assumption $A6$.
So we denote the limit by 
$s_{z,j}(x;\bm{\beta})$  for each $x$ and $\beta$, 
and denote $s_{z,j}(x;\bm{\beta}_0)$ as $s_{z,j}(x)$
for notational briefness.
By Slutsky's theorem, 
we can further denote $\overline{z}(x) = s_{z,1}(x)/ s_{z,0}(x)$ as the limit
of $\overline{Z}(x; \bm{\beta}_0)$. 
Conditions $C1$-$C5$ are regular conditions commonly adopted in the modulated renewal process studies \citep{Pons1988cox, Lin2009pseudo, Lin2013robust}, 
and here used to construct asymptotic properties of $\mathcal{\widehat{P}}_X$ under model \ref{model-cox} in SubSection \ref{ss:modulated}.  Conditions $C6$ ensures the existence of $\sigma_{\pi, \mathcal{G},2}$.

 \begin{itemize}
 \item[C1]  There exists a compact closure $\overline{\mathcal{B}}$ surrounding $\bm{\beta}_0$, 
 such that $s_{z,0}(\beta, x)$ is bounded away from zero on 
 $\overline{\mathcal{B}} \times [0, \tau]$.
 \item[C2] 
 $\bm{\Omega}_z = 
 \int_0^{\tau} 
 	\{ 	s_{z,0}^{-1}(x) s_{z,2}(x)
		- 
		\overline{z}^{\otimes 2}(x)
	\}
	s_{z,0}(x)
	d\Lambda_0(x)
	$
	is positive definite.
\item[C3] 
$n_K^{1/2}\{ E S_{z,j}(x, \bm{\beta}_0) - s_{z,j}(x) \}$ is finite 
for $ j = 0, 1, 2$ and for each $x$ in a dense subset of $[0,\tau]$ including 
0 and $\tau$.
\item[C4] 
For any $\epsilon > 0 $, 
$ n_K^{-1} 
\sum_{i,k}  \|\bm{Z}_{i,k} \|_2^{2} 
		 I\{ \| \bm{Z}_{i,k}  \|_2^2 > n_K \epsilon \} 
		 \exp( \bm{\beta}_0^{\top} \bm{Z}_{i,k})
		\Lambda_0(\tau) 
		 \stackrel{p}{\rightarrow} 0\,.
$
\item[C5]
$n_K^{1/2} \{ S_{z,j}(x, \bm{\beta}_0) - s_{z,j}(x, \bm{\beta}_0) \}$ is tight on $[0,\tau]$
for $j =0,1,2$ and $n_K \geq 1$.
\item[C6]
$\bm{\Omega}_{\pi,2}$ defined in \eqref{R2_3}, 
$\bm{\Omega}_{\pi,3}$ defined in \eqref{R3_2},
and $\bm{\Omega}_{\pi,4}$ defined in \eqref{R4_1}
are positive definitive. 
 \end{itemize}

 {\bf Proof of Theorem 1: }
 
The proof of Theorem 1 is similar to Appendix E.1 in \cite{Shi2021statistical}, 
so we omit the technical details and give an outline of the proof.

Firstly, 
by 
Lemma 1 of \cite{Shi2021statistical}
and Section 2.2 of \cite{Huang1998projection}, 
there exists 
$ \bm\theta^{*}_{\pi} \in \mathbb{R}^{mL}$ 
that satisfies
\begin{align}
\sup_{s \in \mathbb{S},x\in \mathbb{R}^+, a\in \mathbb{A}} 
| Q^{\pi}(s,x,a) - \xi(s,x,a)^{\top} \bm\theta_{\pi}^* |
\leq 
C_1  L^{-p/(d+1)}
\label{R1_1}
\end{align}
for some constant $C_1$. 
Then 
by the definition of $\widehat{\bm{\theta}}_{\pi}$, 
we can rewrite $\widehat{\bm{\theta}}_{\pi}-{\bm{\theta}}_{\pi}^*$ as
\begin{align*}
& 
\widehat{\bm{D}}_{\pi}^{-1} 
\big[
 \frac{1}{n_K}  \sum_{i,k} 
	 	\bm{\xi}_{i,k} 
		 \left\{ 
		 	   \gamma^{ X_{i,k+1}}R_i(T_{i,k+1})
			- (\bm{\xi}_{i,k} -  \gamma^{X_{i,k+1}} \bm{\zeta}_{\pi,i,k+1})^{\top}              
			   {\bm{\theta}}_{\pi}^* 
		\right\}
\big]
\\
=& 
\widehat{\bm{D}}_{\pi}^{-1} 
\big[	\frac{1}{n_K}  \sum_{i,k}
		 \bm{\xi}_{i,k} 
		 \{ 
		 \underbrace{ 
		 	\gamma^{ X_{i,k+1}}R_i(T_{i,k+1})
	           	-  
	           	Q^{\pi}(S_{i,k},X_{i,k}, A_{i,k} ) 
		 	 +  
		  	\gamma^{X_{i,k+1}} 
	 	      	Q^{\pi}(S_{i,k+1},X_{i,k+1}, \pi )
		  }_{\phi_{i,k,1}}
		  +  			
		  \epsilon_{i,k,1}
		 \}
\big]
\\
=&  
	\bm{D}_{\pi}^{-1} 
	\bm{A}_{\xi}^{\top} \bm{\phi}_1 
	+
	( \widehat{\bm{D}}_{\pi}^{-1} - \bm{D}_{\pi}^{-1} )
	\bm{A}_{\xi}^{\top} \bm{\phi}_1 
	+
	\widehat{\bm{D}}_{\pi}^{-1} 
	\bm{A}_{\xi}^{\top}  \bm{\epsilon}_1\,, 
\end{align*}
where 
{\small 
$
	\epsilon_{i,k,1}  
= 
	Q^{\pi}(S_{i,k},X_{i,k},A_{i,k}) 
	-
 	\bm{\xi}(S_{i,k},X_{i,k},A_{i,k})^{\top} \bm\theta_{\pi}^* 
         -
	\gamma^{ X_{i,k+1}} 
     	\{  
	   Q^{\pi}(S_{i,k+1},X_{i,k+1},\pi) 
	   -
	   \bm{\xi}(S_{i,k+1},X_{i,k+1},\pi)^{\top} \bm\theta_{\pi}^* 	 
    	 \}\,,
$}
$
 	 \bm{\phi}_1 
  	= 
 	array(\{ \phi_{i,k,1}\}_{1\leq i\leq n, 1\leq k \leq K_i}) \,,
$
$
 	 \bm{\epsilon}_1
	  =
	 array(\{ \epsilon_{i,k,1}\}_{1\leq i\leq n, 1\leq k \leq K_i})\,,
$
and
$
	\bm{A}_{\xi}
	= 
	array(\{ \bm{\xi}_{i,k}\}_{1\leq i\leq n, 1\leq k \leq K_i})  
 \in  \mathbb{R}^{ n_K \times mL }\,,
$
and 
$\bm{D}_{\pi}= E( \widehat{\bm{D}}_{\pi})$.
Lemma A.1 gives the properties of 
$\bm{\xi}_{i,k}$, $\phi_{i,k,1}$, $\epsilon_{i,k,1}$, 
$\bm{D}_{\pi}$ and $\widehat{\bm{D}}_{\pi}$.

{\it Lemma A.1 Under the conditions given in Theorem 1, we have
\begin{itemize}
\item[(i)]
For any $s \in \mathbb{S}, x \in \mathbb{R}^{+}$ and $a \in \mathbb{A}$,
$ 
| Q^{\pi}(s,x,a)| \leq C_Q
$ 
for some constant $C_Q$.
\item[(ii)] 
$
	E(\bm{\xi}_{i,k}^{\top} \bm{\xi}_{i,k})
= 
	O(L)$.
As $n_K \rightarrow \infty$,
$
	\lambda_{max}
	\{ 
	E( \frac{1}{n_K}\sum_{i,k} \bm{\xi}_{i,k} \bm{\xi}_{i,k}^{\top})
	\}
 = 
 	O(1)
$,
$
	\lambda_{max}
	( 
	\frac{1}{n_K}\sum_{i,k} \bm{\xi}_{i,k} \bm{\xi}_{i,k}^{\top}
	) 
= 
	O_p(1)
$, 
$
\lambda_{min} 
	\{ 
	E( \frac{1}{n_K}\sum_{i,k} \bm{\xi}_{i,k} \bm{\xi}_{i,k}^{\top})
	\}
\geq 
	\overline{c}_1/2
$,
$
\lambda_{min} 
	\{ 
	\frac{1}{n_K}\sum_{i,k} \bm{\xi}_{i,k} \bm{\xi}_{i,k}^{\top})
	\}
\geq 
	\overline{c}_1/3 
$ 
wpa 1.
\item[(iii)]
$ 
	\max_{1\leq i\leq n, 1\leq k \leq K_i} 
	|  \phi_{i,k,1} | \leq C_R + 2 C_Q
$,
$
	\max_{1\leq i\leq n, 1\leq k \leq K_i}
	| \epsilon_{i,k,1} | \leq 2 C_1 L^{- \frac{p}{d+1}}
$.
\item[(iv)] 
As $n_K \rightarrow \infty$,
$
	\| \bm{D}_{\pi} \|_2 = O(1)
$,
$
	\| \bm{D}_{\pi}^{-1} \|_2 
	\leq 
	3 \overline{c}^{-1}
$,
$
	\| \widehat{\bm{D}}_{\pi} - \bm{D}_{\pi}\|_2
	=
	O_p\{L^{1/2} n_K^{-1/2} \log(n_K)\}
$,
$
	\| \widehat{\bm{D}}_{\pi}^{-1} - \bm{D}_{\pi}^{-1} \|_2
	=
	O_p\{L^{1/2} n_K^{-1/2} \log(n_K)\}
$,
$
	\| \widehat{\bm{D}}_{\pi}^{-1} \|_2 
	\leq 
	6 \overline{c}^{-1}	
$
wpa 1.
\end{itemize}
}

Together with inequality \eqref{R1_1}, Markov's inequality and Cauchy-schwarz inequality, 
we can  obtain 
$ 
	\|
	\bm{D}_{\pi}^{-1} \bm{A}_{\xi}^{\top} \bm{\phi}_1 
	\|_2
=
	O_p(L^{1/2} n_K^{-1/2} )
$,
$
	\|
	( \widehat{\bm{D}}_{\pi}^{-1} - \bm{D}_{\pi}^{-1} )
	\bm{A}_{\xi}^{\top} \bm{\phi}_1 
	\|_2
	=
	O_p\{L  n_K^{-1}\log(n_K)\}
$,
$
	\|
	\widehat{\bm{D}}_{\pi}^{-1} 
	\bm{A}_{\xi}^{\top}  \bm{\epsilon}_1
	\|_2
	=
	O_p(L^{-\frac{p}{d+1}})
$,
and therefore
 \begin{align}
 \widehat{\bm{\theta}}_{\pi}-{\bm{\theta}}_{\pi}^*
 =
 \bm{D}_{\pi}^{-1} \bm{A}_{\xi}^{\top} \bm{\phi}_1 
 + 
 O_p\{ L n_K^{-1}  \log(n_K)\}
 +
 O_p( L^{-\frac{p}{d+1}})
 \,.
 \label{R1_2}
 \end{align}
 
Next, we study the asymptotic properties of 
$\widehat{V}^{\pi}(\mathcal{G})  - V^{\pi} (\mathcal{G})$.
Recall that 
 $\widehat{V}^{\pi}(\mathcal{G}) = \bm{\zeta}_{\pi, \mathcal{G}}^{\top} \widehat{\bm\theta}$,
define 
 $
	\sigma^2_{\pi, \mathcal{G},1}
	=
  	 \bm{\zeta}_{\pi, \mathcal{G}}^{\top} 
  	\bm{D}_{\pi}^{-1} 
   	\bm{\Omega}_{\pi,1}
   	(\bm{D}_{\pi}^{\top})^{-1}
   	\bm{\zeta}_{\pi, \mathcal{G}}
$
with 
$
\bm{\Omega}_{\pi,1}  
 =   
 E( \bm{A}_{\xi}^{\top} \bm{\phi}_1 \bm{\phi}_1^{\top} \bm{A}_{\xi} ).
 $.
  Lemma A.2 gives the properties of 
$\sigma_{\pi,\mathcal{G},1}^2$
and $\bm{\Omega}_{\pi, \mathcal{G},1}$.

{\it Lemma A.2
Under the conditions given in Theorem 1, we have:
(i) 
$\| \bm{\Omega}_{\pi,1} \|_2 = O(1)$,
(ii)
$\sigma_{\pi, \mathcal{G},1}^2 \geq C_2 \| \bm{\zeta}_{\pi, \mathcal{G}} \|_2^2$
for some constant $C_2$.
}

By \eqref{R1_1}, \eqref{R1_2} and Lemma A.2, 
it can be obtained that
\begin{align*}	
\sigma_{\pi, \mathcal{G},1}^{-1} 
\{ \widehat{V}^{\pi}(\mathcal{G})  - V^{\pi} (\mathcal{G})  \}
=&
\sigma_{\pi, \mathcal{G},1}^{-1} 
\left[
	 \bm{\zeta}_{\pi, \mathcal{G}}^{\top}
	 ( \widehat{\bm\theta}_{\pi} - \bm\theta_{\pi}^*)
	 -
	 \int
	  \{ 
	     Q^{\pi}(s, x,\pi ) - \xi(s, x,\pi)^{\top} \bm\theta_{\pi}^* 
	  \}
	 d \mathcal{G}(s, x) 
\right]
\\ =&
	\sigma_{\pi, \mathcal{G},1}^{-1} 
	\bm{\zeta}_{\pi, \mathcal{G}}^{\top} 
	\bm{D}_{\pi}^{-1} \bm{A}_{\xi}^{\top} \bm{\phi}_1
	+
	O_p\{    L  n_K^{-1} \log(n_K)\}
 	+
 	O_p\{ L^{-\frac{p}{d+1}} (1+ \| \bm{\zeta}_{\pi, \mathcal{G}} \|_2^{-1}) \}
\\=&
	\sigma_{\pi, \mathcal{G},1}^{-1} 
	\bm{\zeta}_{\pi, \mathcal{G}}^{\top} 
	\bm{D}_{\pi}^{-1} \bm{A}_{\xi}^{\top} \bm{\phi}_1
	+o_p(n_K^{-1/2})
\,,
\end{align*}
when
$L \ll { n_K^{1/2}}/ { \log(n_K)}$
and 
$L^{\frac{2p}{d+1}} 
\gg 
n_K (1+ \| \bm{\zeta}_{\pi, \mathcal{G}} \|_2^{-2} )
$.
Using the martingale central limit theorem 
\citep[corrollary 2.8]{McLeish1974dependent} and 
following arguments in section E.5 of \cite{Shi2021statistical},  
we have
$n_K^{1/2} \sigma_{\pi,\mathcal{G},1}^{-1} 
   	\{ \widehat{V}^{\pi}(\mathcal{G})  - V^{\pi}(\mathcal{G})  \}
\stackrel{d}{\rightarrow } N(0,1)
$.
$\blacksquare$

\newpage
{\bf Proof of Theorem 2: }

Theorem 2 will be proved by three steps. 
In the first step, we give the asymptotic properties of $ \widehat{\mathcal{P}}_X(x; \bm{z})$ under the modulated renewal process model \eqref{model-cox} in SubSection \ref{ss:modulated}. 
In the second step, we use the properties of 
$ \{ \widehat{\mathcal{P}}_X(x; \bm{z})- \mathcal{P}_X(x; \bm{z}) \}$
to obtain 
$
\widetilde{\bm{\theta}}_{\pi} -{\bm{\theta}}_{\pi}^* 
=  
\bm{D}_{\pi}^{-1} 
( \bm{A}_{\xi}^{\top} \bm{\phi}_2 + \bm{A}_{I}^{\top} \bm{\varphi}_2 )
+ 
 O_p\{ L  n_K^{-1} \log(n_K)\}
 +
 O_p( L^{-\frac{p}{d+1}})
 \,.
$
Lastly, we give the asymptotic properties of 
$n_K^{1/2} \sigma_{\pi,\mathcal{G},2}^{-1} 
   	\{ \widetilde{V}^{\pi}(\mathcal{G})  - V^{\pi}(\mathcal{G})  \}
$
and finishes the proof.

{\it Step1. Asymptotic properties of 
$\widehat{\mathcal{P}}_X$}.

Define 
$
	M_{i,k+1}(x) 
=
	 \int_0^x \{ 
			dN_{i,k+1}(u) - 
			I(X_{i,k+1}\geq u)
			\exp(\bm{\beta}_0^{\top} 
			\bm{Z}_{i,k})d\Lambda_0(u)
		  	\}
$
and 
$
\bm{\phi}_{i,k,z} 
= 
\int_0^{\tau}  \{ \bm{Z}_{i,k} - \bar{\bm{z}}(u) \} dM_{i,k+1}(u)\,.
$
Then Lemma A.3 gives the asymptotic properties for 
$\widehat{\bm{\beta}}$ and $\widehat \Lambda_0(x)$.

{\it Lemma A.3.Under the conditions given in Theorem 2, we have:
(i)
$n_K^{1/2} (\widehat{\bm{\beta}} - \bm{\beta}_0) $
converges weakly to a normal distribution 
$\mathcal{N}(0, \bm{\Omega}_z^{-1} )$, 
and can be asymptotically written as
$$
	n_K^{1/2} (\widehat{\bm{\beta}} - \bm{\beta}_0) 
=  
	n_K^{-1/2}    
	\sum_{i ,k}
	\bm{\Omega}_z^{-1}
	\bm{\phi}_{i,k,z} 
	+ o_p(1)\,.
$$
(ii)
$	n_K^{1/2} \{ \widehat{\Lambda}_0(x) - \Lambda_0(x) \} $
converges weakly to a mean zero Gaussian process, 
and can be asymptotically written as
$$	n_K^{1/2} \{ \widehat{\Lambda}_0(x) - \Lambda_0(x) \} 
=  
  n_K^{-1/2}  \sum_{i,k}   
     \big[
	 \int_0^x  s_{z,0}^{-1}(u)dM_{i,k+1}(u)
	-
	 \int_0^x \overline{\bm{z}}(u)^{\top}d\Lambda_0(u) 
	\bm{\Omega}_z^{-1} \bm{\phi}_{i,k,z}  
    \big]
    + o_p(1)
$$
for each $x \in [0, \tau]$.
}

Furthermore, by Taylor's expansion,  we can rewrite 
$
n_K^{1/2}
      \{  \widehat \Lambda(x; \bm{z})  - \Lambda(x; \bm{z}) \}
$
as 
\begin{align*}
&	
	 n_K^{1/2}
	 \{
	  \widehat \Lambda_0(x) \exp( \widehat{\bm{\beta}}^{\top}\bm{z})
          - 
           \Lambda_0(x) \exp( \bm{\beta}_0^{\top}\bm{z})  	
           \}
\\
=&
	 n_K^{1/2}
	 \{  \widehat \Lambda_0(x)  - \Lambda_0(x)   \}	 
   	  \exp( \bm{\beta}_0^{\top} \bm{z} )
   	+ 
  	 \Lambda_0(x)  \exp( \bm{\beta}_0^{\top} \bm{z})
   	n_K^{1/2}
 	(\widehat{\bm{\beta}} - \bm{\beta}_0)^{\top} \bm{z}
	+ 
	O_p(n_K^{-1/2})
\\
=&
	n_K^{-1/2}
	\sum_{i,k} 
	\exp(\bm{\beta}_0^{\top}\bm{z}) 
	\big[
		\int_0^x s^{-1}_{z,0}(u) dM_{i,k+1}(u) 
		+
		\int_0^x \{  \bm{z} - \overline{\bm{z}}(u) \}^{\top} d\Lambda_0(u)  
		\bm{\Omega}_z^{-1} \bm{\phi}_{i,k,z}  
     	\big]
     	+
     	o_p(1)
\,,
   \end{align*}
   which is also $O_p(1)$ for each fixed $x$ and $\bm{z}$.
Recall that $\mathcal{P}_X(x; \bm{z}) = 1 - \exp\{ - \Lambda(x; \bm{z}) \}$.
Therefore
$
n_K^{1/2} \{ \widehat{\mathcal{P}}_X(x; \bm{z}) - \mathcal{P}_X(x; \bm{z})\}
$
is asymptotically equivalent with
\begin{align}
	n_K^{-1/2}  \sum_{i,k} 
	\exp\{- \Lambda(x; \bm{z}) +\bm{\beta}_0^{\top} \bm{z} \} 
	[
	 \int_0^x s_{z,0}^{-1}(u) dM_{i, k+1}(u) 
	 +
	  \int_0^x  \{ \bm{z} - \overline{\bm{z}}(u)\}^{\top} d\Lambda_0(u)
	   \bm{\Omega}_z^{-1} 
	   \bm{\phi}_{i,k,z} 
	]\,.
 \label{pf_Px}
\end{align}

{\it Step 2.  Asymptotic properties of $\widetilde{\bm\theta}_{\pi}$}

Denote
$
\widetilde{\bm{D}}_{\pi}=
	           		\frac{1}{\sum_i K_i } 
				\sum_{i=1}^n \sum_{k= 0}^{K_i-1}
	 				\bm{\xi}_{i,k}
					 (\bm{\xi}_{i,k}- \widetilde{\mathcal{U}}_{\pi,i,k+1}
					 )^{\top} 
$
and define					 
$$\mathcal{U}_{\pi,i,k+1} = 
 \int
           	  		\gamma^{x} 
            	 		 \bm{\xi}(S_{i,k+1}, x, \pi)
          	 		d\mathcal{P}_X( x; S_{i,k+1},S_{i,k},X_{i,k}, A_{i,k})\,.
$$

By \eqref{R1_1} and 
the definition of $\widetilde{\mathcal{U}}_{\pi,i,k+1}$ given in 
section 3.3, 
$(\widetilde{\bm{\theta}}_{\pi}-{\bm{\theta}}_{\pi}^*) $
can be rewritten as 
\begin{align*}
 &  
 \	\widetilde{\bm{D}}_{\pi}^{-1}
	\big[ 
		\frac{1}{n_K}  \sum_{i,k}
	 	\bm{\xi}_{i,k} 
		\{ 
		 	   \gamma^{ X_{i,k+1}}R_i(T_{i,k+1})
			- 
			(\bm{\xi}_{i,k} - \widetilde{\mathcal{U}}_{\pi,i,k+1})^{\top}
			{\bm{\theta}}_{\pi}^*		
		\}
	 \big]
\\
=&  
\	\widetilde{\bm{D}}_{\pi}^{-1}
	\big[ 
		\frac{1}{n_K}  \sum_{i,k}
		 \bm{\xi}_{i,k} 
		\{  
			\gamma^{ X_{i,k+1}}R_i(T_{i,k+1})
		       - 
			Q^{\pi}(S_{i,k},X_{i,k},A_{i,k}) 
			+
			 \int 
	        		 \gamma^{x} 
            	  	Q^{\pi}(S_{i,k+1}, x, \pi)
          	  	d \widehat{\mathcal{P}}_X( x; \bm{Z}_{i,k})
		   	+
			\epsilon_{i,k,2}
		\}
	\big]
\\
=& 
\	\widetilde{\bm{D}}_{\pi}^{-1}
	\big[ 
		\frac{1}{n_K}  \sum_{i,k}
		 \bm{\xi}_{i,k} 
		\{  
		\underbrace{
			\gamma^{ X_{i,k+1}}R_i(T_{i,k+1})
		       - 
			Q^{\pi}(S_{i,k},X_{i,k},A_{i,k}) 
			+
			 \int 
	        		 \gamma^{x} 
            	  	Q^{\pi}(S_{i,k+1}, x, \pi)
          	  	d \mathcal{P}_X( x; \bm{Z}_{i,k})
		 }_{\phi_{i,k,2}}
		   	+
			\epsilon_{i,k,2}
		\}
\\
&     + 
	\frac{1}{n_K}  \sum_{i,k}
		 \bm{\xi}_{i,k} 
		  \int
		  \gamma^{x} 
		  Q^{\pi}(S_{i,k+1}, x, \pi)	
		  d\{  \widehat{\mathcal{P}}_X(x; \bm{Z}_{i,k}) 
		  	-
			\mathcal{P}_X(x;\bm{Z}_{i,k}) 
		    \}
	\big]
\,,
\end{align*}
where 
$
\epsilon_{i,k,2}
  = 
      \{    
      Q^{\pi}(S_{i,k},X_{i,k},A_{i,k}) 
      - 
      \bm{\xi}_{i,k}^{\top} \bm\theta_{\pi}^* 
       \}
+
\int 
	          \gamma^{x} 
     \{  \bm{\xi}(S_{i,k+1},x,\pi)^{\top} \bm{\theta}_{\pi}^*
        -
          Q^{\pi}(S_{i,k+1},x,\pi) 
     \}
     d\widehat{\mathcal{P}}_X( x; \bm{Z}_{i,k})
  \,.
$
Lemma A.4 gives the properties of 
$\phi_{i,k,2}$, $\epsilon_{i,k,2}$ and $\widetilde{\bm{D}}_{\pi}$.

{\it Lemma A.4 Under the conditions given in Theorem 2, 
we have: 
(i)
$ 
	\max_{1\leq i\leq n, 1\leq k \leq K_i} 
	|  \phi_{i,k,2} | \leq C_R + 2 C_Q
$,
$
	\max_{1\leq i\leq n, 1\leq k \leq K_i}
	| \epsilon_{i,k,2}  | \leq  3 C_1 L^{- \frac{p}{d+1}}
$
wpa 1.
(ii)
$
	\| \widetilde{\bm{D}}_{\pi} - \bm{D}_{\pi}\|_2
	=
	O_p\{L^{1/2} n_K^{-1/2} \log(n_K)\}
$,
$
	\| \widetilde{\bm{D}}_{\pi}^{-1} \|_2 
	\leq 
	6 \overline{c}^{-1}	
$,
$
	\| \widetilde{\bm{D}}_{\pi}^{-1} - \bm{D}_{\pi}^{-1} \|_2
	=
	O_p\{L^{1/2} n_K^{-1/2} \log(n_K)\}
$.
}

Similar to the proof of Theorem 1, 
the first term in the above equation can be written as 
$
\bm{D}_{\pi}^{-1} \bm{A}_{\xi}^{\top} \bm{\phi}_2 
 + 
 O_p\{ L n_K^{-1}  \log(n_K)\}
 +
 O_p( L^{-\frac{p}{d+1}})
$
with 
$
 	 \bm{\phi}_2
  	= 
 	array(\{ \phi_{i,k,2}\}_{1\leq i\leq n, 1\leq k \leq K_i}) \,.
$
For the second term, 
by the asymptotic expression
given in \eqref{pf_Px},
we have 
{\small
\begin{align*}
& \
	\frac{1}{n_K}  \sum_{j,l}
		 \bm{\xi}_{j,l} 
		 \int_0^{\tau}
			 \gamma^{x} 
            	 	 Q^{\pi}(S_{j,l+1}, x,\pi )
        	 \ d
	\big\{ \widehat{\mathcal{P}}_X(x; \bm{Z}_{j,l})
                 -{\mathcal{P}}_X(x; \bm{Z}_{j,l})
			\big\}		 
\\
=
&\ 
\int_0^{\tau} 
	\underbrace{
		{n_K}^{-1}  \sum_{j,l}
			\bm{\xi}_{j,l} 
			\gamma^{x} 
            	 	 Q^{\pi}(S_{j,l+1}, x,\pi )
			 \exp\{- \Lambda(x; \bm{Z}_{j,l}) +\bm{\beta}_0^{\top} \bm{Z}_{j,l}\} 		
	}_{
		\bm{G}_{\pi,1}(x)
	    }
	s_{z,0}^{-1}(x)  
	n_K^{-1}  \sum_{i,k}  dM_{i, k+1}(x) 
\\
&+
\underbrace{
	{n_K}^{-1}  \sum_{j,l}
		\int_0^{\tau} 
			\bm{\xi}_{j,l} 
			\gamma^{x} 
            	 	 Q^{\pi}(S_{j,l+1}, x,\pi )
			 \{ \bm{Z}_{j,l} - \overline{\bm{z}}(x)\}^{\top}
		d P_x(x; Z_{j,l})
	}_{
		\bm{G}_{\pi,3} 
	    }
	\bm{\Omega}_z^{-1} 
	 n_K^{-1}  \sum_{i,k}  \bm{\phi}_{i,k,z} 
\\  
&-
	\int_0^{\tau} 
		\int_x^{\tau}
		  	\underbrace{
				{n_K}^{-1}  \sum_{j,l}
				\bm{\xi}_{j,l} 
				\gamma^{u} 
            	 	 	Q^{\pi}(S_{j,l+1}, u,\pi )
				 \exp\{	- \Lambda(u; \bm{Z}_{j,l}) 
				 		+ 2 \bm{\beta}_0^{\top} \bm{Z}_{j,l}
					\} 		
			}_{
				\bm{G}_{\pi,2}(u)
		   	    }
	  	d\Lambda_0(u)
	   	s_{z,0}^{-1}(x)  
		n_K^{-1}  \sum_{i,k}  
	dM_{i, k+1}(x)
\\
& -
\underbrace{
	{n_K}^{-1}  \sum_{j,l}
		\int_0^{\tau} 
			\bm{\xi}_{j,l} 
			\gamma^{x} 
            	 	 Q^{\pi}(S_{j,l+1}, x,\pi )
			 \exp( \bm{\beta}_0^{\top} \bm{Z}_{j,l})
			 \{
			 	\bm{Z}_{j,l} \Lambda_0(x)
				-
				\int_0^x  \overline{z}(u) d\Lambda_0(u)
			\}^{\top}
		dP_X(x; \bm{Z}_{j,l})		
	}_{
		\bm{G}_{\pi,4} 
	   }
\\
& \times 	\bm{\Omega}_z^{-1} 
	n_K^{-1}  \sum_{i,k}  \bm{\phi}_{i,k,z} 
\\
=
&
\int_0^{\tau} 
	\{
		\bm{G}_{\pi,1}(x)
		-
		\int_x^{\tau} \bm{G}_{\pi,2}(u) d\Lambda_0(u)
	\}
	s_{z,0}^{-1}(x)  
	n_K^{-1}  \sum_{i,k}  
	dM_{i, k+1}(x)
+
	( \bm{G}_{\pi,3} - \bm{G}_{\pi,4}) 
	\bm{\Omega}_z^{-1} 
	 n_K^{-1}  \sum_{i,k}  \bm{\phi}_{i,k,z} 
\\
=
& \
	n_K^{-1}  \sum_{i,k} 
	\underbrace{
	\big[
		\int_0^{\tau} 
		\{
			\bm{g}_{\pi,1}(x)
			-
			\int_x^{\tau} \bm{g}_{\pi,2}(u) d\Lambda_0(u)
		\}
		s_{z,0}^{-1}(x)  
		dM_{i, k+1}(x)
+
		( \bm{g}_{\pi,3} - \bm{g}_{\pi,4}) 
		\bm{\Omega}_z^{-1} 
		\bm{\phi}_{i,k,z} 
	\big]
	}_{
		\varphi_{i,k,2}
	    }
+
o_p(n_K^{-1/2})
 \,.
\end{align*}}
Here, 
$\bm{g}_{\pi,k} = E( \bm{G}_{\pi,k})$ for $k = 1,2,3,4$, 
and 
the last equation follows from the convergence of 
$\bm{G}_{\pi,k}  $, 
which could be obtained by Assumption A6.
Lastly, 
denote
$
 \bm{\varphi}_2= 
 array(\{ \bm{\varphi}_{i,k,2}\}_{1\leq i\leq n, 1\leq k \leq K_i}) $
 and
 $
\bm{A}_{I}= array(\{ 1 \}_{1\leq i\leq n, 1\leq k \leq K_i}) 
$, 
then it is not hard to obtain  
\begin{align}
\widetilde{\bm{\theta}}_{\pi} -{\bm{\theta}}_{\pi}^* 
=
\bm{D}_{\pi}^{-1} 
( \bm{A}_{\xi}^{\top} \bm{\phi}_2 + \bm{A}_{I}^{\top}\bm{\varphi}_2 )
+ 
 O_p\{ L n_K^{-1}  \log(n_K)\}
 +
 O_p( L^{-\frac{p}{d+1}})\,.
\label{R2_2}
\end{align}

 {\it Step 3.  Asymptotic properties of $ \widetilde{V}^{\pi}(\mathcal{G}) $}

Define  
$
   \sigma^2_{\pi, \mathcal{G},2}
=
   \bm{\zeta}_{\pi, \mathcal{G}}^{\top} 
  \bm{D}_{\pi}^{-1} 
   \bm{\Omega}_{\pi,2} 
   (\bm{D}_{\pi}^{\top} )^{-1}
   \bm{\zeta}_{\pi, \mathcal{G}}
$
with 
\begin{align}
 \bm{\Omega}_{\pi,2}  
 = 
  E\{ (\bm{A}_{\xi}^{\top} \bm{\phi}_2 + \bm{A}_{I}^{\top} \bm{\varphi}_2 )
        (\bm{A}_{\xi}^{\top} \bm{\phi}_2 + \bm{A}_{I}^{\top} \bm{\varphi}_2 )^{\top} \}
 \,.
 \label{R2_3}
 \end{align}
Then similar with Lemma A.2, we can obtain
$\| \bm{\Omega}_{\pi,2} \|_2 = O(1)$
and 
$ 	\sigma_{\pi, \mathcal{G},2}^2 
	\geq 
	C_4 \| \bm{\zeta}_{\pi, \mathcal{G}} \|_2^2$
for some constant $C_4$.
Together with \eqref{R2_2}, 
$
\sigma_{\pi, \mathcal{G},2}^{-1} 
\{ \widetilde{V}^{\pi}(\mathcal{G})  - V^{\pi} (\mathcal{G})  \}
$
can be written as
\begin{align*}	
&
\sigma_{\pi, \mathcal{G},2}^{-1} 
\left[
	 \bm{\zeta}_{\pi, \mathcal{G}}^{\top}
	 ( \widetilde{\bm\theta}_{\pi} - \bm\theta_{\pi}^*)
	 -
	 \int
	  \{ 
	     Q^{\pi}(s, x,\pi ) - \xi(s, x,\pi)^{\top} \bm\theta_{\pi}^* 
	  \}
	 d \mathcal{G}(s, x) 
\right]
\\ =&
	\sigma_{\pi, \mathcal{G},2}^{-1} 
	\bm{\zeta}_{\pi, \mathcal{G}}^{\top} 
	\bm{D}_{\pi}^{-1} 
	( 
		\bm{A}_{\xi}^{\top} \bm{\phi}_2
		+
		\bm{A}_{I}^{\top} \bm{\varphi}_2
	)
	+
	O_p\{    L  n_K^{-1} \log(n_K)\}
 	+
 	O_p\{ L^{-\frac{p}{d+1}} (1+ \| \bm{\zeta}_{\pi, \mathcal{G}} \|_2^{-1}) \}
\\=&
	\sigma_{\pi, \mathcal{G},2}^{-1} 
	\bm{\zeta}_{\pi, \mathcal{G}}^{\top} 
	\bm{D}_{\pi}^{-1} 
	( 
		\bm{A}_{\xi}^{\top} \bm{\phi}_2
		+
		\bm{A}_{I}^{\top} \bm{\varphi}_2
	)
	+o_p(n_K^{-1/2})
\,,
\end{align*}
when
$L \ll { n_K^{1/2}}/ { \log(n_K)}$
and 
$L^{\frac{2p}{d+1}} 
\gg 
n_K (1+ \| \bm{\zeta}_{\pi, \mathcal{G}} \|_2^{-2} )
$.
Similar with the  proof of Theorem 1, 
we can use martingale central limit theorem 
\citep[corrollary 2.8]{McLeish1974dependent} to obtain that
$n_K^{1/2} \sigma_{\pi,\mathcal{G},2}^{-1} 
   	\{ \widetilde{V}^{\pi}(\mathcal{G})  - V^{\pi}(\mathcal{G})  \}
	\stackrel{d}{\rightarrow} N(0,1)
	\,.
$
$\blacksquare$

\newpage

{\bf Proof of Theorem 3(i): }

We first give the following lemma which is required for the Kernel-based estimates.
{\it Lemma A.5.
For any function $g(x)$, define 
$
g^*(x) 
= 
\int_0^{\tau} 
	\frac{1}{b_n} K(\frac{u-x}{b_n}) g(u) 
du
$.
Then $ | g^*(x) - g(x) | = O(b_n^2) $
if  $g(u)$  has a bounded second-order derivative in a neighborhood of $x$.
}

Theorem 3 will be proved by two steps. 
In the first step, we obtain the asymptotic properties of 
$\widehat \lambda_0(x)$ and $\widehat \lambda(x; \bm{z})$.
Then, using the properties of $\widehat \lambda(x; \bm{z})$, 
to derive the asymptotic properties of 
$\widehat{\bm\theta}_{\pi,\mathcal{I}}$ 
and  
$ \widehat{V}^{\pi}_{\mathcal{I}}(\mathcal{G})$.
\\

{\it Step 1.  Asymptotic properties of $ \widehat \lambda(x; \bm{z})$}

Rewrite 
\begin{align*}
 \widehat{\lambda}(x) - \lambda(x) 
=&
		\widehat{\lambda}_0(x)
		\exp(\widehat{\bm{\beta}}^{\top} \bm{z} )
	 	-
		 \lambda_0(x)
	         \exp(\bm{\beta}_0^{\top} \bm{z} )
\\
=&
	\{ 
		\widehat{\lambda}_0(x) 
		- 
		\lambda_0(x)
	\}
	\exp\{\bm{\beta}_0^{\top} \bm{z}\}
	+
	\lambda_0(x)
	\{  
		\exp(\widehat{\bm{\beta}}^{\top} \bm{z} )
	 	-
	         \exp(\bm{\beta}_0^{\top} \bm{z} )
	 \}
	+
	r_n(x; \bm{z})
\end{align*}
with 
$r_n(x;\bm{z}) 
= 	\{ 
		\widehat{\lambda}_0(x) 
		- 
		\lambda_0(x)
	\}
	\{  
		\exp(\widehat{\bm{\beta}}^{\top} \bm{z} )
	 	-
	         \exp(\bm{\beta}_0^{\top} \bm{z} )
	 \}
$.

For the first term, 
we first rewrite 
\begin{align*}
 \widehat{\lambda}_0(x) - \lambda_0(x)
 =&
 \int_0^{\tau} 
 	\frac{1}{b_n} K(\frac{u-x}{b_n}) 
 d \{ \widehat\Lambda_0(u) - \Lambda_0(u)\} 
 +
  \underbrace{
  \int_0^{\tau} 
 	\frac{1}{b_n} K(\frac{u-x}{b_n}) 
 	d\Lambda_0(u)
}_{\lambda_0^*(x) }
-
\lambda_0(x)
\\
=&
 \{ \widehat\lambda_0(u) - \lambda_0^*(u)\} 
 +
 \{ \lambda_0^*(u) - \lambda_0(u) \}
 \end{align*}

Following Theorem 4.2.2 of  \cite{Ramlau1983smoothing},
we can show that 
$(n_K b_n)^{1/2} \{\widehat{\lambda}_0(x) - \lambda_0^*(x) \}$ 
converges weakly to a normal distribution for each $x\in [0,\tau]$,
and therefore 
$
| \widehat{\lambda}_0(x) - \lambda_0^*(x) | = O_p(n_K^{-1/2} b_n^{-1/2})
$.
Moreover,
by Lemma A.5, 
$ | \lambda_0^*(x) - \lambda_0(x) | = O(b_n^2)$,
if $ \lambda_0(x) $ has a bounded second derivative in a neighborhood of $x$.
Therefore, 
together with Lemma A.3(ii) and Lemma A.5, 
$n_K^{1/2} 
\{
	\widehat{\lambda}_0(x)
	-
	\lambda_0(x)
\}
$
is asymptotically equivalent with 
\begin{align}
&
\int_0^{\tau}
	\frac{1}{b_n} K(\frac{u-x}{b_n})
d [n_K^{1/2} \{ \widehat{\Lambda}_0(u) - \Lambda_0(u )\} ]
+
O( n_K^{1/2} b_n^2)
\nonumber 
\\
=&
\int_0^{\tau}
	\frac{1}{b_n} K(\frac{u-x}{b_n})
	n_K^{-1/2} \sum_{i,k}
	[
		s_{z,0}^{-1}(u) dM_{i,k+1}(u)
		-
		\overline{z}^{\top}(u) d\Lambda_0(u)
		\bm{\Omega}_z^{-1}
		\phi_{i,k,z}
	]
+
O( n_K^{1/2} b_n^2 )
\nonumber
\\
=& 
n_K^{-1/2} \sum_{i,k}
\big[
	\int_0^{\tau}
		\frac{1}{b_n} K(\frac{u-x}{b_n})
		s_{z,0}^{-1}(u) 
	dM_{i,k+1}(u)
	-
	\lambda_0(x) \overline{z}^{\top}(x) 
	\bm{\Omega}_z^{-1}
	\phi_{i,k,z}
\big]
+
O( n_K^{1/2} b_n^2 )\,.
\label{R3_1_1}
\end{align}

On the other hand, 
by Taylor expansion and Lemma A.3(i), we have
\begin{align}
n_K^{1/2} \lambda_0(x) 
	\{  
		\exp(\widehat{\bm{\beta}}^{\top} \bm{z} )
	 	-
	         \exp(\bm{\beta}_0^{\top} \bm{z} )
	 \}
=&
	\lambda_0(x) \exp(\bm{\beta}_0^{\top} \bm{z} ) 
	\bm{z}^{\top} 
	n_K^{1/2} (\widehat{\beta} - \beta_0)
+
	O_p( \| \widehat{\beta} - \beta_0 \|_2^2 n_K^{1/2} ) 
\nonumber
\\
=&
	\lambda_0(x; \bm{z} ) 
	\bm{z}^{\top} 
	\bm{\Omega}_z^{-1}
	n_K^{-1/2} \sum_{i,k}
	\phi_{i,k,z}
+
	o_p(1)\,.
\label{R3_1_2}
\end{align}

Combining \eqref{R3_1_1} and \eqref{R3_1_2},
we have $r_n = O_p(n_K^{-1} b_n^{-1/2})$, 
and therefore 
{\small
\begin{align}
n_K^{1/2} \{ \widehat{\lambda}(x; \bm{z}) - \lambda(x; \bm{z}) \}
=& \
n_K^{-1/2} \sum_{i,k}
\big[
	\exp(\bm{\beta}_0^{\top} \bm{z})
	\int_0^{\tau}
		\frac{1}{b_n} K(\frac{u-x}{b_n})
		s_{z,0}^{-1}(u) 
	dM_{i,k+1}(u)
\nonumber 
\\
&+
	\lambda(x; \bm{z} ) 
	\{ \bm{z} - \overline{\bm{z}}(x) \}^{\top}
	\bm{\Omega}_z^{-1}
	\phi_{i,k,z}
\big]
+	O_p( n_K^{1/2} b_n^2 + n_K^{-1/2} b_n^{-1/2})
\,.
\label{R3_1_3}
\end{align}
}

{\it Step 2.  Asymptotic properties of $ \widehat{\bm\theta}_{\mathcal{I}}$}

Similar with \eqref{R1_1}, 
there exists 
$ \bm\theta^{*}_{\mathcal{I},\pi} \in \mathbb{R}^{mL}$ 
that satisfies
$
\sup_{s \in \mathbb{S},x\in \mathbb{R}^+, a\in \mathbb{A}} 
| Q^{\pi}(s,x,a) - \xi(s,x,a)^{\top} \bm\theta_{\pi}^* |
\leq 
C_1  L^{-p/(d+1)}
$
for some constant $C_1$. 
Recall that
$R_{\mathcal{I},i}(T_{i,k+1}) 
=  \frac{R_i(T_{i, k+1})}{ \lambda( X_{i,k+1}; \bm{Z}_{ik}) }
$
and 
$\widehat{R}_{\mathcal{I},i}(T_{i,k+1}) 
 =  \frac{ R_i(T_{i, k+1})}{ \widehat \lambda(X_{i,k+1}; \bm{Z}_{ik})}
$,
we have  
\begin{align*}
&
\widehat{\bm{D}}_{\pi}(\widehat{\bm{\theta}}_{\pi,\mathcal{I}}^*- {\bm{\theta}}_{\pi,\mathcal{I}}^*) 
\\
= &
 n_K^{- 1}  \sum_{i,k} \bm{\xi}_{i,k} 
 \big\{ \gamma^{X_{i,k+1}}  \widehat{R}_{\mathcal{I},i}(T_{i,k+1})  
 	-(\bm{\xi}_{i,k}- \gamma^{X_{i,k+1}} \bm{\zeta}_{\pi,i,k+1})^{\top}
	  {\bm{\theta}}_{\pi,\mathcal{I}}^*
 \big\}
\\
=& 
\underbrace{
n_K^{- 1}  \sum_{i,k} \bm{\xi}_{i,k} 
\big\{  \gamma^{X_{i,k+1}} R_{\mathcal{I},i}(T_{i,k+1})   
	- (\bm{\xi}_{i,k}- \gamma^{X_{i,k+1}} \bm{\zeta}_{\pi,i,k+1})^{\top} 
	  {\bm{\theta}}_{\pi,\mathcal{I}}^*
 \big\}
 }_{ J_1}
\\
&+ 
n_K^{- 1} \sum_{i,k}  
	 \frac{\bm{\xi}_{i,k}  \gamma^{X_{i,k+1}} R_{\mathcal{I},i}(T_{i,k+1})
	        }{\lambda(X_{i,k+1}; \bm{Z}_{i,k}) }
	 \big\{  \widehat \lambda( X_{i,k+1}; \bm{Z}_{i,k})
	 	- \lambda( X_{i,k+1}; \bm{Z}_{i,k})
	\big\}
+ O_p(n_K^{-1} b_n^{-1}) 
\\
=&
	J_1
+
\int_0^{\tau}
\underbrace{
	\frac{1}{n_K}  \sum_{i,k}  
 	\bm{\xi}_{i,k}  
	\gamma^{X_{i,k+1}} 
	R_{\mathcal{I},i}(T_{i,k+1})
	\{\lambda_0(X_{i,k+1})\}^{-1}
	\frac{1}{b_n} K(\frac{u- X_{i,k+1}}{b_n})
}_{	\bm{G}_5(u)  }
	s_{z,0}^{-1}(u) 
	n_K^{-1} \sum_{j,l}
dM_{j,l+1}(u)
\\
&+
\underbrace{
	\frac{1}{n_K}  \sum_{i,k}  
 	\bm{\xi}_{i,k}  
	\gamma^{X_{i,k+1}} 
	R_{\mathcal{I},i}(T_{i,k+1})
	\{ \bm{Z}_{i,k} - \overline{\bm{z}}(X_{i,k+1}) \}^{\top}
}_{\bm{G}_6}
	\bm{\Omega}_z^{-1}
	n_K^{-1} \sum_{j,l}
	\phi_{j,l,z}
+O_p( b_n^2)  + O_p( n_K^{-1} b_n^{-1})\\
=& J_1 + J_2 + J_3 + 
O_p( b_n^2)  + O_p( n_K^{-1} b_n^{-1})
\end{align*}

As shown in the proof of Theorem 1, 
\begin{align*}
J_1 
=&
\frac{1}{n_K}  \sum_{i,k} \bm{\xi}_{i,k} 
	\Big\{ 
		\underbrace{ 
			\gamma^{X_{i,k+1}} R_{\mathcal{I},i}(T_{i,k+1})   
			 -  
	 		Q^{\pi}_{\mathcal{I}}(S_{i,k},X_{i,k}, A_{i,k} ) 
			+  
			\gamma^{X_{i,k+1}} 
			Q^{\pi}_{\mathcal{I}}(S_{i,k+1},X_{i,k+1}, \pi )
		}_{\phi_{i,k,3}}
		  +  			
		  \epsilon_{i,k,3}
		  \Big\}
\\
= & 
	\frac{1}{n_K}  \sum_{i,k} 
	\bm{\xi}_{i,k}  \phi_{i,k,3} 
	+
 	O_p( L^{-\frac{p}{d+1}})
\end{align*}
where\\
{\small
$
	\epsilon_{i,k,3}  
= 
	Q_{\mathcal{I}}^{\pi}(S_{i,k},X_{i,k},A_{i,k}) 
	-
 	\bm{\xi}(S_{i,k},X_{i,k},A_{i,k})^{\top} \bm\theta_{\pi,\mathcal{I}}
         -
	\gamma^{ X_{i,k+1}} 
     	\{  
	   Q^{\pi}_{\mathcal{I}}(S_{i,k+1},X_{i,k+1},\pi) 
	   -
	   \bm{\xi}(S_{i,k+1},X_{i,k+1},\pi)^{\top} \bm\theta_{\pi,\mathcal{I}} 
    	 \}\,.
$}

For the second term $J_2$, 
since
\begin{align*}
 &E \big\{ 
 	 \frac{1}{n_K}  \sum_{i,k}  
 	\bm{\xi}_{i,k}  \gamma^{X_{i,k+1}} R_{\mathcal{I},i}(T_{i,k+1})
	\{\lambda_0(X_{i,k+1})\}^{-1}
	\frac{1}{b_n} K(\frac{u- X_{i,k+1}}{b_n})
      \big\}
\\
=&
  E\big \{  
   	 \frac{1}{n_K}  \sum_{i,k} 
	 \bm{\xi}_{i,k} 
	 \int_0^{\infty} 
        		\frac{1}{b_n} K(\frac{x-u}{b_n}) 
	  	\frac{
		 	 \gamma^{x} 
			 r(S_k,X_k,A_k,S_{k+1},x)
		}{
			 \lambda(x; \bm{Z}_{ik})\lambda_0(x)
		}
	 d\mathcal{P}_X(x; \bm{Z}_{i,k})
 \big\}
\\
=&
  E\big[ 
  	 \frac{1}{n_K}  \sum_{i,k}  
	    	\bm{\xi}_{i,k} 
		\int_0^{\infty} \frac{1}{b_n} K(\frac{x-u}{b_n}) 
 		\gamma^{x} r(S_k,X_k,A_k,S_{k+1},x)
		\{\lambda_0(x)\}^{-1}
	 	\exp\{ - \Lambda(x;\bm{Z}_{ik})\}
	 dx
\big]
\\
=&
 \underbrace{
	E\big[  
    	 \frac{1}{n_K}  \sum_{i,k}  
	\bm{\xi}_{i,k} 
 	 \gamma^{u} 
	 r(S_k,X_k,A_k,S_{k+1},u)
	 \{\lambda_0(u)\}^{-1}
	 \exp\{ - \Lambda(u;\bm{Z}_{ik})\}
   }_{\bm{g}_5(u)}
\big]
+O(b_n^2)\,,
\end{align*}
when 
$r(s,x,a,s',\cdot)$ has a bounded second-order derivative in a neighborhood of $u$.
Then 
$J_2$ is can be asymptotically written as 
\begin{align*}
	n_K^{-1} \sum_{i,k} 
	\int_0^{\tau} 
	\bm{g}_5(u) s_{z,0}^{-1}(u) 
	d M_{i,k+1}(u)
+
	O_p(b_n^2)
\,.
\end{align*}

For the third term $J_3$,
denote 
$ \bm{g}_6 = E(\bm{G}_6 )$,
then
$ J_3 
= n_K^{-1} \sum_{i,k}
\bm{g}_6 \bm{\Omega}_z^{-1}\phi_{i,k,z}
+o_p(n_K^{-1/2})
$.
Define 
$
\varphi_{i,k,3}
=
\int_0^{\tau}  \bm{g}_5(u) s_{z,0}^{-1}(u) 
		d M_{i,k+1}(u)
		+
		\bm{g}_6 \bm{\Omega}_z^{-1}\phi_{i,k,z}
$,
then
$
 	\widehat{\bm{D}}_{\pi}
	(
		\widehat{\bm{\theta}}_{\pi,\mathcal{I}}
		-
		{\bm{\theta}}_{\pi,\mathcal{I}}^*
	) 
$ 
is asymptotically equivalent with 
 $ n_K^{-1} \sum_{i,k} 
 (	
	\bm{\xi}_{i,k} \phi_{i,k,3}
	+
	\varphi_{i,k,3}
 )
 $.
 Together with Lemma A.1(iv), 
we have 
\begin{align*}
	\widehat{\bm{\theta}}_{\pi,\mathcal{I}}^*
	-
	{\bm{\theta}}_{\pi,\mathcal{I}}^*
= 
\bm{D}_{\pi}^{-1}  
(\bm{A}_{\xi}^{\top} \bm{\phi}_3 + \bm{A}_{I}^{\top} \bm{\varphi}_3 )
   +
  O_p( L^{-\frac{p}{d+1}})
  +
  O_p( b_n^2 )
  +
  O_p(b_n^{-1}n_K^{-1})
  \,,
\end{align*}
where 
 $
 \bm{\phi}_3=   array(\{ \phi_{i,k,3}\}_{1\leq i\leq n, 1\leq k \leq K_i})^{\top} 
 $
 and 
 $
 \bm{\varphi}_3
 = 
 array(\{ \bm{\varphi}_{i,k,3}\}_{1\leq i\leq n, 1\leq k \leq K_i})^{\top}  $.
Define  
$
   \sigma^2_{\pi, \mathcal{G},3}
=
   \bm{\zeta}_{\pi, \mathcal{G}}^{\top} 
  \bm{D}_{\pi}^{-1} 
   \bm{\Omega}_{\pi,3} 
   (\bm{D}_{\pi}^{\top} )^{-1}
   \bm{\zeta}_{\pi, \mathcal{G}}
$
with 
\begin{align}
 \bm{\Omega}_{\pi,3}  
 = 
  E\{ (\bm{A}_{\xi}^{\top} \bm{\phi}_3 + \bm{A}_{I}^{\top} \bm{\varphi}_3 )
        (\bm{A}_{\xi}^{\top} \bm{\phi}_3 + \bm{A}_{I}^{\top} \bm{\varphi}_3 )^{\top} \}
 \,.
 \label{R3_2}
 \end{align}
Then 	
$	n_K^{1/2}
	\sigma_{\pi, \mathcal{G},3}^{-1} 
	\{ 
		\widehat{V}^{\pi}_{\mathcal{I}}(\mathcal{G})  
		- 
		V^{\pi}_{\mathcal{I}} (\mathcal{G})  
	\}
$
is asymptotically equivalent  with 
$$
	n_K^{1/2}
	\sigma_{\pi, \mathcal{G},3}^{-1} 
	\bm{\zeta}_{\pi, \mathcal{G}}^{\top} 
	\bm{D}_{\pi}^{-1} 
	(
		\bm{A}_{\xi}^{\top} \bm{\phi}_3
		 +
		 \bm{A}_{I}^{\top} \bm{\varphi}_3
	)
+	o_p(1)\,,
$$
 when 
$L^{\frac{2p}{d+1}} 
\gg 
n_K (1+ \| \bm{\zeta}_{\pi, \mathcal{G}} \|_2^{-2} )
$,
 $n_K b_n^4 \rightarrow 0$ and 
 $n_K b_n^2 \rightarrow \infty$.
Lastly,
by the martingale central limit theorem,
$	n_K^{1/2}
	\sigma_{\pi, \mathcal{G},3}^{-1} 
	\{ 
		\widehat{V}^{\pi}_{\mathcal{I}}(\mathcal{G})  
		- 
		V^{\pi}_{\mathcal{I}} (\mathcal{G})  
	\}
$
converges weakly to a normal variable with mean 0 and variance 1.
$\blacksquare$

{\bf Proof of Theorem 3(ii): }
For $1\leq j \leq 4$, 
let $\bm{G}_{\mathcal{I},\pi,j}$ denote the quantities 
obtained by replacing 
$Q^{\pi}$
in 
$\bm{G}_{\pi,j}(x)$
with 
$Q^{\pi}_{\mathcal{I}}$,
and let 
$\bm{g}_{\mathcal{I},\pi,j}(x)$
denotes the expectations of 
$\bm{G}_{\mathcal{I},\pi,j}(x)$.
Then 
following the proofs of Theorems 1-3, 
$
	\widetilde{\bm{\theta}}_{\pi,\mathcal{I}}^*
	-	{\bm{\theta}}_{\pi,\mathcal{I}}^*
$ can be rewritten as  
\begin{align*}
& \widetilde{\bm{D}}_{\pi}^{-1}
\bigg[
 	\frac{1}{n_K}  \sum_{i,k} 
		\bm{\xi}_{i,k} 
		 \{ 
		 	\gamma^{X_{i,k+1}}  
			\widehat{R}_{\mathcal{I},i}(T_{i,k+1})  
			-
			(	\bm{\xi}_{i,k} 
				- 
				\gamma^{X_{i,k+1}}
				 \widetilde{\mathcal{U}}_{\pi,i,k+1}
			)^{\top}
	  		{\bm{\theta}}_{\pi,\mathcal{I}}^*
 		  \}
\bigg]
\\
=& 
	\widetilde{\bm{D}}_{\pi}^{-1}
	\bigg[
 		\frac{1}{n_K}  \sum_{i,k} 
			\bm{\xi}_{i,k} 
			\{
	 			\gamma^{X_{i,k+1}} 
				R_{\mathcal{I},i}(T_{i,k+1})   
	 			- 
	 			(	\bm{\xi}_{i,k}
	 			 	- 
				  	\gamma^{X_{i,k+1}}  
				  	\mathcal{U}_{\pi,i,k+1}
				 )^{\top} 
				 {\bm{\theta}}_{\pi,\mathcal{I}}^*
			  \}
\\
& + 
 \frac{1}{n_K}  \sum_{i,k}  
	 \frac{\bm{\xi}_{i,k}  \gamma^{X_{i,k+1}} R_{\mathcal{I},i}(T_{i,k+1})
	        }{\lambda(X_{i,k+1}; \bm{Z}_{i,k}) }
	 \big\{  \widehat \lambda( X_{i,k+1}; \bm{Z}_{i,k})
	 	- \lambda( X_{i,k+1}; \bm{Z}_{i,k})
	\big\}
\\
&+
\frac{1}{n_K}  \sum_{i,k}  
	\bm{\xi}_{i,k}
	(\widetilde{\mathcal{U}}_{\pi,i,k+1} - \mathcal{U}_{\pi,i,k+1} )^{\top}
	{\bm{\theta}}_{\pi,\mathcal{I}}^* 
\bigg]
+ O_p(n_K^{-1} b_n^{-1}) 
\\
=&
\bm{D}_{\pi}^{-1}
\{
	\frac{1}{n_K}  \sum_{i,k}  
	(
		\bm{\xi}_{i,k} \phi_{i,k,4} 
		+
		\bm{\varphi}_{i,k,3}
		+ 
		\bm{\varphi}_{i,k,4}
	)
\}
+ 
O_p\{ L n_K^{-1} \log(n_K)\}
+O_p(L^{-\frac{p}{d+1}})
+O_p(b_n^2)
+O_p(b_n^{-1}n_K^{-1})
\,,
\end{align*}
where
$
	\phi_{i,k,4}
=
	\gamma^{ X_{i,k+1}}R_{\mathcal{I},i}(T_{i,k+1})
	-
	Q^{\pi}_{\mathcal{I}}(S_{i,k},X_{i,k},A_{i,k}) 
	+
	\int 
		\gamma^{x} 
            	 Q^{\pi}_{\mathcal{I}}(S_{i,k+1}, x, \pi)
          d\mathcal{P}_X( x; \bm{Z}_{i,k})
\,,
$
and
$
\bm{\varphi}_{i,k,4} 
=
\int_0^{\tau} 
		\{
			\bm{g}_{\mathcal{I},\pi,1}(x)
			-
			\int_x^{\tau} 
			\bm{g}_{\mathcal{I},\pi,2}(u) 
			d\Lambda_0(u)
		\}
		s_{z,0}^{-1}(x)  
		dM_{i, k+1}(x)
+
		( \bm{g}_{\mathcal{I},\pi,3} 
		- \bm{g}_{\mathcal{I},\pi,4}) 
		\bm{\Omega}_z^{-1} 
		\bm{\phi}_{i,k,z} 
$
Here, 
$\bm{\varphi}_{i,k,4}$  takes similar form with $\bm{\varphi}_{i,k,2}$ 
except that 
$\bm{g}_{\pi, j}(x)$ are replaced by $\bm{g}_{\mathcal{I},\pi,j}(x)$.

Furthermore,
define
$
 \bm{\phi}_4= 
 array(\{ \phi_{i,k,4}\}_{1\leq i\leq n, 1\leq k \leq K_i}) 
$,
$
\bm{\varphi}_4= 
array(\{ \bm{\varphi}_{i,k,4}\}_{1\leq i\leq n, 1\leq k \leq K_i}) 
$,
and
$
   \sigma^2_{\pi, \mathcal{G},4}
=
   \bm{\zeta}_{\pi, \mathcal{G}}^{\top} 
  \bm{D}_{\pi}^{-1} 
   \bm{\Omega}_{\pi,4} 
   (\bm{D}_{\pi}^{\top} )^{-1}
   \bm{\zeta}_{\pi, \mathcal{G}}
$
with 
\begin{align}
\bm{\Omega}_{\pi,4}  
 = 
  E[
  	\{ 
		\bm{A}_{\xi}^{\top} \bm{\phi}_4
		 +
		 \bm{A}_{I}^{\top} 
		 ( \bm{\varphi}_3 + \bm{\varphi}_4)
	\} 
	\{ 
		\bm{A}_{\xi}^{\top} \bm{\phi}_4
		 +
		 \bm{A}_{I}^{\top} 
		 ( \bm{\varphi}_3 + \bm{\varphi}_4)
	\}^{\top}
     ]
 \,.
 \label{R4_1}
 \end{align}
Then, 
when 
$L^{\frac{2p}{d+1}} 
\gg 
n_K (1+ \| \bm{\zeta}_{\pi, \mathcal{G}} \|_2^{-2} )
$,
 $n_K b_n^4 \rightarrow 0$ and 
 $n_K b_n^2 \rightarrow \infty$,
$$
	n_K^{1/2}
	\sigma_{\pi, \mathcal{G},4}^{-1} 
	\{ 
		\widetilde{V}^{\pi}_{\mathcal{I}}(\mathcal{G})  
		- 
		V^{\pi}_{\mathcal{I}} (\mathcal{G})  
	\}
=
	n_K^{1/2}
	\sigma_{\pi, \mathcal{G},3}^{-1} 
	\bm{\zeta}_{\pi, \mathcal{G}}^{\top} 
	\bm{D}_{\pi}^{-1} 
	\{ 
		\bm{A}_{\xi}^{\top} \bm{\phi}_4
		 +
		 \bm{A}_{I}^{\top} 
		 ( \bm{\varphi}_4 - \bm{\varphi}_3)
	\}
	+	o_p(1)\,,
$$
converges weakly to a normal distribution with mean 0 and variance 1.
$\blacksquare$

\newpage
\subsection{Proof of Lemmas A.1-A.5}
\label{ssect:proof_lemmaA}

{\bf Proof of Lemma A.1}

 By assumption A3 and Lemma 1 in \cite{Shi2021statistical},
 $Q^{\pi}(\cdot ,\cdot,a)$ is a 
 H$\ddot{o}$lder continuous function with exponent $p_0$ 
with respect to $(s,x) \in \mathbb{S}\times \mathbb{R}^+$, 
and thus (i) holds.
For (ii), 
by  Lemma E.1 in \cite{Shi2021statistical},
there exists some constant $c^*$ such that 
$\sup_{s,x}\| \bm{\Phi}_{L}(s,x)\|_2 \leq  c^*L$.
Therefore,
$
	E(\bm{\xi}_{i,k}^{\top} \bm{\xi}_{i,k}) 
\leq 
	m \sup_{s,x}\| \bm{\Phi}_{L}(s,x)\|_2^2 
\leq 
    c^*mL
$
is $O(L)$.
The other results are proved in Lemma E.3 in \cite{Shi2021statistical}.
For (iii), 
the boundedness of $\phi_{i,k,1}$ 
follows from 
$ | Q^{\pi}(s,x,a)| \leq C_Q$ 
given in Lemma A.1(i) 
and $P(\max_{ k} | R(T_{k}) | \leq c_R) = 1$ given in the conditions of Theorem 1, 
while the boundedness of $\epsilon_{i,k,1}$ follows from \eqref{R1_1}. 
Lastly, the properties of 
$\bm{D}_{\pi}$ 
and 
$\widehat{\bm{D}}_{\pi}$ given in (iv) are consistent with Lemma E.2 in  \cite{Shi2021statistical} 
and can be obtained with similar arguments.
$\blacksquare$\\

{\bf Proof of Lemma A.2}

By Lemma A.1(iii), 
$
	\bm{\Omega}_{\pi,1}  
 =   
	 E( 
 	 \frac{1}{n_K} \sum_{i,k}
 	\bm{\xi}_{i,k} \bm{\xi}_{i,k}^{\top}
	\phi_{i,k,1}^2
 	)
\leq 
	  (C_R + 2C_Q) 
  	E(
  		\frac{1}{n_K} \sum_{i,k}
 		\bm{\xi}_{i,k} \bm{\xi}_{i,k}^{\top}
       	   )
$.
Then $\| \bm{\Omega}_{\pi,1} \|_2 = O(1)$ follows from Lemma A.1(iii).

Now we prove  Lemma A.2(ii).
Recall that
$ 
	\omega_{\pi}(s,x,a) 
	=
	E( 
		\phi_{i,k,1}^2
		|
		S_k = s, X_k = x, A_k = a 
	)\,.
$
Then by $\omega_{\pi}(s, x, a) > c_R^{-1}$, 
it can be shown that
\begin{align*}
\bm{ \Omega}_{\pi,1}  
 &= 
 n_K^{-1} \sum_{i,k} E( \phi_{i,k,1}^2 \xi_{i,k} \xi_{i,k}^{\top} )
 = 
 n_K^{-1} \sum_{i,k} 
 	E\{
		\omega_{\pi}(S_{i,k}, X_{i,k}, A_{i,k})
		 \xi_{i,k} \xi_{i,k}^{\top} 
	 \}
\\
&\geq 
	c_R^{-1}  
	 n_K^{-1} \sum_{i,k} 
	 E(
	  \xi_{i,k} \xi_{i,k}^{\top}
          )\,.
\end{align*}
By Lemma A.1(ii), 
we have
$
 \lambda_{\min} ( \bm{\Omega}_{\pi,1} ) \geq 
c_R^{-1}  \overline{c}_1/2
$
. 
Moreover, 
by $\| \bm{D}_{\pi}\|_2 = O(1)$ given in Lemma A.1(iii), 
there exists some constant $\overline{C} >0 $
such that 
$
\lambda_{\min} \{ \bm{D}_{\pi}^{-1}(\bm{D}_{\pi}^{\top})^{-1} \} 
\geq \overline{C}
$.
Therefore, 
$
   \sigma^2_{\pi, \mathcal{G},1} 
=
   \zeta_{\pi, \mathcal{G}}^{\top} 
   \bm{D}_{\pi}^{-1} 
   \Omega_{\pi,1} 
   ( \bm{D}_{\pi}^{\top} )^{-1}
   \zeta_{\pi, \mathcal{G}}
\geq 
	C_2
  	\left   \| \bm{\zeta}_{\pi, \mathcal{G}} 
	\right \|_2^2 
$
with  $C_2 = \overline{C} c_R^{-1}  \overline{c}_1/2$.
$\blacksquare$\\

{\bf Proof of Lemma A.3}

We first introduce some notations. 
For 
$1\leq i \leq n$, 
$0 \leq k \leq K_i-1$,
denote 
$\{ X_{i,k+1}, \bm{Z}_{i,k}, N_{i,k+1} \}$ 
as
$\{ X_{(g)}, \bm{Z}_{(g)}, N_{(g)}\}
$
for 
$ g = \sum_{l < i } K_l +k+1$.
Then for $j = 0,1,2$, 
$S_{z,j}(x; \bm{\beta})$ can be rewritten as 
$
S_{z,j}(x; \bm{\beta})
=
n_K^{-1} \sum_{g = 1}^{n_K} 
I(X_{g} \geq x) 
\exp( \bm{\beta}^{\top} \bm{Z}_{(g)})
\bm{Z}_{(g)}^{\otimes j}
$, 
and 
$\widehat{\bm{\beta}}$ is the solution to 
$
\bm{U}_z(\bm{\beta})
=
n_K^{-1} \sum_{g = 1}^{n_K}
\int_0^{\tau} 
\{
	\bm{Z}_{(g)} - \overline{\bm{Z}}(x; \bm{\beta})
\}
d N_{(g)}(x)
=  0
$.
Under assumption A6, conditions C1-C5, 
and by Theorem 3.2 of \cite{Pons1988cox},
$n_K^{1/2}(\widehat{\bm{\beta}} - \bm{\beta}_0)$
converges weakly to a 
Gaussian variable with mean 0 and 
variance $\bm{\Omega}_z^{-1}$, 
where 
$\bm{\Omega}_z$
is defined in condition C2.
Moreover, 
$n_K^{1/2}(\widehat{\bm{\beta}} - \bm{\beta}_0)$
can be asymptotically written as 
$$
n_K^{-1/2} \sum_{g = 1}^{n_K} 
\bm{\Omega}_z^{-1} 
\int_0^{\tau} 
	\{
	  \bm{Z}_{(g)} - \overline{\bm{z}}(x)
	\}
d M_{(g)}(x)
+
o_p(1)\,,
$$
with 
$
M_{(g)}(x) 
=
\int_0^x 
   \{
   dN_{(g)}(u)
   -
   I(X_{(g)} \geq u )
   \exp(\bm{\beta}_0^{\top} \bm{Z}_{(g)})
   d\Lambda_0(u)
   \} 
$.
Rewrite  $M_{(g)}(x)$ as $M_{i,k+1}(x)$,
and 
define
$
\bm{\phi}_{i,k,z} 
= 
\int_0^{\tau}  \{ \bm{Z}_{i,k} - \bar{\bm{z}}(u) \} dM_{i,k+1}(u)
$,
then Lemma A.3(i) follows directly.

Now we derive the asymptotic properties of 
$\widehat{\Lambda}_0(x)$.
The convergence of 
$
n_K^{1/2} 
\{
  \widehat{\Lambda}_0(x)
  -
  \Lambda_0(x)
\}
$
follows from the Theorem 3.3 of \cite{Pons1988cox}.
To obtain the asymptotic expression in Lemma A.3(ii), 
we first rewrite 
$
n_K^{1/2} 
\{
  \widehat{\Lambda}_0(x)
  -
  \Lambda_0(x)
\}
$
as
\begin{align*}
& \
n_K^{1/2} 
	\{
  		\widehat{\Lambda}_0(x; \widehat{\bm{\beta}})
		  -
  		\widehat{\Lambda}_0(x; \bm{\beta}_0)
	\}
+
n_K^{1/2} 
	\{
  		\widehat{\Lambda}_0(x; \bm{\beta}_0)
		-
		 \Lambda_0(x)
	\}
\\
=&
-
	\int_0^{x} \bm{z}(u)^{\top} d\Lambda_0(u)
	n_K^{1/2}
	(\widehat{\bm{\beta}} - \bm{\beta}_0)
+
	n_K^{1/2}
	\{ \widehat{\Lambda}_0(x; \bm{\beta}_0)
		-
		 \Lambda_0(x)
  	\}
+o_p(1)\,
\end{align*}
where
$
 n_K^{1/2}
	\{ \widehat{\Lambda}_0(x; \bm{\beta}_0)
		-
		 \Lambda_0(x)
  	\}
=
n_K^{-1} \sum_{g = 1}^{n_K}
\int_0^x  s_{z,0}(u)^{-1} dM_{(g)} (u) +o_p(1)\,.
$
Then together with Lemma A.3(i), 
we have 
\begin{align*}
    n_K^{1/2} 
    \{
	\widehat{\Lambda}_0(x)
	-
	\Lambda_0(x)
     \}
=
    n_K^{-1/2} \sum_{i,k}
    \big[
    	\int_0^x 
		s_{z,0}(u)^{-1}
	dM_{i,k+1}(u)
        -
        \int_0^x
        		\overline{z}(u)^{\top}
	d\Lambda_0(u)
	\bm{\Omega}_z^{-1}
	\phi_{i,k,z}
    \big]
   +
   o_p(1)\,.
\end{align*}
$\blacksquare$\\


{\bf Proof of Lemma A.4}

We first prove  Lemma A.4(i).
The boundedness property of $\phi_{i,k,2}$ 
is similar with $\phi_{i,k,1}$, 
and therefore can be similarly obtained by the arguments i 
the proof of  Lemma A.1(iii).
For the boundedness of 
$\epsilon_{i,k,2}$, 
rewrite it as
$\epsilon_{i,k,2,1} + \epsilon_{i,k,2,2}
$, 
where
$
\epsilon_{i,k,2,1}
  = 
      Q^{\pi}(S_{i,k},X_{i,k},A_{i,k}) 
      - 
      \bm{\xi}_{i,k}^{\top} \bm\theta_{\pi}^* 
     +
\int 
	          \gamma^{x} 
     \{  \bm{\xi}(S_{i,k+1},x,\pi)^{\top} \bm\theta_{\pi}^*
        -
          Q^{\pi}(S_{i,k+1},x,\pi) 
     \}
     d\mathcal{P}_X( x; \bm{Z}_{i,k})
  \,,
$
and
$
\epsilon_{i,k,2,2}
= \int
		  \gamma^{x} 
		  \{            	 	
		   	 \bm{\xi}(S_{i,k+1}, x,\pi )^{\top}
		   	 {\bm{\theta}}_{\pi}^*
			 -
			 Q^{\pi}(S_{i,k+1}, x,\pi ) 
		  \} 
		  d  
		  \{  \widehat{\mathcal{P}}_X(x; \bm{Z}_{i,k})  
		  	-
			\mathcal{P}_X(x;\bm{Z}_{i,k})  
		    \}
$.
By \eqref{R1_1}, 
$\epsilon_{i,k,2,1}$ is bounded by 
$2C_1 L^{-\frac{p}{d+1}}$.
For $\epsilon_{i,k,2,2}$, 
since 
$
n_K^{1/2}
\{ 
	\widehat{\Lambda}_0(x)
	-
	\Lambda_0(x)
\}
$
converges to a Gaussian process with zero mean and 
continuous sample paths 
(see Theorem 3.3 in \cite{Pons1988cox}),
then,
together with the convergence of 
$n_K^{1/2}\{ \widehat{\bm{\beta}} - \bm{\beta}_0\}$,
it is not hard to obtain that 
$
n_K^{1/2}
\{ 
	\widehat{P}_X(x; \bm{z})
	-
	P_X(x;\bm{z})
\}
$
also converges weakly to a stochastic process with zero mean 
and continuous sample paths for each fixed $\bm{z}$.
This implies that 
$
\sup_{x\in [0,\tau]} 
|
\widehat{P}_X(x;\bm{z})
-
P_X(x; \bm{z})
|
\stackrel{p}{\rightarrow}
0
$
for each fixed $\bm{z}$.
Then 
$\epsilon_{i,k,2,2}$ is also bounded by 
$C_1 L^{-\frac{p}{d+1}}$ when 
$n_K$ is large enough, 
and thus we finish the proof.

To prove Lemma A.4(ii),
rewrite 
\begin{align*}
\widetilde{\bm{D}}_{\pi} 
=&
	n_K^{-1} \sum_{i,k} 
	\bm{\xi}_{i,k} 
	( \bm{\xi}_{i,k} - \widetilde{\bm{U}}_{\pi, i, k+1} )^{\top}
\\
=&
\underbrace{
	n_K^{-1} \sum_{i,k} 
	\bm{\xi}_{i,k} 
	( \bm{\xi}_{i,k} - \bm{U}_{\pi, i, k+1} )^{\top}
}_{\widetilde{\bm{D}}_{\pi,1}}	
+
\underbrace{
	n_K^{-1} \sum_{i,k} 
	\bm{\xi}_{i,k} 
	( \bm{U}_{\pi, i, k+1}  - \widetilde{\bm{U}}_{\pi, i, k+1} )^{\top}
}_{\widetilde{\bm{D}}_{\pi,2}}\,.
\end{align*}
Note that 
$
E(\widetilde{\bm{D}}_{\pi,1}) = \bm{D}_{\pi}
$,
then similar with Lemma A.1(iv), we have 
$
\| 
\widetilde{\bm{D}}_{\pi,1} - \bm{D}_{\pi}
\|
=
O_p\{ L^{1/2} n_K^{-1/2} \log(n_K) \}
$
and 
$
\| 
\widetilde{\bm{D}}_{\pi,1} 
\| = O(1)
$.

On the other hand, 
rewrite 
$\widetilde{\bm{D}}_{\pi,2} $ as 
$
n_K^{-1} \sum_{i,k} 
	\bm{\xi}_{i,k}
 	\int_0^{\tau}
		\gamma^{x}
		\bm{\xi}(S_{i,k+1}, x, \pi)^{\top}
		\bm{\theta}_{\pi}^*
	d
	\{
		\widehat{P}_X(x;  \bm{Z}_{i,k})
		-
		P_X(x;  \bm{Z}_{i,k})
	\}
$.
Then 
following \eqref{pf_Px} 
and 
with arguments similar to the proof of Theorem 2 (Step 2),
we can obtain 
$
\|
\widetilde{\bm{D}}_{\pi,2} 
\|
= O_p(n_K^{-1/2})
$
and therefore,
$
\|
	\widetilde{\bm{D}}_{\pi}
	-
	\bm{D}_{\pi}
\|_2
\leq 
\|
	\widetilde{\bm{D}}_{\pi,1}
	-\bm{D}_{\pi}
\|_2
+
\|
	\widetilde{\bm{D}}_{\pi,2}
\|
=
O_p\{ L^{1/2} n_K^{-1/2} \log(n_K)\}
$.

Furthermore, 
since 
$ \| \widetilde{\bm{D}}_{\pi} - \bm{D}_{\pi} \|_2 = o_p(1)$,
we have 
$
\| \widetilde{\bm{D}}_{\pi} - \bm{D}_{\pi} \|_2 
\leq 
\overline{c}/6
$
wpa 1. 
Together with 
$
\|
\bm{D}_{\pi}^{-1}
\|_2
\leq 
3 \overline{c}^{-1}
$
given in  Lemma A.1(iv), 
we have
\begin{align*}
\bm{v}^{\top } \widetilde{\bm{D}}_{\pi} \bm{v}
= &
\bm{v}^{\top } \bm{D}_{\pi} \bm{v}
+
\bm{v}^{\top } (\widetilde{\bm{D}}_{\pi}  - \bm{D}_{\pi}) \bm{v}
\geq 
\frac{\overline{c}}{ 3 } \| \bm{v}\|^2
- 
\frac{\overline{c}}{ 6 } \| \bm{v}\|^2
=
\frac{\overline{c}}{ 6 } \| \bm{v}\|^2 \,,
\end{align*}
for any $\bm{v} \in \mathbb{R}^{mL} $. 
This implies 
$
\| \widetilde{\bm{D}}_{\pi}^{-1} \|_2 \leq 6 \overline{c}^{-1}
$
wpa 1.
Therefore, 
\begin{align*}
\|
	\widetilde{\bm{D}}_{\pi}^{-1} 
	 -
	\bm{D}_{\pi}^{-1}
\|_2
=& 
\|
	\widetilde{\bm{D}}_{\pi}^{-1} 
	(	\widetilde{\bm{D}}_{\pi}
		-
		\bm{D}_{\pi}	 
	)
	\bm{D}_{\pi}^{-1}
\|_2
\leq 
\|
	\widetilde{\bm{D}}_{\pi}^{-1} 
\|_2
\|	
	\widetilde{\bm{D}}_{\pi}
		-
		\bm{D}_{\pi}	 
\|_2 
\|
	\bm{D}_{\pi}^{-1}
\|_2
\leq 
18 \overline{c}^{-2} 
\|	
	\widetilde{\bm{D}}_{\pi}
		-
		\bm{D}_{\pi}	 
\|_2
\end{align*}
is also  
$O_p\{ L^{1/2} n_K^{-1/2} \log(n_K)\}$.
$\blacksquare$\\

{\bf Proof of Lemma A.5 }

Similarly with \cite{Ramlau1983smoothing},
we assume the kernel 
has support within $[-1,1]$ and $0 \leq b_n \leq 1/2 $
to simplify mathematics.
Then by Taylor expansion, 
\begin{align*}
g^*(x) 
=&
\int_0^{\tau}
\frac{1}{b_n} 
K(\frac{u-x}{b_n}) g(u) du
\\
=&
\int_{-1}^1 K(t) g( x + t b_n) dt
\\
=&
\int_{-1}^{1} K(t) 
\{ g(x) + g^{'}(x)t b_n + g^{''}(x+ v_t b_n)t^2 b_n^2\}
dt  
\\
=&
g(x) 
+ 
\int_{-1}^{1} K(t) g^{''}(x+ v_t b_n)t^2 dt 
b_n^2
\,,
\end{align*}
for $0 \leq v_t \leq t$.
Therefore, 
$
|
g^*(x) - g(x)
|
\leq 
\int_{-1}^{1} K(t) t^2 dt 
\sup_{ |u-x|\leq 1/2 } | g^{''}(u)|
b_n^2
$
is 
$O(b_n^2)$ 
when 
$g(u)$ has a bounded second-order derivative in the neighborhood of $x$.
$\blacksquare$

\newpage
\section{Additional Simulation Results}
\label{sect:addsimulation}

In this section, 
we present the simulation studies under scenario 4 to 
investigate the robustness of the proposed estimates. 
Specifically, 
the dataset is generated following scheme 2, 
the estimates are calculated by the methods developed under scheme 1,
and therefore model (6) in Section 3.2 
is mis-specified.
The results are reported in Table \ref{tab-Sim3}.
From the results, 
it is not hard to see that the naive method remains biased, 
the standard estimator under cumulative reward $\widehat{V}^{\pi}(\mathcal{\mathcal{G}})$ still performances well in all settings, 
and the estimates $\widetilde{V}_{\pi}(\mathcal{\mathcal{G}})$, $\widehat{V}_{\pi, \mathcal{I}}(\mathcal{\mathcal{G}})$ and $\widetilde{V}_{\pi, \mathcal{I}}(\mathcal{\mathcal{G}})$, which are based on model (6),
demonstrate negligible bias and provide satisfactory variance estimation.
Hence, the proposed methods are fairly robust to mild misspecification of the observation process model. 

\begin{table} [!ht]
\caption{ \label{tab-Sim3} 
Policy evaluation results when the observation process model is misspecified, 
including the bias and standard deviation of the naive estimators (Bias$_N$, SD$_N$), and 
the bias, standard deviation, estimated standard error, and empirical coverage probability of $95\%$ confidence intervals for standardized estimators (Bias$_S$, SD$_S$, SE$_S$, CP$_S$) and 
modulated estimators 
(Bias$_M$,SD$_M$,SE$_M$,CP$_M$).}
\begin{center}
\def\temptablewidth{1 \textwidth}
\begin{tabular*}
{\temptablewidth}{@{\extracolsep{\fill}}ccrcrcccrccc}
 \hline
     && \multicolumn{2}{c}{Naive}&  
          \multicolumn{4}{c}{Standard}&
         \multicolumn{4}{c}{Modulated } \\
 \cline{3-4} \cline{5-8} \cline{9-12}
      $n$  & K   &$Bias_N$& $SD_N$& $Bias_S$ &   $SD_S$ & $SE_S$ & $CP_S$ & $Bias_M$ &  $SD_M$ & $SE_M$  & $CP_M$\\            
 \hline 
 && \multicolumn{10}{c}{Scenario 4 \quad \quad Cumulative Rewards \quad \quad  True value = 0.594 }\\
  \cline{3-4} \cline{5-8} \cline{9-12}
 100 &  10 & -0.426 & 0.103 & 0.006 & 0.082 & 0.081 & 0.924 & -0.017 &  0.080 & 0.078 & 0.922 \\
 200 &  10 & -0.424 & 0.064 & 0.001 & 0.055 & 0.057 & 0.940 & -0.013 &  0.053 & 0.055 & 0.930 \\
 400 &  10 & -0.428 & 0.042 & 0.001 & 0.041 & 0.040 & 0.940 & -0.009 &  0.041 & 0.039 & 0.938 \\[2mm]
  10 & 100 & -0.413 & 0.142 & 0.012 & 0.123 & 0.099 & 0.910 & -0.008 &  0.109 & 0.093 & 0.922 \\
  10 & 200 & -0.430 & 0.087 & 0.001 & 0.073 & 0.065 & 0.920 & -0.014 &  0.071 & 0.062 & 0.924 \\
  10 & 400 & -0.425 & 0.067 & 0.000 & 0.051 & 0.046 & 0.930 & -0.011 &  0.045 & 0.044 & 0.926 \\
\hline 
 &&\multicolumn{10}{c}{ Scenario 4 \quad \quad Integrated Rewards  \quad \quad  True value =  0.569}
\\
\cline{3-4} \cline{5-8} \cline{9-12}
 100 &  10 & -0.172 & 0.077 &  0.001 & 0.084 & 0.074 & 0.944 & -0.015 & 0.077 & 0.069 & 0.930 \\
 200 &  10 & -0.176 & 0.056 &  0.000 & 0.056 & 0.051 & 0.935 & -0.013 & 0.057 & 0.049 & 0.919 \\
 400 &  10 & -0.181 & 0.042 & -0.003 & 0.041 & 0.036 & 0.924 & -0.014 & 0.039 & 0.035 & 0.910 \\[2mm]
  10 & 100 & -0.170 & 0.068 &  0.000 & 0.073 & 0.072 & 0.938 & -0.015 & 0.071 & 0.068 & 0.938 \\
  10 & 200 & -0.165 & 0.051 &  0.002 & 0.054 & 0.050 & 0.942 & -0.010 & 0.052 & 0.048 & 0.938 \\
  10 & 400 & -0.168 & 0.034 &  0.001 & 0.037 & 0.035 & 0.929 & -0.009 & 0.035 & 0.034 & 0.927 \\
\hline
\end{tabular*}
\end{center}
\end{table}

\newpage  
\section{Additional Application Results}
\label{sect:addapplication}
\begin{figure}
\centering
\begin{tabular}{cc}
(a) \quad & (b) \quad  \\
\includegraphics[height= 5cm]{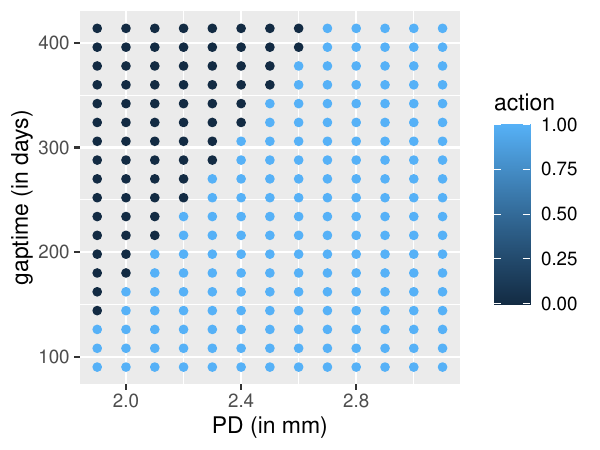}&
\includegraphics[height= 5cm]{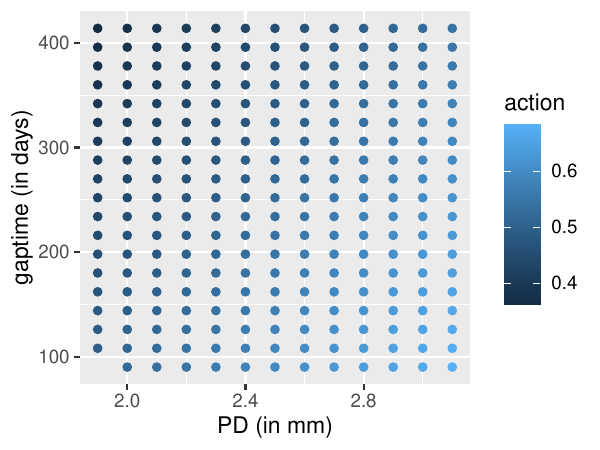}
\\
(c) \quad  & (d) \quad  \\
\includegraphics[height= 5cm]{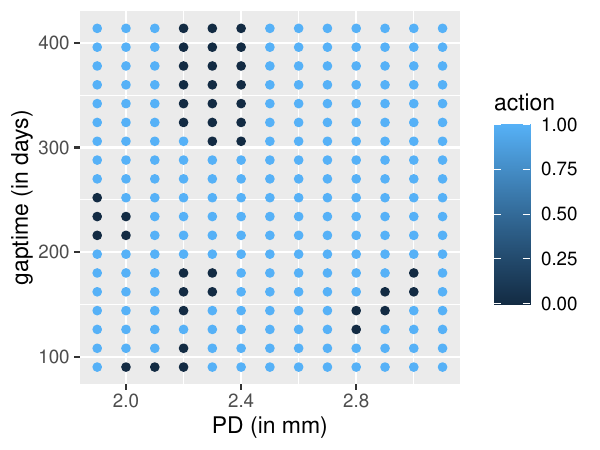}&
\includegraphics[height= 5cm]{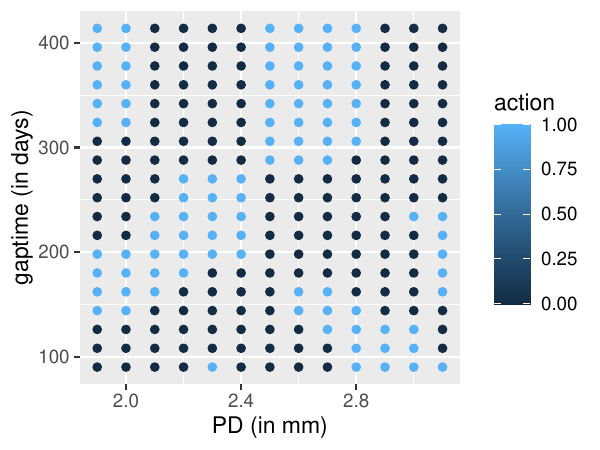}
 \end{tabular} 
 \caption{  \label{policies}	
Plots of policies, including (a) observed fix policy $\pi_{of}$; 
(b) observed stochastic policy $\pi_{os}$;
(c) estimated optimal policy based on discrete reward $\pi_{op}$;
and 
(d) estimated optimal policy based on continuous reward $\pi_{op,\mathcal{I}}$.
 }
\end{figure}

In this section, we present additional application results 
to compare the observed stochastic policy, 
observed fix policy, 
and 
the optimal policy.
The policies are estimated by 2488 visit records from 352 patients (approximately $33\%$ of the total number of study subjects) who are randomly selected from the full dataset, while the policy evaluation procedure used the remaining 5166 records from 704 patients. 
To obtain the observed stochastic policy, 
we first fitted a logistic regression model with the (binary) response $A_k$ and the explanatory variables $S_k$ and $X_k$. 
Then, 
let the estimated probability of $A_k=a$ given $(s, x)$
be $\pi_{os}(a; s, x)$, and denote $\pi_{os}$ as the observed stochastic policy.
Moreover, 
by setting the probability threshold as 0.48,
we further obtain a deterministic function mapping from $(s, x)$ to $\{0,1\}$, 
and denote it as the observed fixed policy $\pi_{of}(a; s, x)$. 
Finally, 
the optimal policy is estimated by double-fitted Q-learning \citep{Shi2021statistical}, 
where the Bellman equation for the optimal policy can be obtained by a trivial extension of Lemma 2.
Plots of the obtained policies are shown in Figure \ref{policies}. 
From the plots, it is not hard to see that, 
under the observed policies $\pi_{of}$ and $\pi_{os}$, 
patients with deeper pocket depth and more frequent recalls 
are more likely to be recommended a shorter recall interval,
while under the estimated optimal policies, 
there is no such linear relationship among actions, states, and gap times.

\begin{figure}		
\centering
 \begin{tabular}{cc}
 	\includegraphics[height= 5cm]{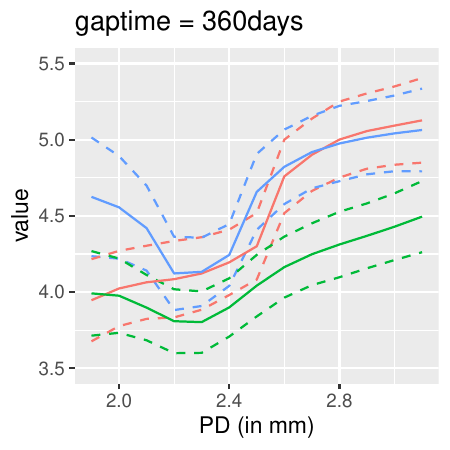}&
         \includegraphics[height= 5cm]{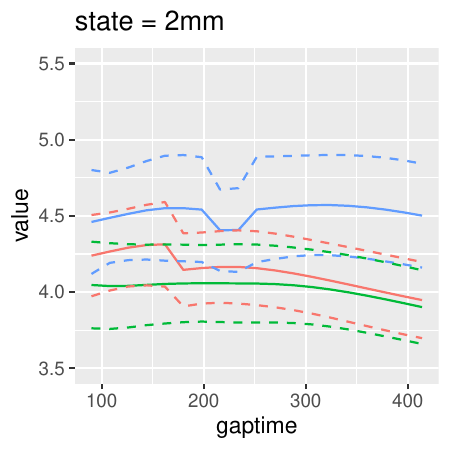}\\
	\includegraphics[height= 5cm]{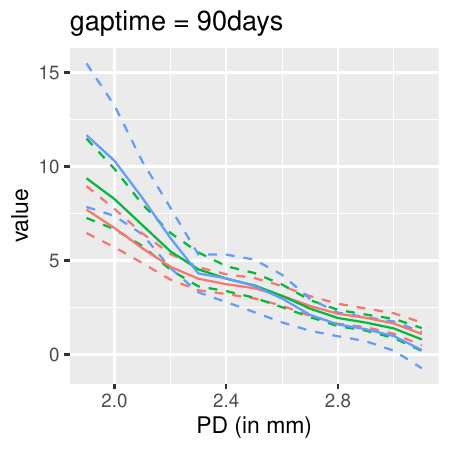}&
         \includegraphics[height= 5cm]{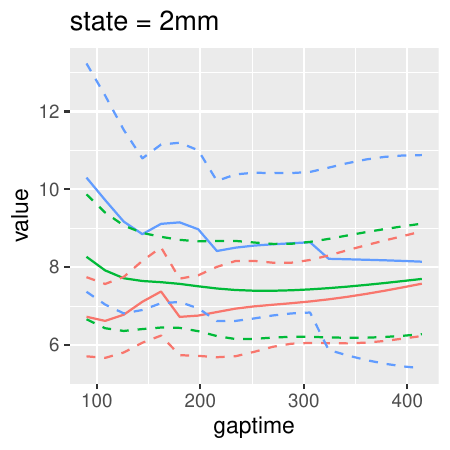}
 \end{tabular} 
 \caption{  \label{values_policies}	
Under the modulated method, 
the estimated value (solid line) and confidence intervals (dashed lines) of 
the observed fix policy (red line), 
the observed stochastic policy (green line),
and the estimated optimal policy (blue line) 
under the modulated method 
when the reward is defined as a PD reduction indicator (upper panels) and 
PD measurement (lower panels), respectively. 
 }
 \end{figure}

Furthermore, 
under the modulated method, we compared the confidence intervals of the value functions 
for these policies in cases where the value function is defined by cumulative reward and integrated reward, respectively. 
The estimated results are presented in Figure \ref{values_policies}.
Together with Figures \ref{policies}, 
we found that changes in the action assignment of the optimal policy, 
compared with observed policies,
indeed improve the value of the policy.
For example,
when the reward is defined as PD measurement,
the optimal policy suggests that patients with a gap time of  three months and PD values less than 2.7 mm may schedule their subsequent visit at six months or later, 
and correspondingly, 
the value of the optimal policy at a gap time $=$ 90 days and PD$<$2.2 mm is significantly higher than that of the two observed policies.

\bibliographystyle{ECA_jasa}
\bibliography{OPEwithODO.bib}

\end{document}